\documentclass[12pt,a4paper]{article}
\usepackage{latexsym}
\usepackage{fleqn}
\usepackage{epsf}
\newcommand{\al}{\alpha}
\newcommand{\Gm}{\Gamma}
\newcommand{\gm}{\gamma}

\newcommand{\dl}{\delta}
\newcommand{\Dl}{\Delta}

\newcommand{\veps}{\varepsilon}
\newcommand{\fie}{\varphi}

\newcommand{\lm}{\lambda}
\newcommand{\Lm}{\Lambda}
\newcommand{\sg}{\sigma}

\newcommand{\om}{\omega}
\newcommand{\Om}{\Omega}
\newcommand{\Ss}[1]{\mbox{$\cal #1$}}
\newcommand{\beq}{\begin{equation}}
\newcommand{\bea}{\begin{eqnarray}}
\newcommand{\beas}{\begin{eqnarray*}}
\newcommand{\eeq}{\end{equation}}
\newcommand{\eea}{\end{eqnarray}}
\newcommand{\eeas}{\end{eqnarray*}}
\newcommand{\ebar}{\bar{e}}
\newcommand{\etabar}{\bar{\eta}}
\newcommand{\psibar}{\bar{\psi}}
\newcommand{\sgbar}{\bar{\sg}}
\newcommand{\Ahat}{\hat{A}}

\newcommand{\What}{\hat{W}}
\newcommand{\chat}{\hat{c}}
\newcommand{\dhat}{\hat{d}}

\newcommand{\sphlap}{\widehat{\Dl}}
\newcommand{\Atil}{\tilde{A}}
\newcommand{\Ltil}{\tilde{L}}
\newcommand{\ltil}{\tilde{l}}
\newcommand{\tautil}{\tilde{\tau}}
\newcommand{\Wtil}{\tilde{W}}
\newcommand{\lmtil}{\tilde{\lm}}
\newcommand{\Jvec}{\vec{J}}
\newcommand{\Jkw}{\vec{J}^2}
\newcommand{\kvec}{\vec{k}}
\newcommand{\Kvec}{\vec{K}}
\newcommand{\Kkw}{\vec{K}^2}
\newcommand{\Lvec}{\vec{L}}
\newcommand{\Leen}{{\vec{L}_1}}
\newcommand{\Leenkw}{\vec{L}_1^2}
\newcommand{\Lkw}{\vec{L}^2}
\newcommand{\Ltw}{{\vec{L}_2}}
\newcommand{\Ltwkw}{\vec{L}_2^2}
\newcommand{\nvec}{\vec{n}}
\newcommand{\prvec}{\vec{\pr}}

\newcommand{\Skw}{\vec{S}^2}
\newcommand{\Svec}{\vec{S}}
\newcommand{\Tvec}{\vec{T}}
\newcommand{\hf}{\frac{1}{2}}
\newcommand{\hfe}{\frac{\veps}{2}}
\newcommand{\half}{\sfrac{1}{2}}
\newcommand{\halfje}{{1/2}}
\newcommand{\halfi}{\sfrac{i}{2}}
\newcommand{\halfe}{\sfrac{\veps}{2}}
\newcommand{\thalf}{\sfrac{3}{2}}
\newcommand{\fhalf}{\sfrac{5}{2}}
\newcommand{\shalf}{\sfrac{7}{2}}
\newcommand{\nhalf}{\sfrac{9}{2}}
\newcommand{\ehalf}{\sfrac{11}{2}}

\newcommand{\Lpi}{\sfrac{L}{2 \sqrt{\pi}}}
\newcommand{\Tpi}{\sfrac{T}{2 \sqrt{\pi}}}

\newcommand{\ad}{\mbox{\,ad\,}}

\newcommand{\mydim}{\mbox{\,dim\,}}
\newcommand{\Imag}{\mbox{\,Im\,}}
\newcommand{\Ker}{\mbox{\,Ker\,}}

\def\inv{\mathop{\rm inv}\nolimits}
\def\tr{\mathop{\rm tr}\nolimits}

\newcommand{\trees}[2]{\,\mbox{tr}_{#1} \left( #2 \right)}

\newcommand{\Vcl}{V_{\mbox{cl}}}

\newcommand{\cee}{{\small C}}
\newcommand{\form}{{\small FORM}}
\newcommand{\fort}{{\small FORTRAN}}
\newcommand{\mathem}{{\small MATHEMATICA}}

\newcommand{\binom}[2]{\left( \begin{array}{c} #1 \\ #2 \end{array} \right)}
\newcommand{\cl}{{\mbox{cl}}}
\newcommand{\ddt}{\frac{d}{dt}}

\newcommand{\drs}{{three-sphere}}
\newcommand{\dV}{d \cdot V}
\newcommand{\eff}{{\mbox{eff}}}
\newcommand{\fm}{\mbox{fm}}
\newcommand{\id}{{\unitmatrix{.6}}}
\newcommand{\ie}{{i.e.\ }}
\newcommand{\eg}{{e.g.\ }}

\newcommand{\MS}{{\small MS}}

\newcommand{\Next}{\nonumber \\ }
\newcommand{\pr}{\partial}
\newcommand{\Sdr}{{S$^3$}}
\newcommand{\SUtw}{{SU(2)}}
\newcommand{\sutw}{{su(2)}}
\newcommand{\EUC}{{\mbox{\scriptsize E}}}
\newcommand{\Skl}{{\mbox{\scriptsize S}}}

\newcommand{\terug}{\hspace{-7mm}}
\newcommand{\un}{\underline}
\newcommand{\unitmatrix}[1]{
        \thinlines
	\setlength{\unitlength}{#1mm}
	\begin{picture}(2.2,5)
	\put(0.7,0){\line(0,1){4.2}}
	\put(1.5,0){\line(0,1){5}}
	\bezier{15}(0.0,3.5)(0.75,4.25)(1.5,5.0)
	\put(0  ,0){\line(1,0){2.2}}
	\end{picture} }
\newcommand{\Order}[1]{\Ss{O}\left(#1\right)}
\newcommand{\real}{\relax{\rm I\kern-.18em R}}

\newcommand{\sfrac}[2]{\mbox{\large $\frac{#1}{#2}$}}
\newcommand{\wdrie}{\sqrt{3}}
\newfam\meuffam
\font\tenmeuf=eufm10
\font\sevenmeuf=eufm7
\font\fivemeuf=eufm5

\textfont\meuffam=\tenmeuf
\scriptfont\meuffam=\sevenmeuf
\scriptscriptfont\meuffam=\fivemeuf

\font\tenmsy=msbm10 

\newfam\msyfam
\textfont\msyfam=\tenmsy

\def\Bbb{\ifmmode\let\next\Bbb@\else
\def\next{\errmessage{Use \string\Bbb\space only in math mode}}\fi\next}
\def\Bbb@#1{{\Bbb@@{#1}}}
\def\Bbb@@#1{\fam\msyfam#1}

\begin{document}
\vskip-1cm
\hfill INLO-PUB-16/96
\vskip5mm
\begin{center}
{\LARGE{\bf{\underline{Glueballs on the \drs}}}}\\
\vspace*{5mm}
\vspace*{1cm}{\large Bas van den Heuvel} \\
\vspace*{1cm}
Instituut-Lorentz for Theoretical Physics,\\
University of Leiden, PO Box 9506,\\
NL-2300 RA Leiden, The Netherlands.\\ 
\end{center}
\vspace*{5mm}{\narrower\narrower{\noindent
\underline{Abstract:}  
We study the non-perturbative effects of the global features
of the configuration space for \SUtw\ gauge theory on the \drs.
The strategy is to reduce the full problem
to an effective theory for the dynamics of the low-energy modes.
By explicitly integrating out the high-energy modes, 
the one-loop correction to the effective hamiltonian is obtained.
Imposing the $\theta$ dependence
through boundary conditions in configuration space incorporates
the non-perturbative effects
of the non-contractable loops in the full configuration space.
After this we obtain the glueball spectrum of the 
effective theory with a variational method.
}\par}

\section{Introduction}
\label{ch-intro}

One of the methods to gain understanding of non-perturbative
phenomena in non-abelian gauge theory is by using the
finite-volume approach~\cite{lus2,baa2}.
The mechanism of asymptotic freedom~\cite{pol} renders the coupling constant
small at high energies or, equivalently, at small distances.
Taking the finite volume to be small results in a small coupling constant: 
one can use perturbation theory.
Gradually increasing the volume then allows one to study
the onset of non-perturbative phenomena and to see, hopefully,
the setting in of confinement.
In this regime one can employ 
a hamiltonian formulation~\cite{chr} of the problem.
The strategy in studying the dynamics of the Yang-Mills field
is based on two observations. First,
quantum field theory can in a sense be regarded as ordinary
quantum mechanics, but with an infinite number of degrees of freedom.
For the case of non-abelian gauge field theory, the configuration
space of this quantum mechanical problem is complicated, which
gives rise to non-perturbative behaviour.
Second, coming from the perturbative regime, these complications 
first show up in the low-energy modes of the gauge field: one can
capture the influence of the topology of the configuration space
by studying the effective theory of
the finite number of affected modes.
\par
Let us first reformulate quantum field theory as quantum mechanics.
In the hamiltonian picture we 
are dealing with wave functionals in 
the configuration space. Speaking generally, each point
in the configuration space corresponds to a configuration 
of the fields $\varphi(\vec{x})$ .
This means we 
have fields $\fie(\vec{x})$, conjugated momenta $\Pi(\vec{x})$,
a hamiltonian $H[\fie,\Pi]$ and wave functionals 
$\Psi[\fie]$. 
If we decompose the fields $\fie(\vec{x})$ in orthogonal modes,
we can reformulate the hamiltonian
problem in terms of the generalized Fourier coefficients $q_k$
and their conjugate momenta $p_k$. This leads to a hamiltonian
$H[q_k,p_k]$ and wave functionals $\Psi[q_k]$. A typical
hamiltonian would look like
\beq
 H[q_k,p_k] = \sum_k \left(p_k^2 + \om_k q_k^2 \right) 
 + \mbox{interaction terms}.
\eeq
\par
At small coupling, the modes will not interact and the 
potential will rise quadratically in all directions.
As a consequence, the wave functions will be the familiar 
harmonic oscillator wave functions.
One subsequently uses perturbation theory to take the interactions
between this infinity of modes into account.
What we have just described is the ordinary perturbative approach
to quantum field theory.
This perturbation scheme breaks down when at larger coupling
the behaviour of the wave function for some of the modes is dictated by 
the properties of the configuration space, as explained below.
We then have to replace these wave functions by functions that
have the behaviour required by the configuration space.
In this regime the hamiltonian method is superior to 
the covariant path integral approach of Feynman.
\par
An essential feature of the non-perturbative behaviour is that the
wave functional spreads out and becomes
sensitive to the global features of the configuration space.
This spreading out of the wave functional will occur first in 
those directions of the configuration space where
the potential energy is lowest, \ie in the direction of the
low-energy modes of the gauge field (small $\om_k$). 
If the configuration space has
non-contractable circles, the wave functionals are drastically
affected by the topology when the support of the wave
functional extends over the entire circle (\ie  bites in its own tail).
This is what typically happens in non-abelian gauge theories.
The Yang-Mills configuration space is the space of gauge orbits
$\cal{A}/\cal{G}$ ($\cal{A}$ is the collection of 
gauge fields or connections, $\cal{G}$ the group of local gauge transformations).
We know
from Singer~\cite{sin} that the topology of this configuration space
is highly non-trivial when \Ss{G} is non-abelian.  
The configuration space also has a Riemannian geometry~\cite{bab} 
that can be made explicit, once explicit coordinates are chosen on
$\cal{A}/\cal{G}$. This geometry also influences the wave
functionals, if the latter are no longer
localized within regions much smaller than the inverse curvature
of the field space.
\par
One of the most prominent properties of a non-abelian gauge theory, 
is the multiple vacuum structure.
This means that the gauge degrees of freedom cannot be completely 
removed~\cite{gri} by imposing a simple condition like the Coulomb gauge.
After gauge fixing there will remain Gribov copies, that is, 
gauge equivalent gauge field configurations that all satisfy
the gauge fixing condition. 
Intimately related to this
is the existence of instantons~\cite{bel}.
They can be shown to describe tunnelling 
between gauge copies of the vacuum, that is, tunnelling from 
one Gribov copy to another. The 
configuration sitting at the top of the tunnelling path 
with lowest energy barrier is called the sphaleron~\cite{kli}:
it is a saddle-point of the potential with one unstable mode.
Let us suppose that, coming from the vacuum, the potential energy in the
direction of the sphaleron(s) grows slowly. This
means that at increasing volume the wave functionals
around the different gauge copies of the vacuum
will start to flow out over the instanton barriers
and will develop a substantial overlap with each other.
We then have the situation described above of probing the
non-contractable loops in configuration space.
\par
\begin{figure}
\centering
\epsfxsize=0.8\textwidth
\leavevmode
\epsfbox{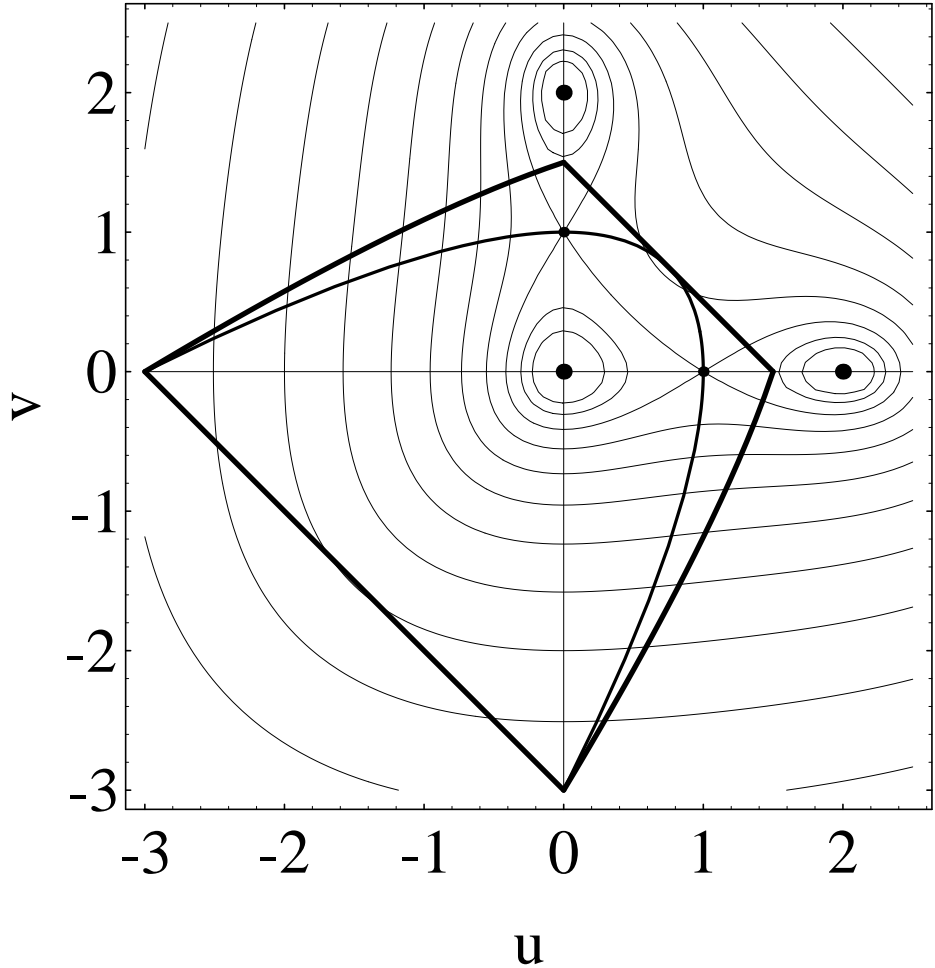} 
\caption{
The $(u,v)$ plane: a two-dimensional subspace of \Ss{A}.
Location of the classical vacua, sphalerons, lines of equal potential.
The boundary of the fundamental domain lies between the Gribov horizon 
(fat sections) and the
lower bound $\tilde{\Lm}$ (drawn parabola): $\tilde{\Lm} \subset
\Lm \subset \Om$, with $\Om$ the Gribov region.}
\label{uvplane}
\end{figure}

 
We impose gauge invariance by restricting the dynamics to a fundamental
domain, that is, a set of gauge fields that is in one-to-one correspondence
with the physical configuration space $\cal{A}/\cal{G}$~\cite{sem,zwan,baa6}.
Explicitly, we first define $\Lm'$ to be the set of gauge field 
configurations $A$ with the property that $|| ^g \! A|| \geq ||A||$
for all gauge transformations $g$. We will take this gauge field 
to represent the orbit on which it lies. One shows that $\Lm'$ is 
a convex subset of the set of transverse gauge fields, and
that $\Lm' \subset \Omega$, where $\Omega$ is the Gribov region which
is characterized by the condition that the Faddeev-Popov operator
is positive:  $\mbox{FP}(A) \geq 0$. 
As $\Lm'$ is a convex subset of a linear vector space, it is topologically
trivial. In order to get a one-to-one correspondence
between $\Lm'$ and $\cal{A}/\cal{G}$, certain points on the boundary of $\Lm'$
(these are the points where the minimum of the norm of the
gauge field on the orbit is degenerate) must be identified, after which
we denote $\Lm'$ by $\Lm$.
Only after these boundary identifications, $\Lm$ has
the non-trivial topological structure of the
physical configuration space.
If we restrict the dynamics to the fundamental domain $\Lm$,
the boundary identifications give rise to boundary conditions
on the wave functional.
See fig.~\ref{uvplane} to get an idea of 
how the potential landscape might look in relation to
the various spaces defined above. This figure displays 
a two-dimensional subspace
(consisting of low-energy modes) of the configuration space.
\par
The use of the hamiltonian formulation
is especially fruitful when the non-per\-tur\-ba\-tive effects 
manifest themselves appreciably only in a small number of the modes
of the gauge field. For this limited number of modes one defines an
effective hamiltonian that takes the perturbative
effect of all the other modes into account~\cite{lus2}. 
By studying the resulting ordinary quantum mechanical problem, one 
can go beyond purely perturbative results~\cite{baa2}.
\par
Apart from offering the possibility to go beyond perturbation theory, 
the finite-volume approach sketched above, which is sometimes called
'analytical', has the merit of being directly
comparable to the numerical results of lattice calculations.
For this the finite volume of the analytic approach must be chosen 
to be the finite volume of the lattice calculations.
Putting a theory on a finite lattice necessarily means 
that it is put in a finite volume with certain boundary conditions. 
For the standard lattice geometry, this finite volume is cubic with
periodic boundary conditions, \ie a torus.
\par
To make the reduction of an infinite to a finite number of degrees
of freedom, we use the Born-Oppenheimer approximation, which
has its origin in the quantum mechanical treatment of molecules.
In field theory one integrates out the fast modes from the path
integral and obtains an effective hamiltonian in a 
finite number of slow modes.
Using these methods in the intermediate-volume regime 
for \SUtw\ gauge theory
on the torus resulted 
in good agreement with the lattice results~\cite{baa2,kro}.
In case of periodic boundary conditions on the torus
 and already at small volumes, 
one should take into account non-perturbative effects
in the effective theory due to tunnelling through quantum induced barriers.
These barriers are contributions to the potential coming from
the integrating out of the fast modes.
This tunnelling
sets in much before the instanton effects play any role. 
Koller $\&$ Van Baal~\cite{baa2} have shown that one can derive 
very accurate results, beyond the semiclassical analysis, by imposing 
non-trivial boundary conditions on the wave functional at the 
corresponding sphalerons.
\par
However one would like to extend these results to
larger volumes in an attempt to get as close as possible to 
the confinement domain, and in this way get some intuition about how
confinement sets in. 
The first thing to do, as argued above, 
is to include contributions coming from instantons:
one wants to take into account the sphaleron directions in 
configuration space, because there the
potential will show the strongest deviation from gaussian behaviour.
This is not an easy task first of all because the instanton
solutions on a torus are only known numerically~\cite{gar}.
\par
To avoid this problem, we have shifted our attention to the
case where the finite volume is \Sdr, (the surface of) a \drs~\cite{baa1}. 
Here the instantons are known analytically, and the tunnelling paths
that connect gauge copies of the vacuum lie within the space of
low-energy modes.
Another simplification is that
the structure of the perturbative vacuum is much simpler than
in the case of the torus: there is no vacuum valley, that is, there is
no continuous set of minima of the classical potential.
A drawback of the \drs\ is that we lose the possibility to 
check against lattice results.
Also, the approach to infinite-volume results will go as powers of $1/R$
as compared to the exponential behaviour in $L$ for the torus~\cite{lus4}.
Thus, due to the intrinsic curvature of \Sdr, it will be even
less easy to derive infinite-volume results from our calculations.
Nevertheless, within this model one can study non-perturbative phenomena
like the effects of large gauge transformations and especially of the
$\theta$ angle on the glueball spectrum.
A gauge transformation is called 'large' if it cannot be continuously
deformed into the identity. Typically, two Gribov copies of the vacuum
are related by a large gauge transformation.
The $\theta$ angle, which will be properly defined in section~\ref{ch-eff},
is a free parameter of the theory
that one has to allow when one implements gauge invariance under
large gauge transformations.
The conditions at the boundary of the fundamental domain will
depend on the $\theta$ angle, thus incorporating
the non-perturbative effects of the non-contractable loops 
in the full configuration space.
Previous research into gauge theory on \Sdr~\cite{cut1} did not
include the study of the effects of large gauge transformations.
\par
This paper is set up as follows.
In section~\ref{ch-eff} we define the effective model.
In section~\ref{ch-oneloop} we explicitly integrate out
the high-energy modes using a background field method~\cite{heu2}.
This gives us the one-loop correction to the effective potential. 
In section~\ref{ch-RR} we construct a basis of functions
that respect the ($\theta$-dependent) boundary conditions. 
Using this basis we
perform a Rayleigh-Ritz analysis to approximate the spectra of both
the lowest order hamiltonian and of the one-loop corrected 
hamiltonian. Summaries of these results appeared in
respectively~\cite{heu1} and~\cite{heu2}.
From the eigenvalues one can obtain the masses of the 
different excitations (the glueballs) in this model.
Section~\ref{ch-concl} contains a discussion of the
results. We argue that the effect of the instantons is large,
but calculable. We determine the range of validity of the
effective model and show that the results are in 
reasonable agreement with results on the torus.

\section{The effective model}
\label{ch-eff}

\subsection{Introduction}

In this section we will define the effective model.
First, the necessary machinery needed
for the analysis on \Sdr\ is developed. 
In particular, we will write down
bases of functions for scalar and vector functions on \Sdr.
These bases of functions will be used in section~\ref{ch-oneloop}
for the evaluation of various functional determinants.
These functions will also allow us to isolate the space of low-energy modes
on which the effective theory will be defined.
After this we will familiarize ourselves with some
of the topological properties of \SUtw\ gauge theory on the
\drs\ and we show how the instanton degrees of 
freedom are embedded in the space of slow modes.
Then we will describe the derivation of the effective hamiltonian,
and discuss the validity of the adiabatic approximation.
The hamiltonian
itself need to be supplemented with suitable boundary conditions
in field space
to obtain a well-posed quantum mechanical problem. 
These conditions emerge when imposing gauge invariance
through restricting our problem to the fundamental domain.
We will use dynamical considerations to argue that we have enough
freedom to choose tractable boundary conditions.

\subsection{Low-energy modes}

We begin by introducing two framings on \Sdr. These will
lead to a basis of scalar functions.
By taking tensor products of these scalar modes
with eigenfunctions of the spin operator, we will construct
vector functions on \Sdr. By a similar method we will also
construct \sutw-valued functions, where \sutw\ denotes the
Lie algebra corresponding to the Lie group \SUtw.
After this we will write down the space of low-energy modes for 
\SUtw\ gauge theory on \Sdr.
\par
Let $\tau_a$ be the Pauli matrices.
We introduce the unit quaternions $\sg_\mu$ and their
conjugates $\sgbar_\mu = \sg^\dagger_\mu$ by
\beq
  \sg_\mu = ( \id ,i \vec{\tau}), \quad 
  \sgbar_\mu = ( \id,- i \vec{\tau}).
\eeq
They satisfy the multiplication rules
\beq
 \sg_\mu \sgbar_\nu = \eta^\al_{\mu \nu} \sg_\al, \quad
  \sgbar_\mu \sg_\nu = \etabar^\al_{\mu \nu} \sg_\al,
\eeq
where we used the 't Hooft $\eta$ symbols~\cite{tho}, generalized slightly to
include a component symmetric in $\mu$ and $\nu$ for $\al=0$.
The $\eta^i$ and $\etabar^i$ form bases for respectively the
self-dual and anti-self-dual four by four matrices.
The $\eta$ symbols enjoy the following useful relations
\beq
  \eta^\al_{\mu \nu} \eta^\beta_{\rho \nu} = 
  \eta^\nu_{\al \beta} \eta^\nu_{\mu \rho}, \quad
  \etabar^\al_{\nu \mu} \etabar^\beta_{\nu \rho} = 
  \etabar^\nu_{\al \beta} \etabar^\nu_{\mu \rho},\quad
  \eta^\al_{\beta \gm} = \etabar^\gm_{\al \beta}.
\eeq
We can use $\eta$ and $\etabar$ to define orthonormal framings of \Sdr,
which were motivated by the particularly simple form of the instanton 
vector potentials in these framings (see section~\ref{sec-inst}).
We embed \Sdr\ in $\real^4$ by considering the unit sphere parametrized by a 
unit vector $n_\mu$. Using the scale-invariance of the classical
hamiltonian, we can make the restriction to a sphere of radius $R=1$. 
The $R$~dependence can be reinstated on dimensional grounds.
The framing for \Sdr\ is 
obtained from the framing of $\real^4$ by restricting in the following 
equation the four-index $\al$ to a three-index $a$
(for $\al = 0$ one obtains the normal $n_\mu$ on \Sdr):
\beq
  e^\al_\mu = \eta^\al_{\mu \nu} n_\nu , \quad
  \ebar^\al_\mu = \bar{\eta}^\al_{\mu \nu} n_\nu.
  \label{framedef}
\eeq
The orthogonal matrix $V$ that relates these two frames is given by
\beq
  V^i_j = \ebar^i_\mu e^j_\mu = \half 
  \tr ( (n \cdot \sg) \sg_i (n \cdot \bar{\sg})\sg_j ).
  \label{Vdef}
\eeq
Note that $e$ and $\bar{e}$ have opposite orientations.
The parity operation \Ss{P} defined by
\beq
  \Ss{P}: (n_0, \vec{n}) \rightarrow (n_0, - \vec{n}),
  \label{pardef}
\eeq
interchanges the $e$ and the $\ebar$ framing.
\par
A vector field $A_\mu$ can be written with respect
to either framing~(\ref{framedef}). From now on we will, unless
stated otherwise, use the framing $e^i_\mu$ and write
\beq
  A_\mu = A_i e^i_\mu.
\eeq
Since we want vector fields to be tangent to \Sdr\, 
we have that $n_\mu A_\mu = 0$ and the sum over $i$ runs from 1 to 3.
Latin indices refer to this framing, while Greek indices will refer
to the embedding space $\real^4$.
\par
Each of the framings~(\ref{framedef}) defines a differential operator
\beq
  \pr^i = e^i_\mu \frac{\pr}{\pr x^\mu} , \quad
  \bar{\pr}^i = \bar{e}^i_\mu \frac{\pr}{\pr x^\mu},
\eeq
to which belong \sutw\ angular momentum operators, which for
historical reasons will be denoted by $\Leen$ and $\Ltw$:
\beq
  L_1^i = \halfi\,\pr^i , \quad
  L_2^i = \halfi\,\bar{\pr}^i.
\eeq
They are easily seen to commute and to satisfy the condition 
\beq
  \Lkw \equiv \Leenkw = \Ltwkw.
\eeq
\par
The manifold \Sdr\ is isomorphic to the Lie group \SUtw.
The spatial symmetries of \Sdr\ correspond to the 
left and right action of \SUtw\ on itself: 
$\mbox{so}(4) = \mbox{su}(2) \oplus \mbox{su}(2)$.
In appendix~\ref{appbasis} we show that $\Leen$ and $\Ltw$ 
correspond to the generators of the left and right symmetries on \SUtw.
\par
A basis of scalar functions on \Sdr\ is given by the 
set of eigenfunctions of the commuting operators 
$\{ \Lkw, {L_1}_z, {L_2}_z \}$:
\bea
 |l\, m_L\, m_R \rangle, 
 \quad \quad l=0,\half,1,\ldots,
 \quad m_L, m_R = -l, \ldots,l.
  \label{scalbasis}
\eea
Since the laplacian on \Sdr\ is given by $\pr_i \pr_i = - 4 \Lkw$,
these modes are eigenfunctions
of the spherical laplacian with eigenvalue $-4 l (l+1)$ 
and degeneracy $(2 l +1)^2$. See appendix~\ref{appbasis}
for other ways to arrive at this basis.
\par
To classify vector fields $A_i$ on \Sdr, we introduce a spin operator
$\Svec$ by $ (S^a A)_i = - i \veps_{a i j} A_j$.
This is an \sutw\ angular momentum operator that 
commutes with $\Leen$ and $\Ltw$.
We also define $\Kvec = \Lvec + \Svec$, which is shorthand for
$\Kvec_{i j} = \Lvec \,\dl_{i j} + \Svec_{i j}$.
When there is no confusion possible, we will write $\Lvec$ for
$\Leen$. The fact that we single out $\Leen$ is related
to the fact that we chose $e^i_\mu$ as the preferred framing
for expressing vector fields.
A basis of vector fields can be obtained by taking tensor products
of the scalar functions~(\ref{scalbasis}) with the basis functions
$| 1 \, m_s \rangle$ of the spin operator. In the standard way one
obtains eigenfunctions of the set $\{\Kvec,\Ltw\}$:
\beq
  | l\,m_R;k\,m_k \rangle,\quad \quad
  l=0,\half,1,\ldots,\quad k= |l-1|,\ldots,l+1.
  \label{vecbasis}
\eeq
The three modes with $(l,k) = (0,1)$ are the three vector fields $e^i_\mu$.
The three vector fields $\ebar^j_\mu = V^j_i e^i_\mu$ correspond to
the modes with $(l,k) = (1,0)$, as can be seen from
$L_1^k V^j_i = i \veps_{k i m} V^j_m$.
To identify transversal and longitudinal modes, first note that
$\pr_\mu A_\mu = \pr_i A_i$. Next we introduce the operator 
$Q$ via $Q_{i j} = L_i L_j$.
Note that $QA = 0$ for $A$ in the Coulomb gauge.
Using the identity
\beq
  \left(\Lvec \cdot \Svec \right)^2 = \Lkw - \Lvec \cdot \Svec - Q,
  \label{Qprop}
\eeq
one concludes that the modes with $k=l \pm 1$ are
transverse, whereas the modes with $k=l$ are purely longitudinal.
\par
To deal with \sutw-valued functions, 
we introduce the isospin operator $\Tvec$ by $T^a = \ad(\hf \tau_a)$,
where $\ad(X)(Y) \equiv [X,Y]$.
This operator is yet another \sutw\ angular momentum operator that 
commutes with all operators defined before.
By taking tensor products of the functions obtained so far with the basis functions $| 1 \, m_t \rangle$ of $\Tvec$, we can
construct scalar and vector modes that take their values in \sutw.
Introducing $\Jvec = \Lvec + \Tvec$ we obtain for the \sutw-valued
scalar functions (\eg infinitesimal gauge transformations on \Sdr)
\beq
  | l\,m_R;j\,m_j \rangle,\quad \quad
  l=0,\half,1,\ldots,\quad j= |l-1|,\ldots,l+1.
  \label{suscalbasis}
\eeq
For the vector functions (\eg gauge fields on \Sdr)
we define $\Jvec = \Kvec + \Tvec$ and obtain
\bea
 | l\,m_R;k;j\,m_j \rangle,&& \quad l=0,\half,1,\ldots,
  \quad k= |l-1|,\ldots,l+1, \Next
&& \quad j= |k-1|,\ldots,k+1.
  \label{suvecbasis}
\eea
Generalizing to the case where $\Svec$ and $\Tvec$ correspond to
respectively a spin-$s$ and a spin-$t$ representation, we
obtain (suppressing the $m_R$ dependence)
\bea
 | t;(l s)k;j\,m_j \rangle, && \quad l=0,\half,1,\ldots,
  \quad k= |l-s|,\ldots,l+s, \Next
&& \quad j= |k-t|,\ldots,k+t.
  \label{genbasis}
\eea
The scalar modes are recovered by the choice $s=0$, $t=1$,
the vector modes correspond to the choice $s=1$, $t=1$.
\par
An \SUtw\ gauge field on the \drs\ can be written as
\beq
   A_\mu= A_i e_\mu^i = A^a_i e_\mu^i \frac{\sg_a}{2}.
\eeq 
The field strength is given by
$F_{i j} = \pr_i A_j - \pr_j A_i + [A_i,A_j] - 2 \veps_{i j k} A_k$,
where the last term is the so-called spin-connection:
it is a consequence of the fact that we used a framing on
the curved manifold \Sdr. Reinstating the radius $R$
would lead to a factor $1/R$ for this last term, which shows that
this term vanishes in the limit of no curvature, \ie for $R \rightarrow
\infty$.
A gauge transformation 
$g: \mbox{S}^3 \times \real \rightarrow \mbox{SU}(2)$
acts as follows:
\beq
  \left(^g \! A\right)_i = g^{-1} A_i g + g^{-1} \pr_i g = g^{-1} D_i(A) g,
\eeq
where $D(A)$ denotes the covariant derivative in the fundamental 
representation. 
As usual, the field strength $F$ transform under the adjoint representation:
$(^g \! F)_{i j} = g^{-1} F_{i j} g$.
In order to isolate the lowest energy levels, we examine
the potential energy
\beq
 V(A) =  - \frac{1}{2 \pi^2} \int_{\Skl^3} \half \tr(F_{ij}^2).
 \label{VAdef}
\eeq
Note the factor $2 \pi^2$ that we absorbed in the potential.
We define the quadratic fluctuation operator \Ss{M} by
\beq
  V(A) = - \frac{1}{2 \pi^2} \int_{\Skl^3} \tr(A_i \Ss{M}_{ij} A_j) 
   + \Order{A^3}. \label{fluctdef}
\eeq
This gives
\beq
  \Ss{M}_{ij} = 2 \Lkw \dl_{ij} + 2 \Kkw_{ij} - 4 Q_{ij}.
  \label{fluct}
\eeq
Using $\Skw = 2$ and eq.~(\ref{Qprop}) one rewrites
\beq
\Ss{M}_{ij} = \left( \Kkw - \Lkw \right)^2_{ij}.
\eeq
Now focus on the modes~(\ref{suvecbasis}).
For $l=0$, we have $k=1$ and $\Ss{M} = 4$. 
The $9$ eigenmodes are $A^a_i = c^a_i$ with $c^a_i$ constant.
For $l=\frac{1}{2}$, the modes with $k=\frac{1}{2}$ are pure gauge 
($\Ss{M} = 0$ and $Q \neq 0$), whereas the modes with $k = \frac{3}{2}$ are
physical with $\Ss{M} = 9$.
For $l \geq 1$, the modes with $k = l-1$ have $\Ss{M} = (2 l)^2$,
the modes with $k = l+1$ have $\Ss{M} = (2 (l +1))^2$ and the modes
with $k=l$ are again pure gauge.
In particular the 9 modes with $l=1,\,k=0$ have $\Ss{M} = 4$ and are
given by $A^a_i = d^a_m V^m_i$ with $d^a_m$ constant. The lowest
eigenspace of \Ss{M} is thus 18 dimensional and given by
\beq
    A_\mu(c,d) = 
  \left(c^a_i + d^a_j V^j_i \right) e^i_\mu \frac{\sg_a}{2} =
  \left(c^a_i  e^i_\mu+d^a_j\bar{e}^j_\mu \right) 
  \frac{\sg_a}{2}.
\label{Bcddef}
\eeq
This is the space on which we will build the effective theory.

\subsection{Winding numbers and instantons}
\label{sec-inst}

Before writing down the instantons, we first
introduce the notions of winding number and topological charge.
The reader is referred to~\cite{jac} for more details.
\par
In the $A_0 = 0$ gauge, the set of gauge transformations 
\Ss{G} consists of mappings
$g: \mbox{S}^3 \rightarrow \mbox{SU}(2) \cong \mbox{S}^3$.
As a consequence, \Ss{G} splits up in distinct classes
that can be labelled by the winding number of the gauge
transformations in that class:
\beq
n[g] = \frac{-1}{ 24 \pi^2} \int_{\Skl^3} \veps_{i j k} 
\tr \left( (g^{-1} \pr_i g) (g^{-1} \pr_j g) (g^{-1} \pr_k g) \right).
\eeq
Gauge transformations with zero winding number can be continuously
deformed into the identity mapping, whereas mappings with non-zero
winding number cannot. The latter mappings are sometimes referred to
as 'large' or 'homotopically non-trivial' gauge transformations.
\par
Related to this, we define the standard Chern-Simons functional
that measures the topological charge of a gauge
field configuration:
\beq
Q[A] = \frac{1}{16 \pi^2} \int_{\Skl^3}  \veps_{i j k} 
\tr \left( A_i F_{j k} - \sfrac{2}{3} A_i A_j A_k \right).
\eeq
The sign of $n[g]$ was chosen such that we have
$Q[^g \! A] = Q[A] + n[g]$.
\par
In the hamiltonian picture, gauge invariance is implemented
through Gauss' law, which implies invariance of the
wave functional under small gauge transformations.
For large gauge transformations,
we have to allow for a possible $\theta$ angle:
\beq
\psi[^g \!A] = e^{ i n[g] \theta} \psi[A].
\eeq
\par
The (anti-)instantons~\cite{bel} in the framings~(\ref{framedef}), 
obtained from those on $\real^4$ by interpreting the radius in 
$\real^4$ as the
exponential of the time $t$ in the geometry $\mbox{S}^3 \times \real$, become
($\vec\veps$ and $\vec A$ are defined with respect to the 
framing $e^a_\mu$ for instantons and with respect to the framing
$\bar{e}^a_\mu$ for anti-instantons)
\beq
  A_0 = \frac{\vec{\veps} \cdot \vec{\sg}}{2 ( 1 + \veps \cdot n )}
  , \hspace{1.5cm}
  A_a = \frac{\vec{\sg} \wedge \vec{\veps} -( u + \veps \cdot n ) 
  \vec{\sg}} {2 ( 1 + \veps \cdot n )},\label{vecA}
\eeq
where 
\beq
   u = \frac{ 2 s^2}{1 + b^2 + s^2} , \hspace{1.5cm}
   \veps _\mu = \frac{2 s b_\mu}{1 + b^2 + s^2} , \hspace{1.5cm}
   s = \lm e^t.
\eeq
The instanton describes tunnelling from $A = 0$ (Q=0) at $t = - \infty$ to
$A_a = - \sg_a$ (Q=1) at $t = \infty$, over a potential barrier that is
lowest when $b_\mu \equiv 0$. This configuration (with $b_\mu = 0$, $u = 1$)
corresponds to a sphaleron~\cite{kli}, i.e.\ the vector potential $A_a = 
\frac{- \sg_a}{2}$
is a saddle point of the energy functional with one unstable mode, 
corresponding to the direction ($u$) of tunnelling. At $t = \infty$, 
$A_a = - \sg_a$ has zero energy and is a gauge copy of $A_a = 0$ by a
gauge transformation $g = n \cdot \sgbar$ with winding number one, since
\beq
  n \cdot \sg \pr_a n \cdot \sgbar = - \sg_a.
\eeq
\par
We will be concentrating our attention to the $c$ and $d$ modes
of eq.~(\ref{Bcddef}). These modes are degenerate
in energy to lowest order with the modes that describe tunnelling
through the sphaleron and "anti-sphaleron". The latter corresponds
to the configuration with the minimal barrier height separating $A = 0$
from its gauge copy by a gauge transformation $g = n \cdot \sg$
with winding number $-1$. The anti-sphaleron is actually a copy of the
sphaleron under this gauge transformation, as can be seen from 
eq.~(\ref{vecA}), since
\beq
  n \cdot \sgbar e^a_\mu \sg_a n\cdot\sg = - \bar{e}^a_\mu \sg_a.
\eeq
The two-dimensional space containing the tunnelling paths through the
sphalerons is consequently parametrized by $u$ and $v$ through
\bea
  A_\mu(u,v)&=&\left(-u e^a_\mu-v\bar{e}^a_\mu \right)\frac{\sg_a}{2}
  =A_i(u,v)e^i_\mu,\\
  A_i(u,v)&=&\left(-u\dl^a_i-v V^a_i\right)\frac{\sg_a}{2}=
  -u\frac{\sg_i}{2}+v \, n\cdot\bar{\sg}\frac{\sg_i}{2}n\cdot\sg.
\eea
The gauge transformation with winding number $-1$ is easily seen to
map $(u,v) = (w,0)$ into $(u,v) = (0,2-w)$. In particular,
as discussed above, it maps the sphaleron $(1,0)$ to the
anti-sphaleron $(0,1)$. 
Comparing with eq.~(\ref{Bcddef}) shows that we obtain the 
$(u,v)$ space from the $(c,d)$ space by the choice 
$c^a_i = - u \dl^a_i$
and $d^a_i = - v \dl^a_i$.

\subsection{Reduction to a quantum mechanical problem}

Let us start with the naive derivation of the hamiltonian for
the $(c,d)$ modes. From the lagrangian
\beq
  L = - \frac{1}{4 g^2} \int_{\Skl^3} F^a_{\mu\nu} F^{a\mu\nu} 
    = \frac{1}{2 g^2} \int_{\Skl^3} F^a_{0i} F^a_{0i} - \frac{2 \pi^2}{g^2} V(A),
\eeq
we obtain for the $(c,d)$ space
\beq
  L = \frac{2 \pi^2}{2 g^2}
    \left( \dot{c}^a_i \dot{c}^a_i +\dot{d}^a_i \dot{d}^a_i \right) - 
    \frac{2 \pi^2}{g^2} V(c,d).
\eeq
This leads to the hamiltonian
\beq
   \Ss{H} = -\frac{f}{2} \left( 
  \frac{\pr^2}{\pr c^a_i \pr c^a_i} + \frac{\pr^2}{\pr d^a_i \pr d^a_i} 
   \right) + \frac{1}{f} V(c,d), \quad f = \frac{g^2}{2 \pi^2},
\eeq
where the potential $V(c,d) = \Vcl(c,d)$ can be obtained from
eq.~(\ref{VAdef}):
\bea
  V(c,d) &=& V(c) + V(d) + \sfrac{1}{3}
   ( \tr(X) \tr(Y) - \tr(X Y) ), \label{Vcddef} \\
  V(c) &=& 2 \tr(X) + 6 \det c + 
  \sfrac{1}{4} \left( \mbox{tr}^2(X) - \tr(X^2) \right),
\eea
with  the symmetric matrices $X= c c^T$ and $Y = d d^T$.
\par
The correct way to obtain the effective hamiltonian for the
$(c,d)$ modes must start from the full theory.
In the case where the finite volume is the three-torus all
zero momentum states become degenerate with the ground state
at $g=0$. This infinite degeneracy allows one to derive an effective 
hamiltonian for these states using degenerate hamiltonian perturbation
theory starting from the full hamiltonian $\widehat{\Ss{H}}$~\cite{lus2}.
This full hamiltonian in the Coulomb gauge is given in~\cite{chr}.
Proceeding to higher order in this way is complicated, mainly due
to the non-abelian Coulomb Green function that occurs in the 
kinetic part of the hamiltonian, but it can be done~\cite{baa7}.
\par
For the case of the \drs\ however, the ground state at $g=0$ is 
non-degenerate. The $(c,d)$ modes are in this respect not the
analogue of the zero momentum modes on the torus, but are singled
out just because of the fact that they are the slow modes of the
system. To calculate the effective hamiltonian for these modes,
we compute the effective action for the $(c,d)$ degrees of freedom using
a background field method. This explicit calculation will be 
performed in section~\ref{ch-oneloop}.
\par
We will use the hamiltonian formulation of the full problem 
for the discussion of the validity of restricting ourselves 
to the $(c,d)$ space.
Let $x$ denote the 18 $(c,d)$ modes 
and $q$  all the transverse modes orthogonal to the $(c,d)$ space.
We use $p_x$ and $p_q$ to denote the conjugate momenta. The full
hamiltonian in the Coulomb gauge~\cite{chr} is a function of
$x$, $p_x$, $q$ and $p_q$. We define 
\beq
  \Ss{H}_{[x]} = \widehat{\Ss{H}}(x,p_x = 0,q,p_q).
\eeq
Consider the following decomposition of the full wave function
\beq
  \Psi =  \sum_{n=1}^\infty \fie^{(n)}(x) \chi_{[x]}^{(n)}(q),
  \label{decomp1}
\eeq
where the functions $| n \rangle = \chi_{[x]}^{(n)}(q)$ are chosen to be
eigenstates of $\Ss{H}_{[x]}$:
\beq
   \Ss{H}_{[x]} \chi_{[x]}^{(n)}(q) = V^{(n)}(x) \chi_{[x]}^{(n)}(q).
\eeq
The Schr\"{o}dinger equation $\widehat{\Ss{H}} \Psi = E \Psi$
is equivalent to the following set of equations:
\beq
 \langle n | \widehat{\Ss{H}}(x,p_x,q,p_q) 
  \left( \sum_{m=1}^\infty \fie^{(m)}(x) | m \rangle \right)
  =  E \fie^{(n)}(x) \quad \forall n.
 \label{schrod1}
\eeq
\par
As an example we assume the following form for the full hamiltonian,
\beq
  \widehat{\Ss{H}} = -\frac{f}{2} \frac{\pr^2}{\pr x^2_i}
  + \frac{1}{f} \Vcl(x) + \Ss{H}_{[x]}(q,p_q).
\eeq
This would be the form of the full hamiltonian if we neglect the
complications of the non-abelian Coulomb Green function.
The Schr\"{o}dinger equation for this case leads to
\beq
 -\frac{f}{2} \sum_m  \vec{\nabla}^2_{n m} \fie^{(m)}(x)
 + ( \frac{1}{f} \Vcl(x) + V^{(n)}(x) ) \fie^{(n)}(x) 
  =  E \fie^{(n)}(x) \quad \forall n.
 \label{schrod2}
\eeq
Here we introduced the covariant derivative $\nabla$ by
\beq
  \nabla^i_{n m} = \frac{\pr}{\pr x_i} \dl_{n m} 
     +   \langle n | \frac{\pr}{\pr x_i} | m \rangle 
   = \frac{\pr}{\pr x_i} \dl_{n m} - A^i_{n m}.
\eeq
The adiabatic approximation consists of truncating this equation
to $\fie^{(1)}$, that is, we assume the transverse wave function 
to be in its ground state and we assume this ground state 
to decouple dynamically from the excited states. 
This approximation is valid if either the coefficients $A^i_{n m}$
are small, or if we are in the region where the energy difference
$V^{(2)}(x) - V^{(1)}(x)$ is large compared to the
excitation energies of the states we are interested in. 
\par 
Returning to the general problem, the equations for the functions
$\fie^{(n)}(x)$ will not just contain the functions $A_{n m}$ 
defined above, but more general matrix elements of the form
$\langle n | f(p_x,q,p_q) | m \rangle$.
Still, if the excitation energies are small compared to
$V^{(2)}(x) - V^{(1)}(x)$, the higher $\fie$'s will be small
and we can restrict eq.~(\ref{schrod1}) to $\fie^{(1)}$.

\subsection{Boundary conditions}

We impose gauge invariance by restricting the dynamics to the fundamental
domain $\Lm$. The configuration space of our effective model consists
of the intersection of $\Lm$ with the space of low-energy modes, the
$(c,d)$ space. This space was investigated in~\cite{baa6}, and the
information obtained there on the boundary of the fundamental
domain will be sufficient for our present purposes. To see this
focus on the two-dimensional $(u,v)$ plane (fig~\ref{uvplane}), 
which contains the tunnelling paths.
We have to provide boundary conditions at the
boundary of the fundamental domain to obtain a well-defined
quantum mechanical problem. 
At weak coupling, the potential energy at the boundary of the
fundamental domain is
higher than the energy $E$ of the wave function: the
wave function is localized around the perturbative vacuum $c=d=0$
and the boundary conditions are not felt. 
\par
Increasing the coupling
results in the spreading of the wave function over the configuration
space. We are interested in the regime where $E$ is of the order
of the sphaleron energy: we will have
a substantial flow of the wave function over the instanton barrier, 
but at the rest of the boundary the potential is still much higher than
$E$. This means that at most parts of the boundary, the
wave function will have decayed exponentially before reaching it.
As a consequence, the boundary conditions imposed there
will not have a large effect on the spectrum.
By the same token,
the precise location of the boundary in these regions is not
important either. This gives us the freedom to choose tractable
boundary conditions.
At the sphalerons, however, the boundary conditions are fixed.
Since the gauge transformation connecting the two sphalerons has
winding number one, we have to set
\beq
  \Psi(A(\mbox{Sph},0)) = e^{i \theta} \Psi(A(0,\mbox{Sph})),
\label{oerbc}
\eeq
thus introducing the $\theta$ angle.
\par
We define radial coordinates $r_c$ and $r_d$
by
\beq
  r_c = \left[ c^a_i c^a_i \right]^{\hf}, \quad 
  r_d = \left[ d^a_i d^a_i \right]^{\hf}.
\eeq
The sphaleron has radial coordinates $(\sqrt 3,0)$ and angular coordinates
$\chat^a_i = - \dl^a_i$ (with $\chat^a_i = c^a_i/r_c$). It will be connected
with the anti-sphaleron at $(0,\sqrt 3)$.
We will restrict the $(r_c,r_d)$ plane by $r_c < \sqrt{3},~r_d < \sqrt{3}$
and impose boundary conditions at the edges.
When we come to the variational calculation, we will use
basis functions of the form $\phi(r_c,r_d) Y(\chat,\dhat)$.
As argued above, for values of the coupling constant at 
which our approximation will
be valid, only the effect of the boundary conditions at the sphaleron
will be felt. By imposing boundary conditions in the $(r_c,r_d)$ plane 
we pair up two submanifolds, of which only the sphaleron/anti-sphaleron
need belong to the boundary of the fundamental domain.
\par
Consider the following decomposition of the full wave function
\beq
  \Psi = \frac{1}{r_c^4 r_d^4} \sum_{n=1}^\infty 
  \psi^{(n)}(c,d) \chi_{[c,d]}^{(n)}(q),
\eeq
where $q$ denotes all the modes orthogonal to the $c$ and $d$ modes.
This is just eq.~(\ref{decomp1}) with an extra factor $r_c^4 r_d^4$
extracted for technical reasons (\ie $\fie^{(n)} \leftrightarrow
\psi^{(n)} r_c^{-4} r_d^{-4}$).
Under the adiabatic approximation,
we obtain a hamiltonian for $\psi = \psi^{(1)}$ given by
\beq
  \Ss{H} = -\frac{f}{2} \left( 
  \frac{\pr^2}{\pr r_c^2} + \frac{\pr^2}{\pr r_d^2} -
  12 \left(\frac{1}{r_c^2} + \frac{1}{r_d^2}\right) +
  \frac{1}{r_c^2} \Dl_{\chat} + \frac{1}{r_d^2} \Dl_{\dhat}
   \right) + \Ss{V}(c,d), \label{hamrad}
\eeq
with $\Dl_{\chat}$ the laplacian in the angular coordinates.
The extra term $V^{(1)}(c,d)$ that one might expect is contained
within $\Ss{V}(c,d)$.
\par
The boundary condition on $\psi$ follows directly from
eq.~(\ref{oerbc}),
but care must be taken when imposing the condition on the normal 
derivative of $\psi$.
Matching along the sphaleron path across the boundary
of the fundamental domain, we need to compensate for the curvature
with the appropriate jacobian factor. We will make this more
precise.
\par
Let us focus on the tunnelling path $c^a_i = -u \, \dl^a_i$. Note
that this path is equivalent to all paths $c^a_i = -u \, S(\vec{\al})^a_i$,
with $S$ an orthogonal matrix, due to the residual gauge symmetry.
We will first remove this gauge symmetry to obtain a genuine
one-dimensional tunnelling parameter. Introduce the following 
decomposition for $c$:
\beq
  c = S(\al_p) H(h_i), 
  \quad S \in \mbox{SO}(3),~H^T = H.
\eeq
If we write
\beq
  H = \sum_{i=0}^5 h_i H_i,
\eeq
with $\tr (H_i H_j) = \dl_{i j}$ and with $H_0 \propto \id$,
then $h_0$ will play the role of the gauge invariant tunnelling parameter.
The sphaleron is located at $h_0 = \sqrt{3}$, $h_i = 0,~(i=1,\ldots,5)$.
\par
The isolation of the 18 $(c,d)$ modes is appropriate 
close to the perturbative vacuum, because, as we have seen,
they are the slow modes of the theory. 
Close to the sphaleron however,
the only slow mode is the tunnelling mode $w = h_0/ \sqrt{3}$. 
To derive the correct boundary
conditions, we should reduce the dynamics around the sphaleron
to a one-dimensional tunnelling problem, by integrating out all other modes. 
To this end, we consider the following decomposition of $\Psi$:
\beq
\Psi = \sum_n \fie^{(n)} (w)  \chi^{(n)}_{[w]}(q),
\eeq
where $q$ not only denotes the non-$(c,d)$ modes, but also the $d$ modes
and the $h_i$ modes with $i=1,\dots,5$. Note that $\Psi$ does
not depend on the gauge degrees of freedom $\al_p$.
When imposing the boundary conditions, we want to relate the wave 
function at the sphaleron to the one at the anti-sphaleron.
Consider the tranverse modes $q$ within the $(c,d)$ space 
at the sphaleron. The gauge transformation that maps the
sphaleron to the anti-sphaleron will
{\it not} map these modes on modes within the
$(c,d)$ space at the anti-sphaleron: transverse modes inside and outside
the $(c,d)$ space get mixed under the
gauge transformation. 
The transverse wave function at the sphaleron restricted to the
$(c,d)$ modes does therefore not fit naturally
to this transverse wave function at the anti-sphaleron.
The full transverse wave functions do however map to each other.
Another manifestation of this symmetry is the fact that
the effective potential is symmetric around the sphaleron
only when {\it all} transverse modes are integrated out. 
\par
Before going to the effective hamiltonian for $\fie^{(1)}(w)$ we must
remove the residual gauge freedom.
The gauge invariant wave function $\Psi$ is independent of the
coordinates $\al_p$.
We will show that only after a suitable rescaling of the wave function,
the laplacian takes its cartesian form with respect to the
tunnelling parameter $w$. The boundary conditions at the sphaleron
must be imposed on this rescaled wave function.
\par
Consider the laplacian for the $c$ modes.
If we write $u_\mu = (h_0, \ldots, h_5,\al_1,\al_2,\al_3)$ 
the laplacian takes the form
\beq
\Dl \Psi = \frac{1}{J} \pr_\mu \left(J g^{ \mu \nu} \pr_\nu \Psi \right).
\label{Lapldef}
\eeq
Here $J = {\det}^{\halfje}(g)$ and $g$ is the metric which can be
obtained from
$d \sg^2 = \tr (\dot{c} \dot{c}) = g_{\mu \nu} \dot{u}_\mu \dot{u}_\nu$.
For $\Psi$ independent of $\vec{\al}$, one can show
that at the sphaleron path eq.~(\ref{Lapldef})
reduces to
\beq
\Dl \Psi = \frac{1}{h_0^3} \pr_i
\left( h_0^3 \pr_i \Psi \right).
\eeq
To obtain an ordinary one-dimensional tunnelling problem, 
we must rescale the wave function by $w^{3/2} \propto h_0^{3/2}$:
\beq
  \fie(w) = w^{-\frac{3}{2}} \tilde{\fie}(w).
\eeq
The tunnelling problem then looks as
\beq
\left( - \frac{f}{2} \frac{\pr^2}{\pr w^2} + \tilde{V}(w) \right) 
  \tilde{\fie}(w) = E \tilde{\fie}(w).
\eeq
We can derive a similar equation for a function $\tilde{\fie}'$ of the
parameter $w'$ corresponding to tunnelling through the
anti-sphaleron. When matching these tunnelling paths through the
relation $w' = 2 - w$, 
boundary conditions must be imposed on $\tilde{\fie}$ and on its
normal derivative at the sphaleron $w=1$. Combining the factors $w^{3/2}$
and $r^{-4}$ leads to the following boundary conditions on $\psi$:
\bea
  \psi(\mbox{Sph},0) &=& e^{i \theta} \psi(0,\mbox{Sph}),
  \label{bcgen1} \\
  \frac{\pr (r_c^{-\frac{5}{2}}\psi)}{\pr r_c}(\mbox{Sph},0) &=& 
    - e^{i \theta} \frac{\pr (r_d^{-\frac{5}{2}}\psi)}{\pr r_d}(0,\mbox{Sph}).
  \label{bcgen2}
\eea

\section{The one-loop effective lagrangian}
\label{ch-oneloop}

\subsection{Introduction}
In this section we will calculate the influence of the
high-energy modes on the dynamics of the low-energy modes~\cite{heu2}.
This is achieved by integrating out the high-energy modes in
the path integral, which gives us the one-loop effective lagrangian
for the low-energy modes. Denoting the low-energy modes by
$B(c,d)$, we will expand the one-loop effective lagrangian
first up to the same order as the classical potential, \ie
up to fourth order in $B$ and to second order in $\dot{B}$.
This allows us to do the renormalization of the lagrangian
properly. 
\par
Since we expect the physics to be sensitive to the
details of the potential along the tunnelling path,
we use an expansion of the effective potential around
the sphaleron to construct a fit of the effective potential up
to sixth order in the tunnelling parameter $u$.
This allows us to write down the effective lagrangian 
including some fifth and sixth order terms in $B$ so as to
reproduce the behaviour along the tunneling path.
\par
It is our purpose to use the effective hamiltonian in
a variational method to calculate the spectrum.
Although it is possible to calculate the effective potential
exactly along the tunnelling path, a polynomial approximation
in $B$ is much more useful, since this makes
the analytic evaluation of the matrix elements feasible.
\par
The one-loop effective lagrangian will be given in terms of
a path integral over the high-energy modes and over ghost fields.
Using Feynman diagrams to expand these path integrals is the
standard method to proceed, but due to the fact that
the theory is defined on \Sdr, the summation over the
space components of a loop momentum looks rather different
from the more familiar summation over the time component.
To isolate the intricacies of the \drs, we will first
assume $B$ to be time independent. 
This allows us to perform the integrations
over time components of loop momenta.
Within this so-called adiabatic 
approximation the calculation of the effective potential 
as an expansion in $B$ is then a purely algebraic problem.
\par 
After this we return to the use of Feynman diagrams to
evaluate the one-loop contribution to the kinetic term.
We will neglect higher order time derivatives of $B$ and also
terms of the form $B^n \dot{B}^2$.
We choose a renormalization scheme that makes the renormalized
kinetic part assume the form of the classical kinetic term, 
replacing the bare by the renormalized coupling constant. 
The result
of this section is then a finite, renormalized effective action.

\subsection{Gauge fixing}

We will impose the background gauge condition on the high-energy modes.
Consider a general gauge field $A_\mu = \left( A_0, A_i \right)$ on \Sdr,
where $A_0$ is the time component of the gauge field and $A_i$
are the space components with respect to the framing $e^i_\mu$.
We will project out the background field $B(c,d)$. Let $P_S$ be 
the projector on the constant scalar modes, and let $P_V = P_c + P_d$
be the projector on the $(c,d)$-space.
Explicitly we have:
\bea
P_S \psi &=& P^L_0 \psi = \frac{1}{2 \pi^2} \int_{\Skl^3} \psi, \\
\left(P_c A\right)_i &=& \left(P^L_0 A\right)_i 
  = \frac{1}{2 \pi^2} \int_{\Skl^3} A_i, \\ 
\left(P_d A\right)_i &=& \left(P^K_0 A\right)_i 
  = \left( \frac{1}{2 \pi^2} \int_{\Skl^3} A_k V^l_k \right) V^l_i.
\eea
We define the background field $B$ and the quantum field $Q$ by
respectively
\beq
  B_\mu = ( P A )_\mu = \left( P_S A_0, (P_V A)_i \right), \quad
  Q_\mu =  A_\mu - B_\mu.
\eeq
We define the gauge fixing function 
$\chi = \left (1 - P_S \right)D_\mu(PA) A_\mu + P_S A_0$.
We use $\chi$ to impose the background gauge condition:
\beq
  \chi = 0 \Leftrightarrow 
    \left\{ 
      \begin{array}{ll}
         B_0 & = 0 \\
         D_\mu(B) Q_\mu & = 0
      \end{array}
    \right.
\eeq
Introducing the Faddeev-Popov determinant
\beq
\Dl[A] = \left( \int D \! g\,\dl[ \chi( ^gA ) ]\right)^{-1},
\label{FPdet}
\eeq
and performing the standard manipulations with the partition function
leads to
\beq
Z = \int D \! B_k\, D' \! Q_\mu\, \Dl[A]\, \exp\left[ \frac{1}{g_0^2} 
 \int \tr\left(\half F_{\mu \nu}^2(B+Q) + 
 \frac{1}{\xi} (D_\mu(B) Q_\mu)^2 \right) \right].
\eeq
The primed integration means that we have excluded the $l=0$ modes
from the integration over the scalar field $Q_0$ 
and the $l=0,~k=1$ and $l=1,~k=0$ modes
(the $(c,d)$ modes) from the integration over the vector field $Q_k$.
\par
We now focus on the Faddeev-Popov determinant~(\ref{FPdet}).
Under an infinitesimal gauge transformation $\Lm$ the change in
$A_\mu$ is given by $D_\mu(A) \Lm$ and the change in the gauge
fixing function by
\bea
\dl \chi &=& \frac{\dl \chi}{\dl \Lm} \Lm  \Next
&=&  \left (1 - P_S \right) 
  \left\{ D_\mu(PA) D_\mu(A) \Lm + [ (P D(A) \Lm)_\mu, A_\mu] \right\}
   + P_S D_0(A) \Lm \Next
&=& \left (1 - P_S \right) 
  \left\{ D_\mu(B) D_\mu(A) \Lm - [B_\mu, (P D(A) \Lm)_\mu ] \right. \Next 
&& \quad \left. -[Q_\mu, (P D(A) \Lm)_\mu ]\right\} + P_S D_0(A) \Lm \Next
&=& \left (1 - P_S \right) 
  \left\{ D_\mu(B) (1 - P) D_\mu(A) \Lm + \pr_\mu (P D(A) \Lm)_\mu \right. \Next
&& \quad \left. -[Q_\mu, (P D(A) \Lm)_\mu ]\right\} + P_S D_0(A) \Lm \Next
&=& \left (1 - P_S \right) 
 \left\{ D_\mu(B) (1 - P) D_\mu(A) \Lm - 
  [Q_\mu, (P D(A) \Lm)_\mu ]\right\}\Next
&&+ P_S D_0(A) \Lm.
\eea
The Faddeev-Popov determinant is given by
$\Dl[A] = \det ( \dl \chi / \dl \Lm )$.
Introducing ghost fields $\psi$ and $\psibar$, this determinant can
be written as a fermionic path integral. We will evaluate this
path integral in the quadratic approximation, that is, we throw
away terms of order three and higher in the quantum fields $\psi$ and $Q_\mu$.
After this the integration over the $l=0$ modes of $\psi$ can
be performed to give an irrelevant constant and we obtain
\beq
\Dl[A]
\propto
 \int \! D' \! \psi\, D' \! \psibar \exp \left[ \beta \int \tr \left(
\psibar D_\mu(B) (1 - P) D_\mu(B) \psi \right) \right].
\eeq
Substituting this in the partition function with $\beta = -1/g_0^2$, 
expanding the
classical euclidean action up to second order in $Q$,
and choosing the Feynman gauge $\xi = 1$ we obtain
\bea
Z  
&=&
\int D\!B_k\, D'\!Q_\mu\, D'\!\psi\, D'\!\psibar  \exp\left[ \frac{1}{g_0^2} 
 \int \tr \left( \psibar \left\{ - D_\mu(B) (1 - P) D_\mu(B) \right\} \psi 
  \right. \right. \Next
&& \quad + \left. \left.  \half F_{\mu \nu}^2(B) - 
  2 (D_\mu F_{\mu \nu})(B) Q_\nu + 
  Q_\mu W_{\mu \nu}(B) Q_\nu \right) \vphantom{\frac{1}{g_0^2}} \right],
\eea
with
\beq
\left\{ \begin{array}{ll}
W_{0 0} &= -D_\rho^2(B) \\
W_{0 i} &= - W_{i 0} = - 2 \ad (\dot{B}_i) \\
W_{i j} &= -2 \ad(F_{i j}(B)) - (D_\rho^2(B))_{i j}
+ 2 \dl_{i j}
\end{array}
\right.
\eeq
Remember that the covariant derivative $D_i(B)$ acting on vectors (or tensors)
gives extra connection terms (due to \Sdr\ being a curved manifold), e.g.
\bea
  (D_i(B) C)_j &=& \pr_i C_j + [B_i, C_j] - \veps_{i j k} C_k \\ 
 &=& \left(- 2 i L_i + i B_i^a T^a - i S_i \right)_{j k} C_k
\eea
\par
The action contains a term $J_\nu Q_\nu$ with 
$J_\nu = (D_\mu F_{\mu \nu})(B)$. 
Since $B$ need not satisfy the equations of motion,
this term does not vanish.
Upon expanding the path integral in Feynman diagrams, this term
will give rise to extra diagrams, where $J$ acts as a source.
In appendix~\ref{appJ} we will
show that the presence of $J$ will only contribute to
terms in the effective lagrangian that we will consider to
give only small corrections: they are at least of order 
$c^2 d^4$ or $c^4 d^2$.
\par
Dropping for the time being the term linear in $Q_\mu$,
we are left with a gaussian integration over the fields $\psi$ and $Q_\mu$
from which we extract the effective action
\bea
S^\eff_\EUC[B] &\equiv& \int_0^T d\tau \left( \Ss{K} + \Ss{V} \right) \Next
&=& S^\cl_\EUC[B] 
- \ln \det \left( - D_\mu(B) (1 - P) D_\mu(B) \right)
+ \half \ln \det \left( W_{\mu \nu} \right).
\label{Seff}
\eea
Although we will not use the primed notation to denote it, 
the determinant does exclude the same modes that were not integrated over
in the path integral.

\subsection{The effective potential}

For computing the effective potential, $B$ is considered to be 
independent of time.
The assumption $\dot{B} = 0$ directly implies $W_{0 i} = 0$.
This results in a factorization of the
integral over $Q_\mu$, and we obtain
the one-loop contribution to
the effective action
\bea
S^{(1)}_\EUC[B,\dot{B} = 0]
&=& - \ln \det \left( - D_\mu(B) (1 - P) D_\mu(B) \right) \Next
&& + \half \ln \det \left(- D_\mu(B) D_\mu(B) \right) 
+ \half \ln \det \left( W_{i j} \right).
\label{Seffadiab}
\eea
We have to calculate these functional determinants for a
general background $B(c,d)$.
\subsubsection{The operators}
We start by expressing the various operators of eq.~(\ref{Seffadiab})
in terms of $c$ and $d$.
For the scalar operator $W_{0 0} = - D_\mu^2(B)$ we find
\beq
- D_\mu(B) D_\mu(B) =  - \pr_0^2 - D_i(B) D_i(B).
\eeq
As it acts on scalar functions, we have 
$D_i(B) = - 2 i L_i + i B_i^a T^a$ and we obtain
\bea
  - D_i(B) D_i(B) &=&  4 \Lkw - 4 B_i^a L_i T^a + B^a_i B^b_i T^a T^b \Next
&=& 4 \Lkw - 4 c^a_i {L_1}_i T^a - 4 d^a_i {L_2}_i T^a \Next
&&+ \left( c^a_i c^b_i + d^a_i d^b_i + c^a_i d^b_m V^m_i
+  d^a_m V^m_i c^b_i \right) T^a T^b.
\label{scalop}
\eea
The other scalar operator (the ghost operator) can also be written as
\beq
  -D_\mu(B) (1-P) D_\mu(B) = -\pr_0^2 - D_i(B) (1-P_V) D_i(B),
\label{scalopproj}
\eeq
where use was made of the fact that the operator does not act
on constant functions. Carefully analysing the extra term allows us
to write
\bea
D_i(B) P_V D_i(B)
&=& P^L_1 D_i(B) (P_V)_{i j} D_j(B) P^L_1 \Next
&=&  - \sfrac{1}{3} \left\{ 
 \left( c^a_i c^b_i - c^a_i c^b_j {L_1}_j {L_1}_i \right) 
+ \left( d^a_i d^b_i - d^a_i d^b_j {L_2}_j {L_2}_i \right) 
 \right\} T^a T^b P^L_1.  
\eea
The vector operator $W_{i j}$ is more complicated due to
the extra connection terms. We write
\beq 
  W_{i j} = - \pr_0^2 \dl_{i j} + \Wtil_{i j},
\eeq
Using the spin operator $S$ we can suppress the spin-indices and write
\bea
\Wtil(B) &=& 2 \Lkw  + 2 \Kkw 
  - 4 c^a_k {L_1}_k T^a - 4 d^a_k {L_2}_k T^a
  + \left (- 6 c^a_k + 2 d^a_m V^m_k \right) S_k T^a \Next
&& + \left( - \veps_{i j k} \veps_{a b c} c^a_i c^b_j 
   - 2 \veps_{i j k} \veps_{a b c} c^a_i d^b_m V^m_j 
   + \veps_{l m n} \veps_{a b c} d^a_l d^b_m V^n_k \right) S_k T^c
\Next
&& + \left( c^a_k c^b_k + d^a_k d^b_k + c^a_k d^b_m V^m_k
+  d^a_m V^m_k c^b_k \right) T^a T^b. 
\label{vectop}
\eea
\subsubsection{Reduction to spatial parts}
As we have seen, the operators we are interested in are of 
the form
\beq
  A(B) = -\pr_0^2 - \pr_\veps^2 + \Atil(B),
\eeq
where we introduced a laplacian term for an $\veps$-dimensional
torus of size $L$ that we attached to our space \Sdr~\cite{lus1}. 
This allows us to neatly perform the dimensional regularization.
The scale $L$ should of course be chosen proportional to the
radius $R$ of the \drs. A precise choice of $L$ would fix our
regularization procedure completely, and different choices
will be related through finite renormalizations fixing
the relations between the associated $\Lm$-parameters.
\par
Suppose that the spectrum of $\Atil$ is $\{\lmtil_i \}$.
The spectrum of $- \pr_\veps^2$ is
$\{ k_\veps^2 \}$ with $\kvec_\veps = 2 \pi \nvec /L$.
If we also take the time periodic with period $T$, the
spectrum of $-\pr_0^2$ is $\{ k_0^2 \}$ 
with $k_0 = 2 \pi n /T $.
Since $[\pr_0,\Atil(B)] = [\pr_\veps,\Atil(B)] = 0$, the spectrum
$\{ \lm_i \}$ of $A(B)$ follows trivially.
\par
Consider the exponential integral
\beq
E_1(x) = \int_x^\infty \frac{ds}{s} e^{-s} =
   -\gm - \ln(x) - \sum_{n=1}^\infty \frac{(-1)^n x^n}{n \, n!}.
\eeq
This implies
\beq
\int_\dl^\infty \frac{ds}{s} e^{-\lm s} = 
-\ln(\lm) - \ln(\dl) - \gm + \Order{ \lm \dl}.
\eeq
Taking the limit $\dl \downarrow 0$ in this integral,
we can write, up to an irrelevant constant,
\bea
\ln \det (A) &=& \sum_i \ln (\lm_i) \Next
&=& -\sum_{k_0} \sum_{k_\veps} \sum_i \int_0^\infty \frac{ds}{s}
e^{-s(k_0^2 + k_\veps^2 + \lmtil_i)} \Next
&=& -\Tpi \left( \Lpi \right)^\veps
\int_0^\infty ds\, s^{-\frac{3}{2} - \hfe} \tr \left( e^{-s \Atil} \right)
\label{lndetA1}
\eea
where we replaced the summations over $k_0$ and $k_\veps$ by integrals.
The operator $\Atil$ can be expressed in terms of the
angular momentum operators defined above, the functions $V^j_i$ 
and the constants $c^a_i$ and $d^a_i$.
To take the trace, we need a basis of functions (see section~\ref{ch-eff}). 
For the scalar
operators (the ghost operator and $W_{0 0}$), 
we can use $ |l\, m_L\, m_R \rangle \otimes |1\, m_t\rangle$
or equivalently $| l\,m_R; j\,m_j \rangle$, where $m_L$, $m_R$ and
$m_j$ correspond to the z-components of $\Leen$, $\Ltw$ and 
$\Jvec = \Leen + \Tvec$ respectively.
For the vector operator $W_{i j}$, we can use
$ |l\, m_R ; k\, m_k \rangle \otimes |1\, m_t\rangle$ or
$| l\,m_R;k; j\,m_j \rangle$.
Note that the $c$ and $d$ modes correspond to the vector modes with 
$(l,k) = (0,1)$ and $(l,k) = (1,0)$ respectively.
For the scalar operators, the trace
must not be taken over the $l=0$ functions, whereas 
for the vector operator
the trace must not include the $c$ and $d$ modes.
For the case of the ghost operator, the operator $(1-P)$
(see eq.~(\ref{scalopproj})) 
causes some intermediate vector modes to be projected to zero.

\subsubsection{Exact determinants}

Using the appropriate basis,
we can calculate the determinants exactly for the vacuum $u=0$ ($B=0$),
and for the sphaleron configuration $u=1$ ($c_i^a=-\dl_i^a$ and $d=0$). 
After integrating over $s$, the final summation
over $l$ is expressed in terms of the function $\zeta(s,a)$, 
which is defined by
\beq
\zeta(s,a) = \sum^\infty_{k=2} \frac{1}{(k^2 + a)^s}, \quad 
\mbox{Re}(s) > \half,
\label{zetadef}
\eeq
After analytic continuation we have
\beq
\zeta(s,a) = \sum^\infty_{m=0} \frac{(s)_m}{m !}\,  
  (-a)^m\, (\zeta_R(2 s + 2 m) - 1 ),\quad s \neq \half,-\half,\ldots,
\label{zetaexp}
\eeq
where $\zeta_R$ denotes the Riemann $\zeta$-function and
$(s)_m$ is Pochhammer's symbol.
If $s$ approaches one of the poles, this expansion can be
used to split off the divergent term: the remainder of the
series is denoted with $\zeta_F(s,a)$. 
As an example we treat $W(B=0) = - \pr_0^2 + 2 \Lkw + 2 \Kkw$.
From eq.~(\ref{lndetA1}) we obtain
\bea
  \ln \det(W_{i j})
&=& -\Tpi \left( \Lpi \right)^\veps
\Gm(-\half - \halfe) \tr \left(  (2 \Lkw + 2 \Kkw)^{\hf + \hfe} \right)
\Next
&=& -\Tpi \left( \Lpi \right)^\veps
\Gm(-\half - \halfe) \,
3 \sum_{l,k}
  \frac{(2 l + 1) (2 k + 1)}{(2 l (l+1) + 2 k (k+1) )^{-\hf - \hfe}}.
\Next
&=&
  -\Tpi \left( \Lpi \right)^\veps
  \Gm(-\half - \halfe) \left\{
           3\, \zeta(-\thalf - \halfe,-1) 
         + 3\, \zeta(-\half - \halfe,-1)
\right. \Next && \left. \quad
         + 6\, (\zeta_R(-3 -\veps) - 9)  
         - 6\, (\zeta_R(-1 -\veps) - 3)  
 \vphantom{\thalf}\right\}.
\eea
The summation over $(l,k)$ is as follows:
\beq
\sum_{l,k} = \sum_{m=\hf,1,\ldots}^\infty 
\left( \dl_{l m} \dl_{k m} + \dl_{l\,m+1} \dl_{k m}  + \dl_{l m} \dl_{k\,m+1}
\right).
\eeq
Using similar techniques for the other cases,
we find for the
effective potentials
\bea
\Ss{V}^{(1)}(B=0) &=& -18 + 3\,\zeta_R(-3) - 3\,\zeta_R(-1), \\
 \Ss{V}^{(1)}(\mbox{Sph})  &=&
  -1 - 3\,{\sqrt{2}} - 12 \sqrt{3} 
   + \nhalf \sqrt{6} - \fhalf \sqrt{10}
   + \sfrac{11}{4\,\veps} - 
   \sfrac{11}{8}\,\gamma + \sfrac{11}{4} \log (2) 
\Next && 
+  \sfrac{11}{4}\,\log \left(\sfrac{L}{2 \sqrt{\pi}} \right) 
  +{\zeta_F}(-\thalf,-3) + {\zeta_F}(-\thalf,-2) + 
   {\zeta_F}(-\thalf,1) 
\Next && 
  + 3\,{\zeta_F}(-\half,-3)
  +  {\zeta_F}(-\half,-2) - 5\,{\zeta_F}(-\half,1)
\eea
The pole term $\sfrac{11}{4\,\veps}$ has to be absorbed through
a renormalization of the coupling constant. Adding the classical potential
at the sphaleron we get
\beq
\Ss{V}_\eff(\mbox{Sph}) = \frac{ 2 \pi^2}{g^2_0} \frac{3}{2} + \frac{11}{4\,\veps} + \mbox{finite renormalization}.
\eeq
So the infinite part of the renormalization is
\beq
  \frac{1}{g_R^2} =  \frac{1}{g_0^2} + \frac{11}{12\,\pi^2\,\veps}.
  \label{infren}
\eeq
This is the standard result for \SUtw\ gauge theory in $1+3$ dimensions
as of course it should be: the renormalization is an ultra-violet effect
and does not depend on the global properties of the space on which 
the theory is defined.

\subsubsection{General background field}

For a configuration along the tunneling path,
$c^a_i = - u \dl^a_i$, $d=0$, we make an
expansion of the eigenvalues of the operators in $u$.
The operators are:
\bea
 - D_i(u) D_i(u) 
&=& 4 \Lkw + 4 u \Lvec \cdot \Tvec + 2 u^2,\label{ghostu} \\
  - D_i(u) (1 - P_V) D_i(u) 
&=& 4 \Lkw + 4 u \Lvec \cdot \Tvec + 2 u^2 \Next
&&- \sfrac{1}{3} u^2 \left( 2 - (\Lvec \cdot \Tvec)^2 
   - \Lvec \cdot \Tvec\right)  P^L_1, \label{corru} \\
\Wtil(u) &=& 2 \Lkw  + 2 \Kkw + 4 u \Lvec \cdot \Tvec 
+ (6 u - 2 u^2) \Svec \cdot \Tvec + 2 u^2.
\label{fieldu}
\eea
From eq.~(\ref{ghostu}) it directly follows that the eigenfunctions 
of the scalar operator $\Atil = -D_i(B) D_i(B)$
are the $|l\, m_R; j\, m_j \rangle$ with the eigenvalues 
$ 4 l (l+1) + 2 u ( j (j+1) - l (l+1) - 2) + 2 u^2$.
This allows us to write in eq.~(\ref{lndetA1})
\bea
 \tr \left( e^{-s \Atil} \right) &=&  \sum_{l=\hf,1,\ldots}^\infty
( 2 l +1) e^{-s 4 l (l+1)} \sum_{j=|l-1|}^{l+1} (2 j +1)
e^{-s (2 u ( j (j+1) - l (l+1) - 2) + 2 u^2)}, 
\eea
and to expand the second exponential in $u$. After performing
the summation over $j$ and the integration over $s$, the 
$l$-summations can again be expressed in terms of $\zeta$-functions.
Most of the algebra was done using \form~\cite{ver}.
The other scalar operator can be treated similarly.
The eigenvalues can be written down from eq.~(\ref{corru}) as well: 
they differ from the ones above only for $l=1$.
\par
For the vector operator the situation is more complicated since
the eigenvalues are not readily available.
We write
\beq
\Wtil(u) = 2 \Lkw + 2 \Kkw + \What(u),
\eeq
where the precise form of $\What(u)$ can be read off from eq.~(\ref{fieldu}).
Since $\What(0) = 0$ we can treat $\What(u)$ as a perturbation on 
the operator $2 \Lkw + 2 \Kkw$.
As $\Wtil$ commutes with the $\{ \Lkw, {L_2}_z,\Jkw,J_z \}$, 
the basis to use is $|(l 1)k,j \rangle$. The dimension for
the subspace in which we have to find the eigenvalues of $\Wtil$ is
given by the number of possible $k$ values.
Since this dimension is maximally three, it is of course 
possible to obtain the eigenvalues of $\Wtil(u)$ exactly, but
as explained earlier, it is more convenient to expand in $u$.
The unperturbed eigenvalues within each sector are 
$2 l (l+1) + 2 k (k+1)$, so we need non-degenerate perturbation
theory to obtain the expansion in $u$ for the eigenvalues
$\lmtil(j,k,l)$. A method that implements this particularly nice is
due to Bloch~\cite{blo}
\par
Using \mathem~\cite{wol} we computed the expansions for the eigenvalues.
For this we needed the matrix elements of the
operators  
$\Lvec \cdot \Tvec$ and $\Svec \cdot \Tvec$.
To calculate these
matrix elements, we need the 6-$j$ symbols~\cite{edm}. 
For three angular momenta $t,s$ and $l$ we have (cf.\ section~\ref{ch-eff})
\beq
|t,(s l)k,j \rangle = 
\sum_q {(-1)^{t + s + l + j}} [(2 q + 1 ) (2 k + 1)]^{\hf}
\left\{
\begin{array}{ccc}
t & s & q \\
l & j & k
\end{array} 
\right\}
|(t s)q,l,j \rangle.
\eeq
This gives
\bea
&& \terug \langle t,(s l)k',j|\, \Svec \cdot \Tvec \, |t,(s l)k,j \rangle 
= - \half ( s(s+1) + t ( t+1) ) \dl_{k'k}  
\Next
&&  \quad + \half \sum_q q (q+1) (2 q + 1)
[(2 k + 1 ) (2 k' + 1)]^{\hf}
\left\{
\begin{array}{ccc}
t & s & q \\
l & j & k'
\end{array} 
\right\}
\left\{
\begin{array}{ccc}
t & s & q \\
l & j & k
\end{array} 
\right\}.
\eea
The matrix elements for $\Lvec \cdot \Tvec$ follow from this
and the identity 
$\Jkw = \Kkw + \vec{T}^2+ 2\Lvec \cdot \Tvec + 2\Svec \cdot \Tvec$.
As in the case of the scalar operators, we expand the exponential
in $u$ and perform the $s$-integration. The summations over $l$ 
can be expressed in $\zeta$-functions. During the various stages
of these calculations care has to be taken for low values of $l,k$ and
$j$: for certain values the Bloch result for the general 
eigenvalue $\lmtil(j,k,l)$ will not be valid, since
the subspace in which to perform
the diagonalization may have less dimensions than the bulk value
$3 - |l-j|$: some of the $k$ values would be negative.
Moreover, in the $l=1$, $j=1$ sector, which is a priori
three dimensional, 
the $k=0$ modes have to be thrown out explicitly, since they
are the $d$ modes. Exclusion of the $c$ modes is simply
achieved by starting the summation over $l$ with $l = \half$.
The result for the effective potential up to fourth order in $u$ is
\bea
\Ss{V}^{(1)}(u)
 \!\!\!  &=& \! \!\!
 -18 + 3\,{\zeta_R}(-3) - 3\,{\zeta_R}(-1) 
\Next && \! \!\!
+ {\Vcl}(u)\,\left( -{\sfrac{1259}{1152}} + {\sfrac{11}{6\,{\veps}}} - 
     \sfrac{11}{12} \,\gamma  - \sfrac{11}{128}\, \sqrt{2} + 
      \sfrac{11}{6} \, \log (2)  + 
      \sfrac{11}{6} \, \log \left( \Lpi \right) 
\right. \Next && \quad  - 
      \sfrac{23}{6} \,{\zeta}(\thalf,-1) - 
      \sfrac{1}{6}  \,{\zeta}(\fhalf,-1) + 
      \sfrac{53}{3} \,{\zeta}(\shalf,-1) + 
      14\,{\zeta}(\nhalf,-1) 
  \Next && \quad \left. + 
      \sfrac{16}{3} \,{\zeta_R}(3) - 
      \sfrac{19}{6} \,{\zeta_R}(5) - 
      \sfrac{1}{3}  \,{\zeta_R}(7) \vphantom{\sfrac{1259}{1152}}
\right)  
\Next && \! \!\!
  + {u^2}\,\left( \sfrac{179}{192} + \sfrac{33}{64} \,\sqrt{2} + 
      13\,{\zeta}(\thalf,-1) - 9\,{\zeta}(\fhalf,-1)  -  
     106\,{\zeta}(\shalf,-1) 
\right. \Next && \quad \left.- 
       84\,{\zeta}(\nhalf,-1) +
       2\,{\zeta_R}(-1) - 23\,{\zeta_R}(3) + 19\,{\zeta_R}(5) + 
       2\,{\zeta_R}(7)  \vphantom{\sfrac{179}{192}} \right)   
\Next && \! \!\!
   + {u^4}\,\left( \sfrac{863}{768} + \sfrac{1475}{2048}\,\sqrt{2} - 
      \sfrac{53}{4} \,{\zeta}(\thalf,-1) + 
      \sfrac{17}{4} \,{\zeta}(\fhalf,-1) + 
      \sfrac{261}{2}\,{\zeta}(\shalf,-1)
\right. \Next && \quad  + 
      47\,{\zeta}(\nhalf,-1) - 66\,{\zeta}({\ehalf},-1) + 
      \sfrac{63}{4} \,{\zeta_R}(3) - 79\,{\zeta_R}(5) 
 \Next && \quad \left. + 
      \sfrac{227}{4}\,{\zeta_R}(7) + 5\,{\zeta_R}(9) \right).
\label{Veffu4}
\eea
Here we eliminated the $u^3$ term in favour of the classical
potential $\Vcl$ defined by
\beq
  \Ss{V}_\cl = \frac{2 \, \pi^2}{g^2_0} \Vcl,
\eeq
so
\beq
  \Vcl(u) = \thalf u^2 (2-u)^2.
\eeq
One checks that the pole term is absorbed by the coupling constant
renormalization of eq.~(\ref{infren}).
\par
\begin{figure}
\centering
\epsfxsize=0.8\textwidth
\leavevmode
\epsfbox[72 222 540 570]{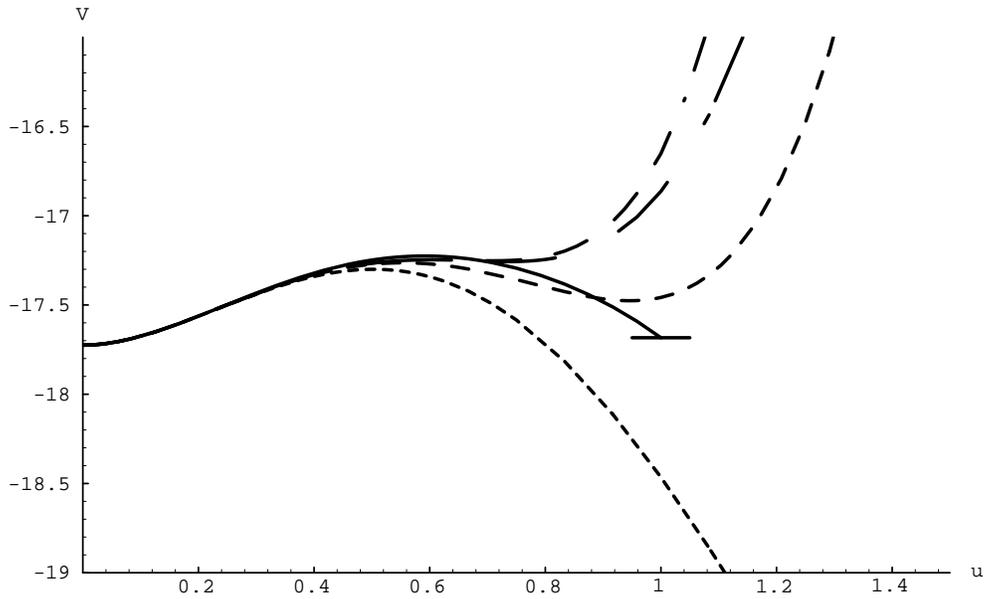}
\caption{Expansion of $\Ss{V}^{(1)}(u)$ in the tunnelling parameter $u$.
We dropped the $\veps$ and $\log (L)$ dependent parts. We have
drawn the expansion up to order $u^4$, $u^6$, $u^8$ and $u^{10}$.
Longer dashes correspond to higher order expansions. The horizontal
line at $u=1$ denotes the exact potential at the sphaleron, the drawn
curve is the sixth order fit explained in the text.}
\label{Veffu}
\end{figure}

Results up to tenth order in $u$ were calculated.
As can be seen from fig.~\ref{Veffu}, the expansions do not converge
to the exact result at $u=1$. This should come as no big
surprise, since we have no reason to expect the radius of convergence
of the expansion to be as large as one. Judging from the picture,
one would estimate a radius of convergence of roughly $0.6$.
\par
To find the effective potential for larger $u$, we 
make a similar expansion of the effective potential around
the sphaleron: $u = 1+a$.
The expansions of the determinants of the scalar operators can be
obtained exactly as described above. The final $l$-summations
can be expressed in terms of $\zeta(s,b)$ with $b=-3,-2,1$.
For the vector operator, we could in principle again use the Bloch 
perturbation method to obtain the eigenvalues $\lmtil(j,k,l)(a)$,
although the levels are now degenerate at $a=0$ . There is however
an easier way. Suppose we have $\Atil = F + \Ahat$ with $F = \Atil(0)$
such that $[F,\Ahat] = 0$.
This allows us to substitute in eq.~(\ref{lndetA1})
\beq
\tr \left( e^{-s \Atil} \right)
= \tr \left( e^{-s F} e^{-s \Ahat} \right)
=  \sum_i e^{-s F_i}
  \sum_{n=0}^\infty \frac{(-s)^n}{n!} \trees{i}{ \Ahat^n}.
\label{lndetA3}
\eeq
The sum over $i$ is a sum over the eigenspaces of $F$, $F_i$ is the
corresponding eigenvalue and
$\trees{i}{}$ denotes a trace within the eigenspace.
For the operator $W$
\beq
\Wtil(a) = 2 \Lkw  + 2 \Jkw - 2
 +  a ( 4 + 4 \Lvec \cdot \Tvec + 2 \Svec \cdot \Tvec )
 +  a^2 ( 2 - 2  \Svec \cdot \Tvec ),
\label{fielda}
\eeq
we indeed have that 
$F = \Wtil(0) = 2 \Lkw  + 2 \Jkw - 2$
commutes with $\What$ and we obtain
\bea
\ln \det (W) &=& 
-\Tpi \left( \Lpi \right)^\veps
\sum_{l,j} (2l+1) (2j+1) \times
\Next &&
 \int_0^\infty ds\, s^{-\frac{3}{2} - \hfe} 
 e^{-s (2 l (l+1) + 2 j (j+1) -2 )} 
  \sum_{n=0}^\infty \frac{(-s)^n}{n!} \trees{l j}{ \What^n }.
\eea
Note that the sum over different $k$ values is now
absorbed in the trace in the $(l,j)$ space. When taking this trace
one has to deal with the same subtleties connected with low values of 
$l$ and $j$ as described before.
The resulting effective potential is an expansion in $a$.
Its behaviour is similar to that of the previous expansion:
we have a radius of convergence of roughly $0.4$ around the sphaleron
$u=1$.
Using the fourth order expansion in $u$ and the first order expansion
in $a$ (\ie the value and the slope of the potential 
at the sphaleron), we can construct a polynomial in $u$ of degree six
that is a good approximation to the effective potential.
\par
Regarding the issue of the radius of convergence, one might think
that also integrating out the $(c,d)$ modes that are orthogonal
to the $u$ mode might result in better convergence. Also one
expects the $u \rightarrow 2-u$ symmetry to be restored in this case.
With the techniques described above, performing
this calculation is straightforward, 
provided one remembers to properly adjust the gauge fixing procedure.
The new expansions in $u$ and $a$
have roughly the same convergence behaviour as before, so there
is no improvement on this point. The symmetry is however manifestly 
restored: the expansion in $a$ contains only even powers of $a$.
\par
For the case of the vacuum and the sphaleron configuration, we could
calculate the spectra of the operators exactly.
For a configuration along the tunnelling path, the
operators have less symmetries, and the evaluation of
the spectra becomes harder. For $B = B(c,d=0)$ and especially
$B = B(c,d)$ we have even fewer operators that commute with
the operators whose determinants we want to calculate.
Nevertheless, we developped methods~\cite{heu2,heu3} to calculate the
functional determinants up to fourth order in $c$ and $d$.
These methods are based on eq.~(\ref{lndetA3}) and a generalization thereof.
We postpone writing down the result of this calculation,
which is the effective potential,
until we have performed the renormalization.

\subsection{The effective kinetic term}

To obtain the one-loop contribution to the operator $\dot{B}^2$,
we perform the usual expansion of the path integral in
Feynman diagrams. The subtleties related to the summation
over the space-momenta were dealt with in the previous section.
The diagrams needed are simple loop diagrams with
one and two insertions of the operator $\Ahat$
defined above. 
The particle in the loop is respectively a ghost, a scalar $Q_0$
or a vector particle.
The $\dot{B}^2$ term comes from the
diagrams with two insertions.
\par
As an example we will treat the diagram $d_2$ with two ghost propagators
and two insertions of the operator $\Ahat$.
The fermionic path integral is given by
\bea
Z &=&
\int D' \! \psi \, D' \! \psibar  \exp\left[ \frac{1}{g_0^2} 
 \int \tr \left( \psibar \left\{ - D_\mu(B) (1 - P) D_\mu(B) \right\} \psi 
  \right) \right] 
\Next &=&
\int D' \! \psi \, D' \! \psibar  \exp\left[ \frac{1}{g_0^2} 
 \int \tr \left( \psibar \left\{ -\pr_0^2 - \pr_\veps^2 + 4 \Lkw
 \right\} \psi + \psibar \Ahat \psi \right) \right].
\label{Zferm}
\eea
The propagator in momentum space is given by
\beq
  \frac{-1}{ k_0^2 + k_\veps^2 + 4 l (l+1) }.
\eeq
We will suppress the manipulations with the integrations over 
$k_\veps$ and $x_\veps$. The dimensional regularization can be
obtained by adding $k_\veps^2$ to the denominators of all propagators
and performing the integration 
\beq
  L^\veps \int  \frac{d^\veps \! k_\veps}{(2 \pi)^\veps}.
\eeq
Suppressing $m_L$, $m_R$ and $m_t$, we find for the diagram with
two insertions
\bea
d_2&=& - \half \int \frac{dk_0}{2 \pi} \frac{dk'_0}{2 \pi} \sum_l \sum_{l'}
\int dt \,  dt'  
\frac{-1}{ k_0^2 + 4 l (l+1) } \,
 \langle l | \, \Ahat(t') \, | l' \rangle \,
e^{-i k_0 ( t - t')} \times
\Next && 
\frac{-1}{ {k'}_0^2 + 4 l' (l'+1) }\,
 \langle l' | \, \Ahat(t) \, | l \rangle \,
e^{-i k'_0 ( t' - t)}.
\eea
Noting that $[\Lkw, \Ahat(t)] = 0$ and substituting $k_0 = p$, $k'_0 = p+q$
we write
\bea
d_2&=& - \half   
\int \frac{dp}{2 \pi} \frac{dq}{2 \pi} 
\sum_l \int dt  \, dt' e^{-i q ( t - t')}  \trees{l}{\Ahat(t) \Ahat(t') } 
 \times \Next &&
\frac{1}{ p^2 + 4 l (l+1) } \,
\frac{1}{ (p+q)^2 + 4 l (l+1) }.
\eea
Using contour integration, one proves
\beq
\int \frac{dp}{2 \pi} \,
\frac{1}{ p^2 + C } \,
\frac{1}{ (p+q)^2 + C} = 
\frac{1}
{\sqrt{C} \left( 4 C + q^2 \right)}.
\label{qint}
\eeq
After performing the $p$ integration and expanding in $q^2$ we obtain
\bea
d_2&=& - \half  
 \int \frac{dq}{2 \pi} 
 \sum_l \int dt \, dt' e^{-i q ( t - t')} \trees{l}{\Ahat(t) \Ahat(t') }
 \times \Next && 
 \frac{1}{4 ( 4 l (l+1) )^{\frac{3}{2}}}
\sum_{n=0}^\infty \frac{(-1)^n}{4^n}
\frac{(q^2)^n}{(4 l (l+1) )^n}.
\eea
Next we substitute
\beq
  q^2 \rightarrow \frac{\pr^2}{\pr t \, \pr t'},
\eeq
and move these time derivatives to
$\Ahat$ by partial integration.
The $q$ and the $t'$ integration thus become trivial, and we obtain
\bea
d_2 &=&
- \sfrac{1}{8} \int dt  \sum_l
 \sum_{n=0}^\infty \frac{(-1)^n}{4^n} \,
\frac{1}{(4 l (l+1) )^{n+\frac{3}{2}}} \,
\trees{l}{\Ahat^{(n)}(t) \Ahat^{(n)}(t) } \Next
&\rightarrow&
- \sfrac{1}{8}   \int dt \sum_l  
   \sum_{n=0}^\infty \frac{(-1)^n}{4^n} \,
\frac{g_{n+\frac{3}{2}}(\veps)}{( 4 l (l+1))^{n + \frac{3}{2} - \hfe}} \, 
\trees{l}{\Ahat^{(n)}(t) \Ahat^{(n)}(t) }.
\label{d2dia}
\eea
The arrow indicates that we have performed the $\veps$-dimensional
integrations that were suppressed in the notation.
We introduced
\beq
 g_n(\veps) 
=  \int d^\veps \! x_\veps \int \frac{d^\veps \! k_\veps}{(2 \pi)^\veps} 
 \frac{1}{(1 + k_\veps^2)^n} = 
 \left( \Lpi \right)^\veps  
  \frac{ \Gm(n- \halfe)}{ \Gm(n)}.
\eeq
The $n=0$ contribution in eq.~(\ref{d2dia}) can be related 
to the $s^2$ term in eq.~(\ref{lndetA3}),
whereas the $n=1$ contribution gives us the $\dot{B}^2$ term
that we were looking for:
\beq
  \int dt \, \sfrac{1}{12} g_{\frac{5}{2}}(\veps) 
\left( \zeta(\half - \halfe,-1) +  \zeta(\thalf - \halfe,-1) \right)
\dot{c}^a_i \dot{c}^a_i.
\eeq
\par
The bosonic path integral is given by
\bea
Z &=&
\int D' \! Q_\mu\,  \exp\left[ \frac{1}{g_0^2} 
 \int \tr \left( 
  -2 (D_\mu F_{\mu \nu})(B) Q_\nu  + Q_\mu W_{\mu \nu}(B) Q_\nu 
\right)  \right]
\Next &=&
\int D' \! Q_0 \, D' \! Q_i \,  \exp\left[ \frac{1}{g_0^2} 
 \int \tr \left( 
  Q_0 \left\{  -\pr_0^2 - \prvec_\veps^2 + 4 \Lkw \right\} Q_0 
\right. \right. 
\Next && \quad  +
  Q_i \left\{  -\pr_0^2 - \prvec_\veps^2 + 2 \Lkw + 2 \Kkw \right\}_{i j} Q_j
  + Q_0 \Ahat Q_0 + Q_i \What_{i j} Q_j  
\Next && \quad \left. \left. - 
4 Q_0 \ad(\dot{B}_j) Q_j - 2 J_0 Q_0 - 2 J_i Q_i
  \right)   \vphantom{\frac{1}{g_0^2}} \right].
\label{Zbos}
\eea
with $J_\nu = (D_\mu F_{\mu \nu})(B)$ (see appendix~\ref{appJ}).
The diagrams where the particle in the loop is a $Q_0$ particle
have the same structure as those with a ghost particle.
Apart from extra terms coming
from the projector $1-P$ in the ghost case, the contribution
is precisely $-\hf$ times the fermionic contribution.
The diagram
with a vector particle gives rise to some extra technical problems,
but it can also be dealt with.
There is also a diagram with an explicit dependence on $\dot{B}$:
it is the diagram with two insertions of the operator 
$W_{0 i} \propto \ad(\dot{B}_i)$,
one $Q_0$ and one $Q_i$ propagator.
\par
Adding up the different contributions, we obtain the
one-loop contribution to the kinetic term
$\dot{c}^a_i \dot{c}^a_i + \dot{d}^a_i \dot{d}^a_i$
in the lagrangian:
\bea
   \Ss{K}^{(1)}(c,d)  &=&
   \left( \dot{c}^a_i \dot{c}^a_i + \dot{d}^a_i \dot{d}^a_i \right) \,
\left(   
     \sfrac{1247}{1152} + 
     \sfrac{5}{16} \, {\sqrt{2}} +
     {\sfrac{11}{12 \, \veps}} - 
     \sfrac{11}{24} \, {\gamma} + 
     \sfrac{11}{12}\,\log (2) \right. 
\Next && \quad
+ \sfrac{11}{12}\,\log (\sfrac{L}{2 \, \sqrt{\pi}})
 + \sfrac{1}{6}\,{\zeta}(\thalf,-1) + 
     11\,{\zeta}(\fhalf,-1) + 
     \sfrac{49}{6}\,{\zeta}(\shalf,-1)
 \Next && \quad \left. - 
     \sfrac{4}{3}\,{\zeta_R}(3) - 
     \sfrac{5}{12}\,{\zeta_R}(5) - 
     \sfrac{8}{3}\,{\zeta_F}(\half,-1)  
  \vphantom{\sfrac{1247}{1152}}\right). 
\label{kineff}
\eea
The divergency will be absorbed by the same coupling constant
renormalization of eq.~(\ref{infren}) that also made the 
potential part finite.

\subsection{Renormalized results}
We will use a renormalization scheme such that the
renormalized kinetic part looks just like the classical
term:
\beq
 \Ss{K}_\eff = \frac{2 \pi^2}{2 g^2_R} 
 \left( \dot{c}^a_i \dot{c}^a_i + \dot{d}^a_i \dot{d}^a_i \right).
\label{schemedef}
\eeq
Using eq.~(\ref{kineff}) this gives us the finite part of the
renormalization~(\ref{infren}):
\beq
   \frac{1}{g^2_R} = \frac{1}{g^2_0} + 
    {\frac{11}{12 \pi^2  \veps}} +
     \frac{11}{12 \pi^2 }\,\log (\sfrac{L}{2 \, \sqrt{\pi}}) + \kappa_0,
\label{renorm}
\eeq
where $\kappa_0$ can be found in table~\ref{kappadat}.
This renormalization scheme can easily be related to other 
schemes like the 
\MS\ or $\overline{\mbox{\MS}}$ scheme.
\par
The finite, renormalised effective potential becomes
\beq
  \Ss{V}_\eff = \frac{2 \pi^2}{g_R^2} \, V_\cl(c,d) + V_\eff^{(1)},
\eeq
with
\bea
V_\eff^{(1)}(c,d) &=&   
V_\eff^{(1)}(c) + V_\eff^{(1)}(d) + 
  \kappa_7 \, {\tr}(X)\,{\tr}(Y) +  \kappa_8 \, {\tr}(X\,Y), \Next
V_\eff^{(1)}(c) &=&   
   \kappa_1 \, {\tr}(X)  + 
   \kappa_2 \, {\det}(c) +
   \kappa_3 \, {\mbox{tr}^2}(X) +
   \kappa_4 \, {\tr}({X^2}) 
\Next &&  +
   \kappa_5 \, {\det}(c)\,{\tr}(X) +
   \kappa_6 \, {\mbox{tr}^3}(X),
\eea
with the numerical values for $\kappa_i$ in table~\ref{kappadat}.
\begin{table}[t]
\centering
\begin{tabular}{|l@{ = }r@{.}l|} \hline
$\kappa_0$ & $ 0$&$0566264741439181  $\\ \hline
$\kappa_1$ & $-0$&$2453459985179565 $\\
$\kappa_2$ & $ 3$&$66869179814223   $\\
$\kappa_3$ & $ 0$&$500703203096610  $\\
$\kappa_4$ & $-0$&$839359633413003  $\\
$\kappa_5$ & $-0$&$849965412245339  $\\
$\kappa_6$ & $-0$&$06550330854836428$\\
$\kappa_7$ & $-0$&$3617122159967145 $\\
$\kappa_8$ & $-2$&$295356861354712  $\\ \hline
\end{tabular}
\caption{The numerical values of the coefficients.}
\label{kappadat}
\end{table}
Note that the $u^5$ term in the effective
potential along the tunnelling path uniquely determines the
coefficient of the $\tr(X) \det(c)$ term.
The $u^6$ term can be obtained from
combinations of the three independent invariants $\mbox{tr}^3(X)$, 
$\tr(X) \, \tr(X^2)$ and $\tr(X^3)$.
We choose to replace the $u^6$ term by $\mbox{tr}^3(X)$,
which is the simplest of these from the viewpoint of the
variational calculation.
Note that not all of these coefficients are small, which means
that the one-loop correction to the spectrum, to be calculated in
section~\ref{ch-RR}, may be substantial.

\section{Calculating the spectrum}
\label{ch-RR}

\subsection{Introduction}

In this section we will approximate the spectrum of the effective 
hamiltonian by applying the Rayleigh-Ritz method~\cite{ree}.
This is a variational method that consists of truncating
some suitably chosen basis of functions and calculating
the matrix of the hamiltonian \Ss{H} with respect to this basis.
The results of the numerical diagonalization are the
energy levels in various sectors. 
Using these, we find the excitation energies which correspond to the
masses of the low-lying glueball states. We also estimate
the window of validity of our effective model.
\par
Consider a hamiltonian \Ss{H} whose spectrum $\sg(\Ss{H})$ 
is $\{\mu_1,\mu_2,\ldots\}$.
The Rayleigh-Ritz method gives upper bounds $\tilde{\mu}_n \geq \mu_n$.
We will use the generalized Temple inequality~\cite{ree,kat} to arrive at
lower bounds for the levels. Let $\psi$ be a normalized trial wave function
and define
\beq
\gm = \langle \psi | \Ss{H} | \psi \rangle, \quad
\eta = \langle \psi | (\Ss{H} - \gm)^2 | \psi \rangle = 
  \langle \psi | \Ss{H}^2 | \psi \rangle - \gm^2.
\eeq
Note that $\eta \geq 0$ and that $\eta = 0$ implies that
$\psi$ is a true eigenfunction of \Ss{H}.
With the help of the values of $\eta$ we can derive lower
bounds.
We assume our variational basis to be so accurate that
$\mu_{n+1} > (\tilde{\mu}_{n+1} + \tilde{\mu}_{n})/2$.
Under this and some other mild assumptions we can conclude~\cite{ree,kat}
\beq
  \tilde{\mu}_n - \frac{2 \eta}{\tilde{\mu}_{n+1} - \tilde{\mu}_{n}}
  < \mu_n < \tilde{\mu}_n.
\label{temple}
\eeq
\par
We can treat the lowest order and the one-loop
case on the same footing by writing
\bea
  \Ss{H} &=& -\frac{f}{2} \left( 
  \frac{\pr^2}{\pr c^a_i \pr c^a_i} + \frac{\pr^2}{\pr d^a_i \pr d^a_i} 
   \right) + \Ss{V}(c,d), \quad f = \frac{g^2}{2 \pi^2}, \\
 \Ss{V}(c,d) &=& \Ss{V}(c) + \Ss{V}(d) + 
    \lm_7 r_c^2 r_d^2 + \lm_8 \inv_2(c,d),  \label{inv2ref} \\
 \Ss{V}(c) &=& 
  \lm_1 r_c^2 + \lm_2 \inv_3(c) + \lm_3 r_c^4 + \lm_4 \inv_4(c)
    + \lm_5 r_c^2 \inv_3(c) + \lm_6 r_c^6.
\eea
The coefficients $\lm_i$ (which are functions of $f$ and $\kappa_j$)
as well as the definitions of the
used invariants can be found in table~\ref{hamdat}. We have
also given the definitions of the invariants that occur in 
$\Ss{V}^2$.
For the lowest order effective hamiltonian, we just set $\kappa_i =0$,
whereas for the one-loop effective hamiltonian we need to use
the values for $\kappa_i$ of table~\ref{kappadat}.
\begin{table}[t]
\centering
\begin{tabular}{||l@{ = }l||} \hline
$\lm_1$ & $ \frac{2}{f} + \kappa_1$\\
$\lm_2$ & $ \frac{2}{f} + \frac{1}{3} \kappa_2$\\
$\lm_3$ & $\kappa_3 + \kappa_4$\\
$\lm_4$ & $ \frac{2}{f} - 8 \kappa_4$\\
$\lm_5$ & $\frac{1}{3} \kappa_5$\\
$\lm_6$ & $\kappa_6$\\
$\lm_7$ & $\kappa_7 + \kappa_8$\\
$\lm_8$ & $\frac{2}{f} - 6 \kappa_8$\\ \hline
\end{tabular} \hspace{2 cm}
\begin{tabular}{||l@{ = }l||} \hline
$\inv_3(c)$ & $ 3 \det(c) $\\
$\inv_4(c)$ & $ \frac{1}{8} \left( \tr^2(X) - \tr(X^2) \right) $\\
$\inv_2(c,d)$ & $ \frac{1}{6} \left( \tr(X) \tr(Y) - \tr(X Y) \right) $\\
$\inv_6(c)$ & $\inv^2_3(c)$ \\
$\inv_7(c)$ & $\inv_3(c) \inv_4(c)$ \\
$\inv_8(c)$ & $\inv^2_4(c)$ \\ \hline
\end{tabular}
\caption{The coefficients for the hamiltonian and the definition
of several invariants.}
\label{hamdat}
\end{table}
A sketch of our methods and some results for the truncated model
were published before~\cite{heu1}.

\subsection{The variational basis}

We decided to use functions of the form
$\phi(r_c,r_d) Y(\chat,\dhat)$ and to incorporate the boundary
conditions in the $(r_c,r_d)$ plane. Apart from the boundary
conditions, we must also respect the symmetries of the
hamiltonian as much as possible to obtain an optimal
block diagonalization. Most of these symmetries,
including the residual gauge symmetry, will be incorporated
in the functions $Y(\chat,\dhat)$.
After this, we will use the exact solution of the 
strong coupling limit of our hamiltonian problem as
a guide to obtain a useful set of radial functions.

\subsubsection{Symmetries}

The hamiltonian $\Ss{H}(c,d)$ is invariant under
the transformation $ c \rightarrow S c R_1,~d \rightarrow S d R_2$
with $S,R_1,R_2 \in \mbox{SO}(3)$ and under the 
interchange $c \leftrightarrow d$.
The generators of left- and right 
multiplication are $\vec{L}^R_c$, $\vec{L}^S_c$, $\vec{L}^R_d$ 
and $\vec{L}^S_d$. These are \sutw\ angular momentum operators and
we have for instance
\beq
\left(L^R_c\right)_i = - i \veps_{i j k} c^a_j \frac{\pr}{\pr c^a_k}.
\eeq
The following set of operators commutes:
\beq
 \left\{\Ss{H},\vec{J}^S,\vec{J}^R,(\vec{L}^R_c)^2 + (\vec{L}^R_d)^2,
  \Ss{P}\right\}.
\label{setcomm}
\eeq
\Ss{P} is defined by $\Ss{P} f(c,d) \equiv f(d,c)$. On \Sdr\ it 
corresponds to the parity $(n_0,\vec{n}) \leftrightarrow (n_0,-\vec{n})$
(cf.\ eq.~(\ref{pardef})). 
The operator $\vec{J}^S \equiv \vec{L}^S_c + \vec{L}^S_d$ implements
constant gauge transformations: we have to demand $(\vec{J}^S)^2 = 0$
for physical wave functions. 
The operator $\vec{J}^R \equiv \vec{L}^R_c + \vec{L}^R_d$ is 
the rotation operator.
The spatial symmetry group is SO(4) and these symmetries
cannot be simply divided in independent sets of translations and rotations.
The operator on \Sdr\ that corresponds to $\vec{J}^R$ is $\Leen + \Ltw$,
and the corresponding symmetry on \SUtw\ (cf.\ appendix~\ref{appbasis})
is $g \rightarrow g_1^{-1} g g_1$.
Writing $g = n \cdot \sg$ and $g_1 = n' \cdot \sg$ we have
\beq
  n \cdot \sg ~\rightarrow~ n_\mu (n' \cdot \sgbar) \sg_\mu (n' \cdot \sg)
  = n_0 \sg_0 - n_i V^j_i(n') \sg_j.
\eeq
The symmetry on \Sdr\ is thus seen to leave $n_0$ invariant and to perform
an SO(3) rotation on $\vec{n}$. It is hence a rotation 
for points around the north and south pole of \Sdr. 
The operators $\Leen$ and $\Ltw$ do not
leave any point invariant and cannot be interpreted as rotation operators.
Only their sum has this property and the different sectors
$(j=0,1,\ldots)$ under the symmetry $\vec{J}^R$ correspond 
to scalar glueballs, vector glueballs, etc.
\par
To prepare the ground for the block diagonalization of the 
hamiltonian, we divide the function space in sectors characterized
by the quantum numbers $j$, $m$, $l_1(l_1+1) + l_2 ( l_2 + 1)$ and $p$
corresponding to the operators of eq.~(\ref{setcomm}).
Note that for low values of $l_1$ and $l_2$ there is a one-to-one
correspondence between unordered pairs $(l_1,l_2)$ and the
numbers $l_1(l_1+1) + l_2 ( l_2 + 1)$. Since the spectrum
of \Ss{H} is independent of the azimuthal quantum number $m$,
we use the notation $l_1 l_2 j$-even and $l_1 l_2 j$-odd
to denote the various sectors.
The lowest energy level in the scalar ($j=0$) sector 
is the vacuum. Energy differences with respect to this level
will be the glueball masses. 

\subsubsection{The angular sector}

We begin with constructing functions of $\chat$ that are
eigenfunctions of the following set of commuting operators
\beq
 \left\{ \Dl_{\chat}, \vec{L}^R_c, \vec{L}^S_c \right\}.
\eeq
Note that the space of the $\chat$ makes up an S$^8$.
From appendix~\ref{appbasis} we know that the
eigenfunctions of the spherical laplacian $\Dl_{\chat}$ will
be the homogeneous harmonic polynomials in $\chat$. 
An orthonormal basis of functions of $\chat$ is given by the set 
$\left\{\langle \chat|L;l_s,l_r,\tau;m_s,m_r \rangle \right\}$. 
Each of these functions is a homogeneous harmonic polynomial 
of degree $L$ in $\chat$.
Its eigenvalues under the various symmetries are collected in
table~\ref{angtabelc}.
We used the operators $\vec{L}^R_c$ and $\vec{L}^S_c$ to further
classify these functions, but for higher values of $L$
we need the extra label $\tau$ for the remaining degeneracy.
This degeneracy is absent for low values of $L$, shows up first for
$(L,l_s,l_r) = (4,2,2)$ and proliferates for high values of $L$.
\begin{table}[t]
\begin{center}
\begin{tabular}{||l||c|c|c|c|c||} \hline
Operator & $(\vec{L}^S_c)^2$ & $(L^S_c)_3$ &$(\vec{L}^R_c)^2$ &
$(L^R_c)_3$ & $\Dl_{\chat}$ \\ \hline
Eigenvalue & $l_s ( l_s + 1)$& $m_s $ & $l_r ( l_r + 1)$ & 
$m_r $ & $ - L ( L + 7)$  \\ \hline
\end{tabular}
\end{center}
\caption{Behaviour of the functions 
$\langle \chat | L;l_s,l_r,\tau;m_s,m_r \rangle$.}
\label{angtabelc}
\end{table}
\par
Explicitly, we have that $\langle \chat|0;0,0,1;0,0 \rangle$ is
just the constant function. For $L=1$ we have the nine
functions $c^a_i$. 
With the help of the raising and lowering operators
$L^S_\pm$ and $L^R_\pm$
we arrive at the nine functions $\langle \chat|1;1,1,1;m_s,m_r \rangle$
which are the appropriate complex linear combinations 
$\{c^+_+, c^0_+, c^-_+, c^+_0, \cdots, c^-_- \}$ of the variables $c^a_i$.
We have \eg $L^S_3 c^-_+ = - c^-_+$ and $L^R_3 c^-_+ = c^-_+$.
\par
To construct all representations
$(L,l_s,l_r,\tau)$ for given $(L,l_s,l_r)$, we proceed as follows.
We construct the eigenspace 
$A^{l_s}_{l_r}$ of the operators $L^S_3$ and $L^R_3$.
This is the space span by all monomials of degree $L$ 
in the complex $c$ variables
that have the eigenvalues $l_s$ and $l_r$ under the operators
$L^S_3$ and $L^R_3$. 
Within $A^{l_s}_{l_r}$,
but now regarded as functions of $c$ in stead of $\chat$,
we construct the intersection of the kernels of the operators
$L^S_- L^S_+$, $L^R_- L^R_+$ and $ (c^a_i c^a_i \Dl)$. 
The first two operators leave us with only those combinations
of the monomials that have the prescribed $(\vec{L}^S)^2$
and $(\vec{L}^R)^2$ eigenvalues; the laplacian imposes the
constraint that the polynomials must be harmonic: $\Dl = 0$.
If this intersection consists of more than one function, we
use explicit Gram-Schmidt orthogonalization to
arrive at the functions $\langle \chat|L;l_s,l_r,\tau;l_s,l_r\rangle$.
With the lowering operators, we trivially construct the rest
of the functions $\langle \chat|L;l_s,l_r,\tau;m_s,m_r\rangle$.
\par
Most calculations were done in \mathem, but
we used \cee\ to perform the construction
of the intersection of the kernels of 
$L^S_- L^S_+$ and $L^R_- L^R_+$. 
For this we used the explicit projectors on the kernel
\beq
  P^S_{l_s} = \prod_{j = l_s+1}^L \left( \id - 
\frac{1}{j(j+1) - l_s (l_s + 1)} L_-^S L_+^S \right),
\eeq
and the analogous expression for $P^R_{l_r}$.
Since all calculations had to be done exactly, we
used arithmetic with (large) integers. In this way we
explicitly constructed the representations $(L,l_s,l_r,\tau)$
for $L \leq 10$ and $l_r \leq 2$. As will be apparent from
the sequel, we did not need higher values for $l_r$.
\par
We construct functions of both $\chat$ and $\dhat$ by using
the familiar rules of adding angular momenta.
Let $i$ denote a representation $(L;l_s,l_r,\tau)$ and consider the functions
$\langle \chat | i_1 ; m_s,m_r \rangle$ and 
$\langle \dhat | i_2 ; m'_s,m'_r \rangle$.
Using Clebsch-Gordan coefficients,
we can define a function $Y^{i_1i_2}(\chat,\dhat)$ which is an eigenfunction of
$\vec{J}^R$ and of $\vec{J}^S$. We will limit the construction
to functions with $\vec{J}^S = 0$, as required by residual gauge symmetry.
This implies that the functions of $c$ and $d$ 
need to have the same $l_s$, which restricts the possible combinations of 
$i_1$ and $i_2$. The resulting function is given by
\bea
Y^{i_1i_2}(\chat,\dhat) &=& \langle \chat \dhat | j,m,l_s;L_1,l_1,\tau_1;L_2,l_2,\tau_2 \rangle \Next
  &=& \sum_{m_s = -l_s}^{l_s} \sum_{m_1 = -l_1}^{l_1} \sum_{m_2 = -l_2}^{l_2}
  (-1)^{l_1 - l_2 + m} \sqrt{2 j + 1} 
   \left( \begin{array}{ccc}
     l_1 & l_2 & j \\ 
     m_1 & m_2 & -m 
   \end{array} \right) \times \Next
  && \frac{(-1)^{l_s - m_s}}{\sqrt{2 l_s + 1}} 
    \langle \chat | L_1;l_s,l_1,\tau_1;m_s,m_1 \rangle 
       \langle \dhat | L_2;l_s,l_2,\tau_2;-m_s,m_2 \rangle. 
\label{cddef}
\eea
Its eigenvalues under the various symmetries are collected in
table~\ref{angtabelcd}. Note the behaviour under parity:
\beq
  \Ss{P} Y^{i_1 i_2} = (-1)^{l_1 + l_2 + j}  Y^{i_2 i_1}.
  \label{parbehav2}
\eeq
\begin{table}[t]
\begin{center}
\begin{tabular}{||l||c|c|c|c|c||} \hline
Operator & 
$(\vec{L}^S_c)^2$ & $(\vec{L}^S_d)^2$ & $(\vec{J}^S)^2$ & $J^S_3$ &
$\Dl_{\chat}$ \\ \hline
Eigenvalue & 
$l_s ( l_s + 1) $ & $l_s ( l_s + 1) $ & $0$ & $0$  & 
$ - L_1 ( L_1 + 7) $ \\ \hline
\multicolumn{6}{c}{} \\ \hline
Operator & 
$(\vec{L}^R_c)^2$ & $(\vec{L}^R_d)^2$ & $(\vec{J}^R)^2$ & $J^R_3$ & 
$\Dl_{\dhat}$\\ \hline
Eigenvalue & 
$l_1 ( l_1 + 1) $ & $l_2 ( l_2 + 1) $ & $j (j+1)$ & $m$ & 
$ - L_2 ( L_2 + 7) $\\ \hline
\end{tabular} 
\end{center}
\caption{Behaviour of the functions 
$\langle \chat \dhat | j,m,l_s;L_1,l_1,\tau_1;L_2,l_2,\tau_2 \rangle$.}
\label{angtabelcd}
\end{table}

\subsubsection{The radial sector}

The strong coupling limit of our hamiltonian problem consists
of the eigenvalue problem for
the kinetic part of the hamiltonian of eq.~(\ref{hamrad}).
If we assume a solution $\fie(r_c) \fie(r_d) 
Y^{i_1i_2}(\chat,\dhat)$, the reduced one-dimensional eigenvalue
problem becomes
\beq
  \left( \frac{\pr^2}{\pr r^2} + \lm  - \frac{( 12 + L (L+7) )}{r^2} \right)
    \fie(r) = 0,
\eeq
whose regular solution is $\fie^{(L)}_\gm(r) = \gm r j_{3+L} (\gm r)$,
with $j_p(z)$ the spherical Bessel function of order $p$ and $\lm = \gm^2$.
The eigenfunctions of the kinetic part of the hamiltonian are thus given by
\beq
  \fie^{(L_1)}_{\gm_1}(r_c)  \fie^{(L_2)}_{\gm_2}(r_d) 
   Y^{i_1i_2}(\chat,\dhat),
  \label{eigtrue}
\eeq
with the energies given by $E = \frac{f}{2} ( \gm_1^2 + \gm_2^2 )$.
\par
During the variational stage of the calculation, however, 
the use of spherical Bessel functions of different order will lead to a 
large number of integrals. Therefore we take the radial functions to
be independent of $L_1$ and $L_2$ and define
\beq
\psi^{i_1i_2}_{\gm_1\gm_2}(c,d) = 
 \fie_{\gm_1}(r_c)  \fie_{\gm_2}(r_d) Y^{i_1i_2}(\chat,\dhat),
\label{eigappr}
\eeq
with $\fie_\gm(r) = \gm r j_3(\gm r)$. These functions are not
eigenfunctions of the kinetic part of the hamiltonian and they will
have discontinuities of the following kind. For $r \downarrow 0$,
we have that $\fie_\gm(r) \sim r^4$. The wave function $\Psi$ behaves
as
$r^{-4} \fie_\gm(r) \langle \chat |L;l_s,l_r,\tau;m_s,m_r\rangle$,
and this function will be discontinuous at $r = 0$.
The variational functions thus have discontinuities at $r_c = 0$ 
and $r_d = 0$. These discontinuities form a set of measure zero, and a
variational calculation will not feel them.
Since the functions $\fie_\gm(r)$ are the eigenfunctions of
the reduced one-dimensional problem with $L=0$, they still
constitute a complete set of functions.
\par
The behaviour under parity can be obtained from eq.~(\ref{parbehav2})
and is given by
\beq
\Ss{P} \psi^{i_1i_2}_{\gm_1\gm_2} = (-1)^{l_1 + l_2 + j}
   \psi^{i_2i_1}_{\gm_2\gm_1}.
\label{parbehav}
\eeq
Taking even and odd combinations gives
\beq
\psi^{pi_1i_2}_{\gm_1\gm_2}(c,d) = 
 \psi^{i_1i_2}_{\gm_1\gm_2}(c,d) + p~\psi^{i_1i_2}_{\gm_1\gm_2}(d,c).
\eeq
We implement the boundary conditions eq.~(\ref{bcgen1}) 
and~(\ref{bcgen2}) for $\theta = 0$ by imposing
the following conditions on $\gm_1$ and $\gm_2$:
\bea
  p = -1 &:&  \fie_{\gm_1}(\wdrie) 
            = \fie_{\gm_2}(\wdrie) = 0, \\
  p = 1 \hphantom{-} &:&  
             \frac{\pr ( r^{-\frac{5}{2}} \fie_{\gm_1})}{\pr r} (\wdrie) =
             \frac{\pr ( r^{-\frac{5}{2}} \fie_{\gm_2})}{\pr r} (\wdrie) = 0. 
             \label{bcspec}
\eea
These conditions are expected to be accurate as long as the
wave function transverse to the sphaleron path (near the sphalerons) is
predominantly in its ground state.
The case $\theta = \pi$ can be treated along the same lines as $\theta = 0$
by interchanging the boundary conditions for the cases $p = 1$ and $p = -1$.
\par
The exact strong coupling results for the lowest levels
in some sectors for $\theta = 0$ are collected in table~\ref{strong}. 
These values are obtained by imposing the boundary conditions above 
on the true eigenfunctions of eq.~(\ref{eigtrue}):
the values of $\gm_1$ and
$\gm_2$ are hence dependent on $L_1$ and $L_2$ respectively.
\begin{table}[t]
\begin{center}
\begin{tabular}{||r@{-}l|c|c|c|r@{.}l||} \hline
\multicolumn{2}{||l|} {sector} & 
$|j,m,l_s;L_1,l_1,\tau_1;L_2,l_2,\tau_2 \rangle$ &
$\gm_1$  & $\gm_2$ & \multicolumn{2}{|c||}{$E$} \\ \hline
000&even & $|0,0,0;0,0,1;0,0,1 \rangle$ & $1.9786$ & $1.9786$ & $3$&$9149 f$ \\
000&even & $|0,0,0;0,0,1;3,0,1 \rangle$ & $1.9786$ & $4.1215$ & $10$&$4508 f$ \\
000&odd  & $|0,0,0;0,0,1;0,0,1 \rangle$ & $4.0345$ & $6.0143$ & $26$&$2246 f$ \\
112&even & $|2,m,1;1,1,1;1,1,1 \rangle$ & $2.7406$ & $2.7406$ & $7$&$5108 f$ \\
112&odd  & $|2,m,1;1,1,1;2,1,1 \rangle$ & $4.7242$ & $5.4016$ & $25$&$7476 f$ \\
022&even & $|2,m,0;0,0,1;2,2,1 \rangle$ & $1.9786$ & $3.4456$ & $7$&$8936 f$ \\
022&odd  & $|2,m,0;0,0,1;2,2,1 \rangle$ & $4.0345$ & $5.4016$ & $22$&$7272 f$ \\
\hline
\end{tabular}
\end{center}
\caption{Strong coupling limit: lowest energy levels in some sectors.}
\label{strong}
\end{table}
\par
For general $\theta$ we multiply $\psi^{p i_1 i_2}_{\gm_1\gm_2}(c,d)$
with a phase factor $\exp(i \theta \alpha(r_c,r_d))$. The function $\al$
is a kind of Chern-Simons functional that gives the correct behaviour to
the wave function under large gauge transformations.
The resulting
functions no longer have well-defined parity, but they do obey the 
general boundary conditions for suitable $\alpha$.  
Also the hermiticity of the hamiltonian for these
functions can be checked explicitly.
Sufficient conditions
on $\alpha$ are: $\alpha(r_c,r_d) = - \alpha(r_d,r_c)$ and 
$\alpha(\wdrie,0) = \hf$. We choose 
\beq
   \alpha(r_c,r_d) = \frac{1}{2} \left( 
     \left(\frac{r_c}{\wdrie} \right)^\beta - 
     \left(\frac{r_d}{\wdrie} \right)^\beta \right). 
   \label{alphadef}
\eeq
For $\beta \rightarrow \infty$ we approach the situation that
the phase factor over the entire edge is constant
and equal to $e^{i \theta}$. But already for the choice $\beta = 2$,
the boundary conditions at the sphalerons are taken into account properly.

\subsection{Matrix elements}

In this section we will describe the calculation of the
matrix elements of \Ss{H} and of $\Ss{H}^2$. 
The hamiltonian $\Ss{H} = \Ss{K} + \Ss{V}$ is given by
eq.~(\ref{hamrad}). 
We let $| n \rangle$ denote the normalized basis function:
\beq
|n \rangle \propto 
  \left( \psi^{i_1i_2}_{\gm_1\gm_2}(c,d) 
     + p (-1)^{l_1 + l_2 + j} \psi^{i_2i_1}_{\gm_2\gm_1}(c,d)
   \right) \exp(i \theta \alpha(r_c,r_d)),
\eeq
with $\psi^{i_1i_2}_{\gm_1\gm_2}(c,d)$ given by eq.~(\ref{eigappr}).
We take $\alpha$ from eq.~(\ref{alphadef}) with $\beta = 2$:
$\alpha(r_c,r_d) = \sfrac{1}{6} \left( r_c^2 - r_d^2 \right)$.
\par
Consider first the matrix element
$\langle n' | \Ss{H} | n \rangle$. For the potential energy 
\Ss{V} the phase factor $\exp(i \theta \alpha(r_c,r_d))$
cancels against its complex conjugate. For the kinetic operator
we obtain
\bea
&& \terug \exp(-i \theta \alpha(r_c,r_d)) \,
 \Ss{K} \left(\psi^{i_1i_2}_{\gm_1\gm_2}(c,d)
 \exp(i \theta \alpha(r_c,r_d)) \right) = \Next
&&  \quad  \hphantom{+} \frac{f}{2} 
  \left( \gm_1^2  + \gm_2^2 + \frac{L_1 ( L_1 + 7)}{r_c^2}
 + \frac{L_2 ( L_2 + 7)}{r_d^2} \right) \psi^{i_1i_2}_{\gm_1\gm_2}(c,d) \Next
&&  \quad + \frac{f}{2} \left(
 - i \theta \, \frac{2}{3} 
    \left( r_c \frac{\pr \fie_{\gm_1}(r_c)}{\pr r_c} \fie_{\gm_2}(r_d)
  -  \fie_{\gm_1}(r_c) r_d \frac{\pr \fie_{\gm_2}(r_d)}{\pr r_d} \right)
  \right) Y^{i_1i_2}(\chat,\dhat)  \Next
&& \quad + \frac{f}{2} \left( \theta^2 \, \frac{1}{9} (r_c^2 + r_d^2) 
   \fie_{\gm_1}(r_c) \fie_{\gm_2}(r_d)\right) Y^{i_1i_2}(\chat,\dhat).
\eea
\par
  To apply Temple's inequality, we need the matrix elements of $\Ss{H}^2$.
Using the hermiticity of \Ss{H} we write
\bea
\langle n' | \Ss{H}^2 | n \rangle 
&=& \langle (\Ss{K} + \Ss{V}) n' |  (\Ss{K} + \Ss{V}) n \rangle \Next
&=& \half \langle \Ss{K} n' |  \Ss{K} n \rangle 
     +    \langle n' | \Ss{V} \Ss{K} | n \rangle + 
    \half \langle n' | \Ss{V}^2 | n \rangle + \mbox{ hermitian conjugate}.
\eea
From these expressions we can read off which matrix elements we have 
to calculate.
\par
Since the potential $\Ss{V}(c,d)$ is invariant under $\vec{L}^R_c$ and
$\vec{L}^R_d$ as well as under $\Jvec^S$, 
all the terms in $\Ss{V}$ and $\Ss{V}^2$ are of the form
\beq
 \Ss{A} =  \langle \chat \dhat | 
   0,0,\ltil_s;\Ltil_1,0,\tautil_1;\Ltil_2,0,\tautil_2 \rangle.
\label{scrAdef}
\eeq
By virtue of the construction in eq.~(\ref{cddef}) all
the angular integrations over $\chat$ and $\dhat$ can
be reduced to angular integrations in $\real^9$. These
integrals can be expressed in terms of reduced matrix elements.
The details on this procedure can be found in appendix~\ref{appreduced}.
\par
The radial integrals we need to compute are
\bea
  J(n,\gm',\gm) &=& 
  \int_0^{\sqrt{3}} dr \, r^n \hat{\fie}_{\gm'}(r) \hat{\fie}_{\gm}(r), \\
  J_l(n,\gm',\gm) &=& 
  \int_0^{\sqrt{3}} dr \, r^n \hat{\fie}_{\gm'}(r) 
    \left( r \frac{\pr}{\pr r}\hat{\fie}_{\gm}(r)\right), \\
  J_{ll}(n,\gm',\gm) &=& 
  \int_0^{\sqrt{3}} dr \, r^n 
    \left( r \frac{\pr}{\pr r} \hat{\fie}_{\gm'}(r)\right)
    \left( r \frac{\pr}{\pr r}\hat{\fie}_{\gm}(r)\right),
\eea
where $\hat{\fie}_\gm$ is normalized to one, and
\bea
\fie_\gm(r) &=& \gm r j_3( \gm r) = f( \gm r), \\
f(z) &=& \left( 1 - \frac{15}{z^2} \right) \cos(z)
   + \left( -\frac{6}{z} + \frac{15}{z^3} \right) \sin(z).
\eea
Consider the integral for non-normalized $\fie$ functions and
let $\al = \sqrt{3} \gm$ and $\al' = \sqrt{3} \gm'$.
For an explicit value of $n$, but unspecified values of
$\gm'$ and $\gm$, we write
\beq
\tilde{J}(n) 
=  \int_0^{\sqrt{3}} dr \, r^n \fie_{\gm'}(r) \fie_{\gm}(r) 
= (\sqrt{3})^{n+1} \int_0^1 dx \,x^n f( \al' x) f(\al x).
\eeq
We express products of sines and cosines of $\al x$ and $\al' x$
in sines and cosines of $s x$ and $v x$ with $s = \al' + \al$
and $v=\al' - \al$. The integrand as it stands is regular
at $x=0$, but individual terms need not be. We therefore
introduce a cut-off $\veps$, after which the integral is a sum
of functions
\beq
  \cos(m,a) = \int_\veps^1 dx \, x^m \cos( a x), \quad
  \sin(m,a) = \int_\veps^1 dx \, x^m \sin( a x). 
\eeq
Using recursion formulae obtained from partial integration
allows us to eliminate the functions $\cos(m,a)$ and $\sin(m,a)$.
After taking the limit $\veps \downarrow 0$,
the result is an expression in the functions
$\sin$, $\cos$, $\mbox{Si}$, $\mbox{Ci}$ and $\ln$ of the variables
$s$ and $v$. By substituting values for $s$ and $v$ and performing
the normalization we construct tables for $J(n,\gm',\gm)$.
Note that we have to be careful for $\al' = \al$: we have to take the
limit $v \rightarrow 0$ before substituting actual values.
\par
The other integrals $\tilde{J}_l(n)$ and $\tilde{J}_{ll}(n)$
can be obtained as follows:
\bea
\tilde{J}_l(n) 
&=&  \int_0^{\sqrt{3}} dr \, r^n \fie_{\gm'}(r) 
\left(r \frac{\pr}{\pr r} \fie_{\gm}(r) \right) \Next
&=& (\sqrt{3})^{n+1} \int_0^1 dx \, x^n f( \al' x) 
\left(x \frac{\pr}{\pr x} f(\al x) \right) \Next
&=& \al \frac{\pr}{\pr \al} \tilde{J}(n)  \Next
&=& \frac{s-v}{2} 
\left( \frac{\pr}{\pr s}  - \frac{\pr}{\pr v} \right) \tilde{J}(n).
\eea

\subsection{Weak coupling expansion}

In this section we will perform the perturbative calculation of
the energy levels of the hamiltonian. The results will serve
as a check on the variational calculation in the regime of
small coupling constant. 
Starting from the strong coupling basis,
we could also develop a perturbation theory in $\frac{1}{f}$, but there
are two reasons why this is less interesting.
First, we are interested in the region where we just start to
see the deviations from perturbative behaviour in $f$.
Second, since our variational basis is in essence a strong coupling 
basis, the reproduction of the strong coupling limit does not give
a strong check on the variational calculation.
\par
For the weak coupling limit, we rescale $x_c = \sqrt{\frac{2}{f}} r_c$
and $x_d = \sqrt{\frac{2}{f}} r_d$ and we split the hamiltonian
of eq.~(\ref{hamrad}) in $\Ss{H} = \Ss{H}_0 + \Ss{H}_1$, with
\bea
  \Ss{H}_0 &=& - \left( 
  \frac{\pr^2}{\pr x_c^2} + \frac{\pr^2}{\pr x_d^2} -
  12 \left(\frac{1}{x_c^2} + \frac{1}{x_d^2}\right) +
  \frac{1}{x_c^2} \Dl_{\chat} + \frac{1}{x_d^2} \Dl_{\dhat}
   \right) + x^2_c + x^2_d, \\
 \Ss{H}_1(c,d) &=& \Ss{H}_1(c) + \Ss{H}_1(d) + 
    \lm_7 \left(\sfrac{f}{2}\right)^2 x_c^2 x_d^2 
  + \lm_8 \left(\sfrac{f}{2}\right)^2 x_c^2 x_d^2 \inv_2(\chat,\dhat), \\
 \Ss{H}_1(c) &=& 
    \left(\sfrac{f}{2} \lm_1 - 1\right) x_c^2 
  + \left(\sfrac{f}{2}\right)^{\frac{3}{2}} \lm_2 x_c^3 \inv_3(\chat) 
  + \left(\sfrac{f}{2}\right)^{2}   \lm_3 x_c^4 \Next
&& + \left(\sfrac{f}{2}\right)^{2}   \lm_4 x_c^4 \inv_4(\chat)
  + \left(\sfrac{f}{2}\right)^{\frac{5}{2}} \lm_5 x_c^5 \inv_3(\chat)
  + \left(\sfrac{f}{2}\right)^{3}   \lm_6 x_c^6.
\eea
\par
Diagonalizing $\Ss{H}_0$ leads to the following
one-dimensional eigenvalue problem:
\beq
  \left( \frac{\pr^2}{\pr x^2} + 
     \lm  - \frac{12 + L ( L+7)}{x^2} - x^2 \right)
    \fie(x) = 0.
\eeq
The regular solution of this equation is \cite{abr}
\beq
  \fie(x) = e^{-\hf x^2} {x^{L+4}} 
    {_1F_1}(- \frac{\lm - 2 L - 9}{4},L + \nhalf,x^2),
\eeq
with ${_1F_1}$ the confluent hypergeometric function.
As in the strong coupling case, we can use the boundary conditions
at $x=\sqrt{\frac{6}{f}}$ to discretize the possible values
for $\lm$ and hence the energy. 
This would give us an alternative to the strong
coupling basis constructed earlier. Disadvantages of this basis
are that the functions themselves depend on the coupling constant
$f$. Also the matrix elements of operators between two 
confluent hypergeometric functions are hard to 
calculate. For the variational method, we will therefore use
the strong coupling basis.
\par
We can however use the weak coupling basis for perturbation
purposes. For $f \downarrow 0$ the location where the boundary
conditions have to be imposed moves away to infinity, which is equivalent to
saying that the wave function, in the original coordinates, is strongly
localized around $c=d=0$.
We can replace the boundary conditions at $x=\sqrt{\frac{6}{f}}$
by demanding normalizability of the functions  $\fie(x)$.
This means that
the $_1 F_1$ function must reduce to a polynomial, which is the case
if its first argument is $-n$ with $n=0,1,\ldots$.
The $_1 F_1$ function then reduces to a Laguerre polynomial
and we obtain the discretization $\lm = 9 + 2 L + 4 n$.
The weak coupling eigenfunctions are thus
\beq
  \psi^{i_1 i_2}_{n_1 n_2} = 
  L^{(L_1+\frac{7}{2})}_{n_1}(x_c^2)  L^{(L_2 + \frac{7}{2})}_{n_2}(x_d^2) 
  x_c^{(L_1 + 4)} x_d^{(L_2 + 4)} e^{- \hf x_c^2} e^{- \hf x_d^2}
   Y^{i_1i_2}(\chat,\dhat),
\label{weakbasis}
\eeq
and the energies are given by 
\beq
  E = 18 + 2 (L_1 + L_2) + 4 ( n_1 + n_2).
\label{weakenergies}
\eeq
\par
We use Blochs formulation~\cite{blo,lus2} of perturbation theory
to calculate the weak coupling expansion of the
energy levels of the hamiltonian.
For this we need the matrix elements of $\Ss{H}_I$
w.r.t.\ the basis of eq.~(\ref{weakbasis}). 
The angular
integrations are treated in appendix~\ref{appreduced}. 
The radial integrals take the form
\beq
  \int_0^\infty dx \, x^n e^{-x^2} x^{L'+4} x^{L+4} 
 L^{(L'+\frac{7}{2})}_{n'}(x^2) L^{(L+\frac{7}{2})}_{n}(x^2).
\eeq
We calculate these integrals by substituting 
the explicit form of the Laguerre polynomials.
\par
Using orthogonality properties of the Laguerre polynomials
and of the angular functions,
one can show that the number of intermediate levels that
we have to take into account is finite.
To be more precise: given the values $E$ and $L$ for 
an unperturbed level and the requested accuracy in terms
of powers of the
coupling constant, one can calculate maximum values $E_{\mbox{ \scriptsize max}}$ and $L_{\mbox{ \scriptsize max}}$ beyond which there will be
no contribution up to the requested accuracy.
\par
We will give the perturbative results for those levels which
will turn out to correspond to the lowest glueball masses:
\bea
 E_{\mbox{000-even}}^{(0)} &=& 18 +
  f\,\left( \sfrac{9}{8} + 
      \sfrac{9}{2}\,\kappa_1 \right)  + 
   {f^2}\,\left( \sfrac{63}{512} - 
      \sfrac{9}{16}\,{{\kappa_1}^2} - 
      \sfrac{3}{8}\,\kappa_2  
\right. \Next && \quad \left.+ 
      \sfrac{99}{8}\,\kappa_3 + 
      \sfrac{63}{8}\,\kappa_4 + 
      \sfrac{81}{16}\,\kappa_7 + 
      \sfrac{27}{16}\,\kappa_8 \right) + \Order{f^3}, \\
 E_{\mbox{000-even}}^{(1)} &=& 22+
  f\,\left( -\sfrac{9}{8} + 
      \sfrac{11}{2}\,\kappa_1 \right)  + 
   {f^2}\,\left( \sfrac{903}{512} + 3\,\kappa_1 - 
      \sfrac{11}{16}\,{{\kappa_1}^2} - 
      \sfrac{13}{8}\,\kappa_2 
\right. \Next && \quad \left.  + 
      \sfrac{165}{8}\,\kappa_3+ 
      \sfrac{105}{8}\,\kappa_4 + 
      \sfrac{135}{16}\,\kappa_7 + 
      \sfrac{45}{16}\,\kappa_8 \right) + \Order{f^3}, \label{blocheven} \\
 E_{\mbox{000-odd}}^{(0)} &=& 22+
  f\,\left( -\sfrac{13}{8} + 
      \sfrac{11}{2}\,\kappa_1 \right)  + 
   {f^2}\,\left( \sfrac{1063}{512} + 
      \sfrac{13}{4}\,\kappa_1 - 
      \sfrac{11}{16}\,{{\kappa_1}^2} - 
      \sfrac{13}{8}\,\kappa_2 
\right. \Next && \quad \left. + 
      \sfrac{165}{8}\,\kappa_3  + 
      \sfrac{105}{8}\,\kappa_4 + 
      \sfrac{99}{16}\,\kappa_7 + 
      \sfrac{33}{16}\,\kappa_8 \right) + \Order{f^3}, \label{blochodd} \\
 E_{\mbox{112-even}}^{(0)} &=& 22+
  f\,\left( \sfrac{13}{24} + 
      \sfrac{11}{2}\,\kappa_1 \right)   + 
   {f^2}\,\left( \sfrac{3167}{4608} + 
      \sfrac{25}{24}\,\kappa_1 - 
      \sfrac{11}{16}\,{{\kappa_1}^2} - 
      \sfrac{7}{8}\,\kappa_2  
\right. \Next && \quad \left. + 
      \sfrac{143}{8}\,\kappa_3 + 
      \sfrac{91}{8}\,\kappa_4 + 
      \sfrac{121}{16}\,\kappa_7 + 
      \sfrac{47}{16}\,\kappa_8 \right) + \Order{f^3}.
\eea

\subsection{Results}

With the results of the previous sections, we are in a position
to actually perform the variational calculation.
A \fort\ programme constructed the truncated matrix for
the hamiltonian and performed the diagonalization
using routines from the Nag-library and the Eispack library.
The output consisted of the upper bounds $\tilde{\mu}_n$ for
the lowest energy eigenvalues and of the corresponding values for
$\eta_n$ (used for computing lower bounds).
We also produced graphical representations of the wave function.
These can be used in verifying a posteriori whether the
assumption is true that the boundary conditions
are only felt at or near the sphalerons.

\subsubsection{Lowest-order hamiltonian}

\begin{figure}  \centering
  \epsfxsize=0.8\textwidth
  \leavevmode
  \epsfbox[72 222 540 570]{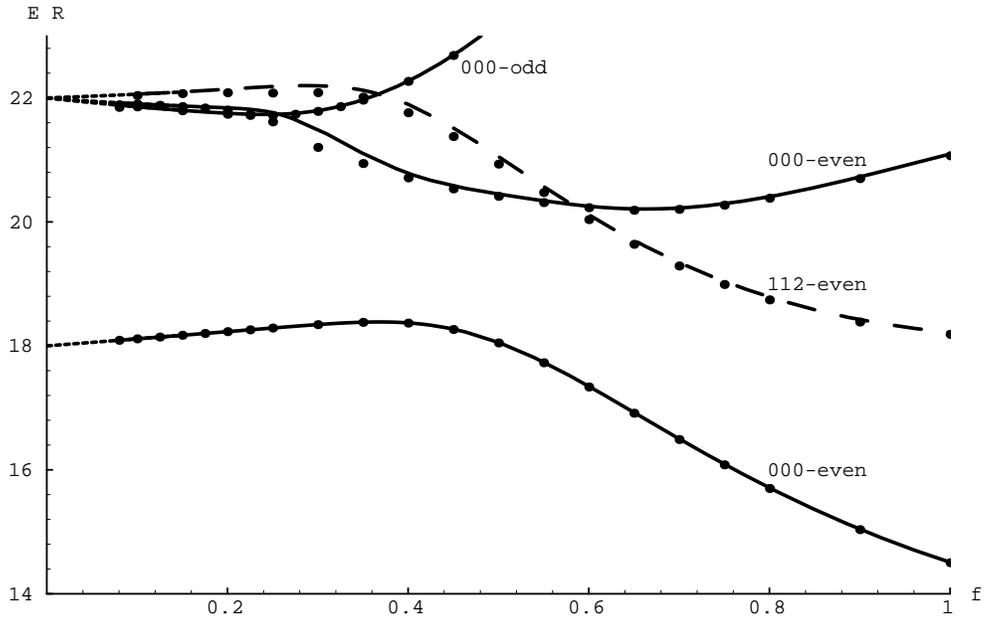}
  \caption{Lowest energy levels for $\theta = 0$. 
  Drawn curves correspond to levels
  in the $(0,0,0)$ sector. The dashed curve denotes the ground level in the
  $(1,1,2)$-even sector. The short-dashed curves are the
  perturbative expansions, and the individual dots are lower
  bounds on the levels as obtained by Temple's inequality.}
  \label{niveaux}
\end{figure}

We start by considering the lowest order results.
After calculating the lowest levels in various sectors, we find
that the lowest glueball masses are to be found
in the sectors 000-even, 000-odd and 112-even for respectively the
scalar and the tensor glueballs.
These levels, including the lower bounds and the perturbative
expansions, are plotted in fig.~\ref{niveaux}.
Note that the crossing of levels only happens
for levels that are in different sectors. States within the same
sector would exhibit avoided level crossing due to level repulsion.
\par
For small coupling, the deviations of the wave function
from the perturbative one will be small. 
To reproduce the energy eigenvalues correctly, we
can suffice with a relatively low number of angular functions,
but to be able to accommodate the strong localization around
$c=d=0$, we need a large number of radial functions.
For larger values of $f$ we can reduce the number of radial functions,
but we must increase the number of angular functions.
The data we used was obtained by combining
different runs with various choices for the truncated basis
as controlled by the parameters 
$N_{\mbox{rad}}$ (number of radial functions) and 
$L_{\mbox{sum}}$ (limit on $L_1 + L_2$).
\par
In the 000-even sector we maximally used 
$N_{\mbox{rad}} = 150 $ and $L_{\mbox{sum}} = 10$ for small $f$, and 
$N_{\mbox{rad}} = 50 $ and $L_{\mbox{sum}} = 14$  for larger values of $f$.
These values correspond to  bases consisting of over 3000
vectors. Increasing the number of basis vectors
becomes quickly limited by the amount of free memory available
in the computer. 
For the case $\theta \neq 0$,
the hamiltonian is complex and we need a factor of two more
memory.
In the 112-even sector, the same number for  $L_{\mbox{sum}}$
resulted in higher numbers of angular functions.
Here we went up to 
$N_{\mbox{rad}} = 150 $ and $L_{\mbox{sum}} = 8$ for small $f$, and 
$N_{\mbox{rad}} = 30 $ and $L_{\mbox{sum}} = 12$  for larger values of $f$.
\par
We did most of the calculations on Sun workstations.
For some values of $f$ we wanted more accurate results
and used computer time on a Cray. We also developed
a pruning technique: after diagonalization, we examine the
variational wave function obtained for, say, the ground state.
We replace those basis vectors (roughly ten to twenty percent) 
that have a small coefficient in this wave function by 
new basis vectors and repeat the diagonalization. 
Repeating this procedure 
allows us to sample a larger variational basis, but we do
not have the guarantee that the results will improve in this way:
it is possible that the true wave function has substantial overlap
with very many vectors in our basis.
\par
\begin{figure}  \centering
  \epsfxsize=0.8\textwidth
  \leavevmode
  \epsfbox[72 222 540 570]{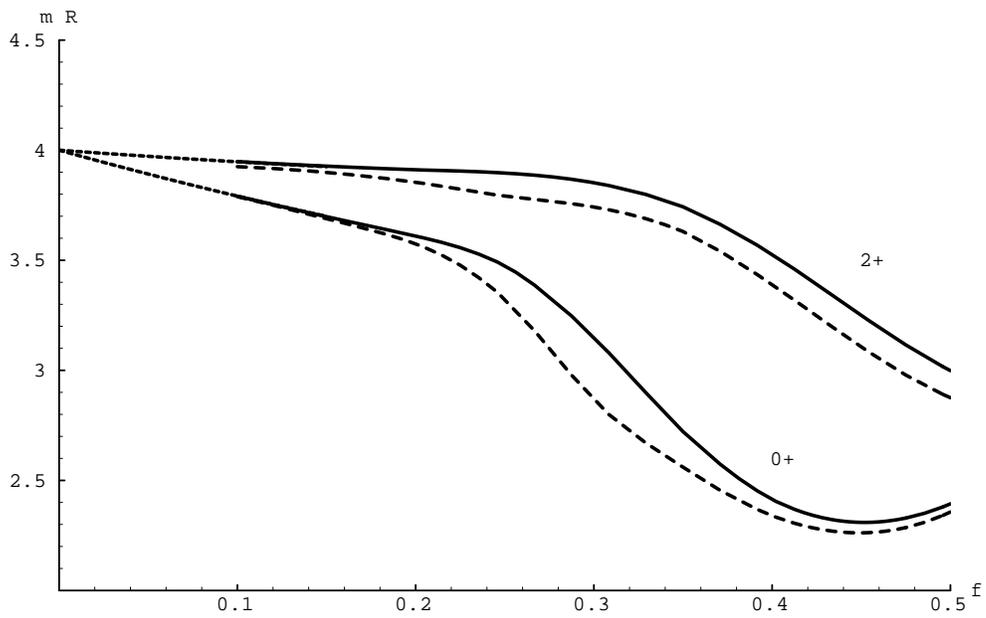}
  \caption{Glueball masses for $\theta = 0$ as a function of 
  the coupling constant. 
  The drawn curves are the masses of the 
  first scalar ($0^+$) and
  tensor ($2^+$) glueball. The dashed lines denote the lower bounds, the dotted
  lines the perturbative results.}
  \label{massas}
\end{figure}
\begin{figure}  \centering
  \epsfxsize=0.8\textwidth
  \leavevmode
  \epsfbox[72 222 540 570]{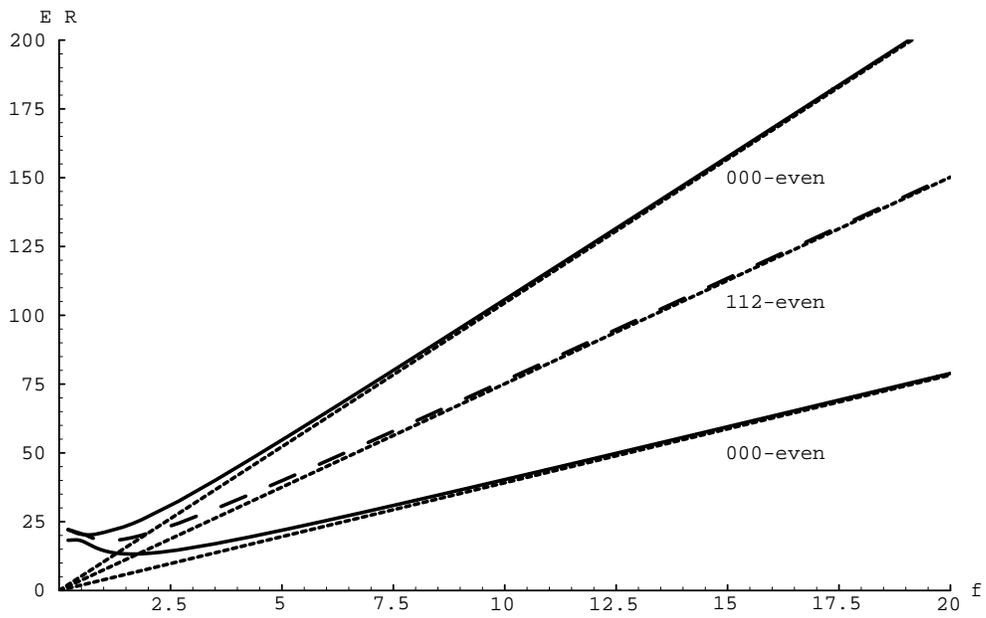}
  \caption{Strong coupling limit, $\theta = 0$: 
  the drawn and dashed lines are
  the variational results for the lowest levels in the sectors
  000-even and 112-even respectively. The dotted
  lines represent the analytic strong coupling limit.}
  \label{niveauxstrong}
\end{figure}

Returning to our calculation,
the lowest-lying scalar ($j = 0$) and tensor ($j=2$) levels 
are found in respectively the sectors 000 and 112.
For $\theta = 0$, the vacuum corresponds to the ground state of 
the 000-even sector.
The scalar glueball $0^+$ can be identified with the 
first excited state in the 000-even sector, the tensor glueball $2^+$
with the ground state in the 112-even sector.
Note that at $f=0.6$ these levels cross, thus making the scalar
glueball heavier than the tensor. We do, however, not expect
our model to be valid anymore for these values of the coupling constant.
Taking differences of the energy levels
with respect to the ground state in the 000-even sector gives us
the masses of the low-lying glueball states (fig.~\ref{massas}).
To check on the strong coupling results, we also 
performed variational calculations for large values of $f$.
The comparison with the results of table~\ref{strong} can be seen
in fig.~\ref{niveauxstrong}. 
\par
\begin{figure}  \centering
  \epsfxsize=0.8\textwidth
  \leavevmode
  \epsfbox[72 222 540 570]{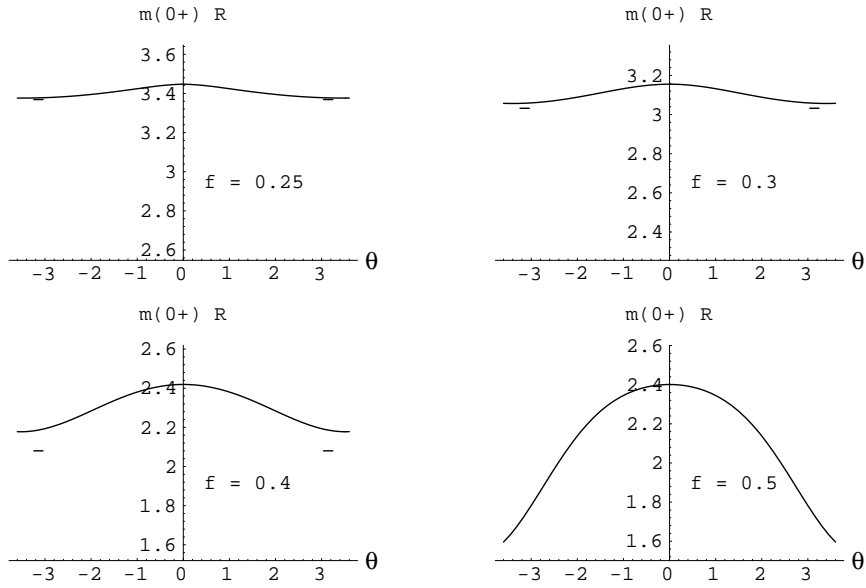}
  \caption{Scalar glueball mass at $f=0.25$, $f=0.3$, $f= 0.4$ and
  $f=0.5$ as a function of $\theta$. The dashes at $\theta = \pm \pi$
  denote the variational result with exact implementation of the
  boundary conditions. Note that the vertical range in these
  four plots is the same.}
  \label{theta}
\end{figure}
\begin{figure}  \centering
  \epsfxsize=0.8\textwidth
  \leavevmode
  \epsfbox[72 272 540 520]{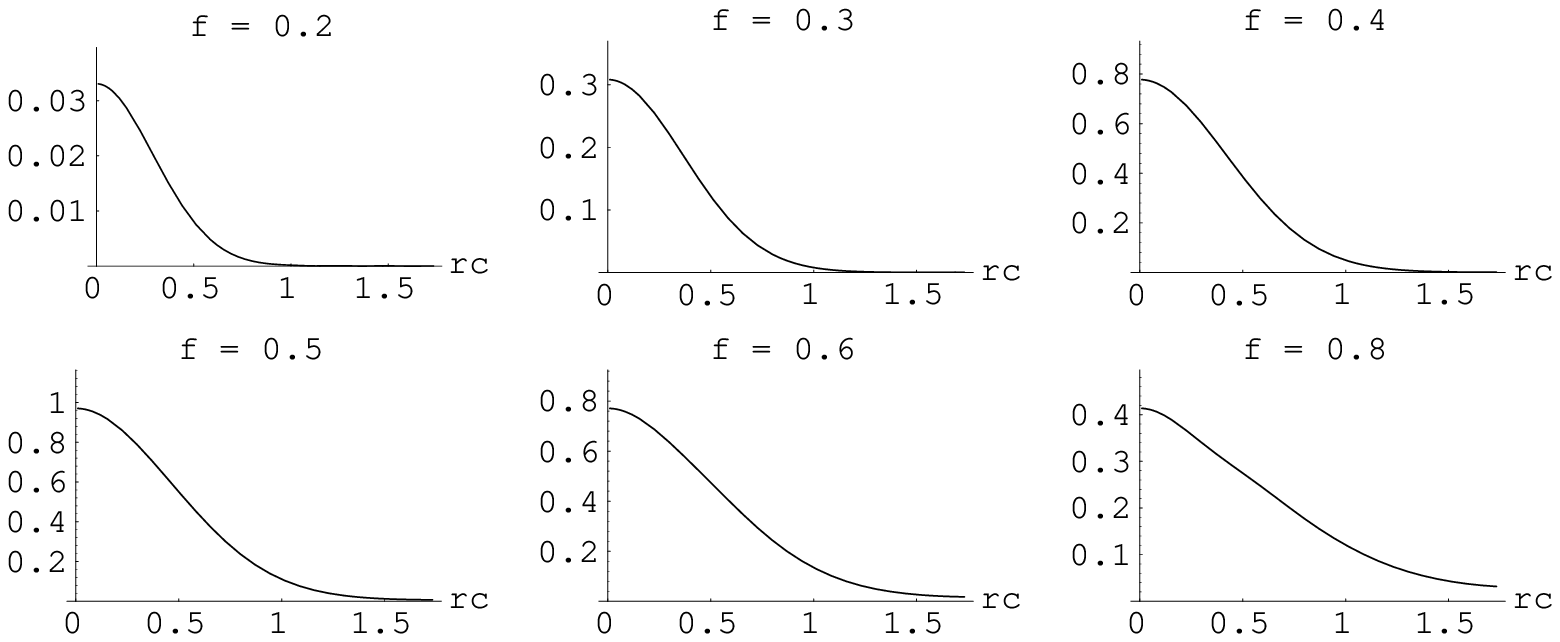}
  \caption{The wave function $\Psi \propto |\psi(r_c,r_d)|/(r_c^4 r_d^4)$
 plotted at $r_d = \sqrt{3}$ for the ground state in the sector 000-even.}
  \label{psi4}
\end{figure}

For small $f$, there is virtually no dependence of the masses on $\theta$,
whereas for larger $f$ the $\theta$ dependence becomes bigger.
This is shown in fig.~\ref{theta}. 
For $\theta = \pi$ we can impose the boundary conditions
exactly, that is, without using the trick with the Chern-Simons
like operator, but with the same method that we imposed the $\theta = 0$
conditions. The variational results obtained in this way 
are also plotted in fig.~\ref{theta}.
Remember that a $\theta$ dependence is a sign that the boundary 
in the $(r_c,r_d)$ plane is felt.
Our model is valid for values of the coupling constant at which
only the boundary conditions at and near the sphalerons are felt. 
Checking this a posteriori with the help of plots of the wave function
(fig.~\ref{psi4}) indicates that $f$ should not be larger than roughly $0.5$.
These plots were obtained as follows. 
Consider the function
\beq
  |\psi(r_c,r_d)|^2 =  \int d \chat~d \dhat~|\psi(c,d)|^2,
\eeq
which is a measure of the
probability distribution in the $(r_c,r_d)$ plane.
Dividing this $|\psi(r_c,r_d)|$ by $r_c^4 r_d^4$, we obtain
a function with the expected behaviour of the true wave function:
it is localized at the sphalerons and decays exponentially in
the transverse directions (see fig.~\ref{psi4}). Note that
the characteristic width grows with increasing $f$.
Although the lowest barrier is the sphaleron,
the measure factor $r_c^4 r_d^4$ causes the configurations that
are close to the sphaleron but have a somewhat higher energy
to make the dominant contribution to the tunnelling. 
The relevant parameter here is the characteristic decay length
of the wave function, which in turn is determined by the rise of the
potential in the transverse directions.
\par
Although we are primarily interested in the case $\theta = 0$,
an appreciable dependence on $\theta$ for a certain value of $f$
shows that the non-perturbative influence of the boundary has
become important.
To explain the fact that the spectrum is not exactly periodic 
in $\theta$, note that our implementation of the 
$\theta$~dependence~(eq.~(\ref{alphadef})) only has this periodicity in
the limit $\beta \rightarrow \infty$. The volume effect
described above implies that the 
relevant distribution $|\psi(r_c,r_d)|$ samples a piece of 
the boundary over which the phase
difference already starts to depart from $e^{i \theta}$.
This also implies that the results using the exact implementation
of the boundary conditions for the case $\theta = \pi$ can
be expected to differ from those using the method of 
eq.~(\ref{alphadef}).
Increasing $\beta$ would improve the periodicity properties, but
would not result in a better effective model: although it would
be less apparent, the wave function would still be spread out
over the boundary.

\subsubsection{One-loop hamiltonian}

We present basically the same plots for the one-loop 
hamiltonian as for the lowest order hamiltonian.
As before, 
the lowest glueball masses are to be found
in the sectors 000-even, 000-odd and 112-even.
These levels, including the lower bounds and the perturbative
expansions, are plotted in fig.~\ref{niveauxone}.
Note the avoidance of level crossing between the ground state and the 
first excited stated in the 000-even sector, which gives rise
(cf.\ eq.~(\ref{temple})) 
to relative large errors for even the ground state.
\par
Especially in the 112-even sector, the values obtained for $\eta$ are
rather large. We like to stress here that the lower bounds
as obtained by Temple's inequality are rather conservative 
and that the actual error is often much smaller. This insight
is gained in studying toy models, but the effect can also be
observed in the case at hand. Going from
$L_{\mbox{sum}} = 12$ to $L_{\mbox{sum}} = 14$ in the 112-even sector, 
the upper bounds shifted very little, whereas a reduction of the 
$\eta$ value by a factor of two was achieved.
\par
It is to be expected that for the one-loop hamiltonian 
a larger number of basis vectors is required, 
since the potential has a more complicated structure in this case.
Also the values of the coefficients in the potential may cause the
large values of $\eta$.
When investigating the influence of $\lm_i$ on the variational results, 
we find that the large value of $\lm_8$ seems to cause the problems.
This coefficient corresponds to the operator $\inv_2(c,d)$
(see eq.~(\ref{inv2ref}) and table~\ref{hamdat}), which is 
the only operator in the potential that can change the $l_s$ value
of a basis function. Apparently, our basis could not be
chosen so large as to yield high-precision results when
the coefficient of this operator becomes large.
\par
\begin{figure}  \centering
  \epsfxsize=0.8\textwidth
  \leavevmode
  \epsfbox[72 222 540 570]{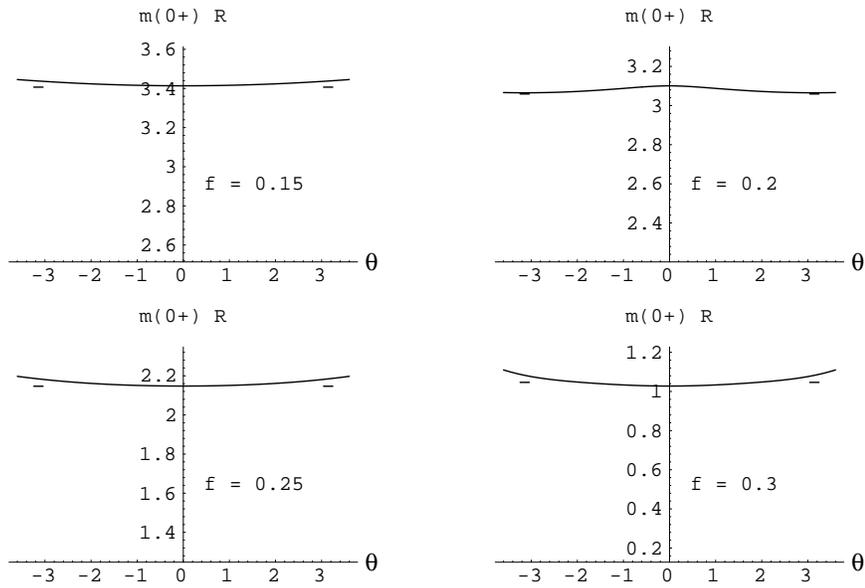}
  \caption{One-loop results.
  Scalar glueball mass at $f=0.15$, $f=0.2$, $f= 0.25$ and
  $f=0.3$ as a function of $\theta$. The dashes at $\theta = \pm \pi$
  denote the variational result with exact implementation of the
  boundary conditions. The vertical range of the plots is
  the same as in fig.~\ref{theta}.}
  \label{thetaone}
\end{figure}
\begin{figure}  \centering
  \epsfxsize=0.8\textwidth
  \leavevmode
  \epsfbox[72 322 540 470]{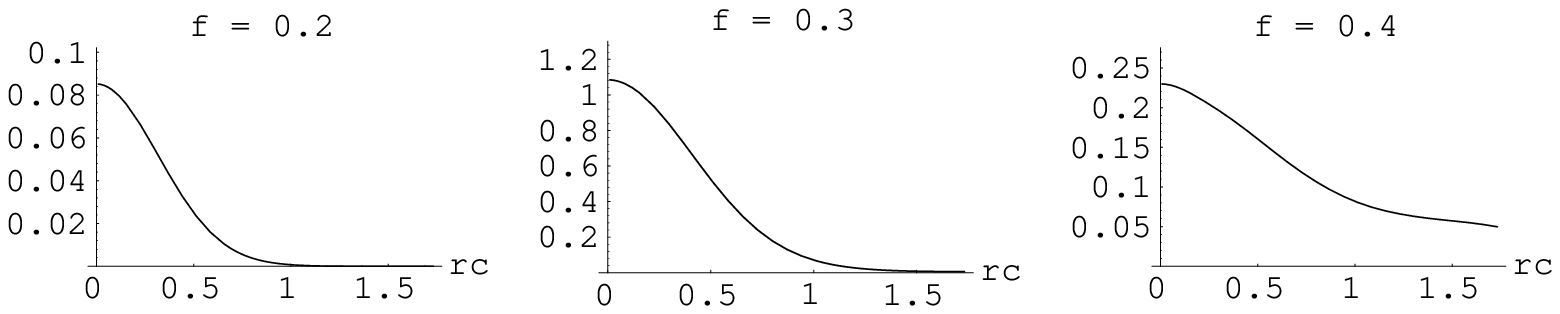}
  \caption{One-loop results. 
 The wave function $\Psi \propto |\psi(r_c,r_d)|/(r_c^4 r_d^4)$
 plotted at $r_d = \sqrt{3}$ for the ground state in the sector 000-even.}
  \label{psione4}
\end{figure}

Taking differences of the energy levels
with respect to the ground state in the 000-even sector again gives us
the masses of the low-lying glueball states: see fig.~\ref{massasone}.
Fig.~\ref{thetaone} shows
the $\theta$~dependence of the scalar glueball.
In view of the remarks above, we left out the lower bounds in these
cases. We can again study the localization properties around the
sphaleron: see fig.~\ref{psione4}.

\subsubsection{Discussion}

Using the Rayleigh-Ritz method we can determine the spectrum of
the effective hamiltonians. The use of Temple's inequality gives 
us confidence that our results are accurate,
especially since experience tells us that the actual error is usually
much smaller than the conservative estimates based on the values
of $\eta$.
The results are also consistent with the strong coupling limit
and with the weak coupling perturbative expansion.
\par
Let us focus first on small values of $f$. Here
the boundary conditions are not felt yet, and
the variational results are in accordance with the results
from perturbation theory.
When $f$ grows,
we see the effect of the boundary conditions in field space set in. For even
larger $f$, the wave function has spread out over the entire 
$(r_c,r_d)$ plane, and the model has lost its validity.
It is however possible that this already happens before this point
because of a break down of the adiabatic approximation.
In order for the adiabatic approximation to be valid,
the wave function for the transverse modes must still be
in its ground state. To check this for the modes outside
the $(c,d)$ space is hard, but transverse to the tunneling
path we can examine the transverse modes within the $(c,d)$
space. Fig.~\ref{psi4} and~\ref{psione4} do not only show
the localization of the transverse wave function
at the sphaleron, but also indicate that this transverse wave
function is in its ground state. We therefore assume that
the adiabatic approximation is valid, and that values
of the coupling constants for which our effective model is useful
are determined by the spreading out over the $(r_c,r_d)$ plane.
\par
From these considerations, we derive the following windows in the
coupling constant.
For the lowest order hamiltonian, the interesting window
is between $f=0.3$ and $f=0.5$. For the one-loop case,
it is between $f=0.2$ and $f=0.3$. 
\par
One of the issues raised above was the level of
localization of the wave function around the sphaleron.
This is related to the assumption that
only the boundary conditions at and near the sphalerons are felt.
We argued that this was determined by the rise of the potential
in the transverse directions. The one-loop correction to the
$\tr(Y)$ term in the potential at the $c$ sphaleron, which can be
expressed in $\kappa_1$, $\kappa_7$ and $\kappa_8$, is such that
it results in a lesser degree of localization. 
In both the lowest order and the one-loop case, a strong localization
of the wave function around the sphaleron is not realized and
the boundary conditions are not felt {\it exclusively} at the sphalerons.
This should not come as a big surprise since it is unnatural for an
eigenvalue problem that the spectrum is determined by boundary conditions
at a finite number of points. Typically the conditions 
at a boundary with codimension one determine the spectrum.
This is illustrated by our problem: we feel the
conditions at a finite part of the boundary, determined by
the localization potential in the transverse directions.
As long as the coupling is low enough, our assumption that
the boundary conditions are only felt close to the sphalerons
is true and the results are valid.

\section{Conclusions}
\label{ch-concl}

The goal of this project was to calculate the
glueball spectrum on \Sdr. 
In particular, we set out to study the effects of the
topology of the configuration space on the spectroscopy.
These effects can be seen as the result of the presence of instantons.
\par
It is important to emphasize one should not expect our results for
the spectrum to be accurate for large volumes. For large volumes
the effects of the non-trivial topology and geometry (curvature of the
configuration space, not to be confused with the curvature of \Sdr)
become too strong to be described by the effective theory.
Within the effective theory we clearly observe
that at large coupling the wave functional will start to 
feel more of the boundary of the fundamental domain than just 
the neighbourhood of the sphaleron. 
However, it has been the main aim of this study
to demonstrate that instanton effects on the low-lying spectrum
are large, but calculable as long as energies remain close to the 
sphaleron energy, where nevertheless semiclassical techniques will
completely fail.
\par
\begin{figure}  
  \centering
  \epsfxsize=0.8\textwidth
  \leavevmode
  \epsfbox[72 222 540 570]{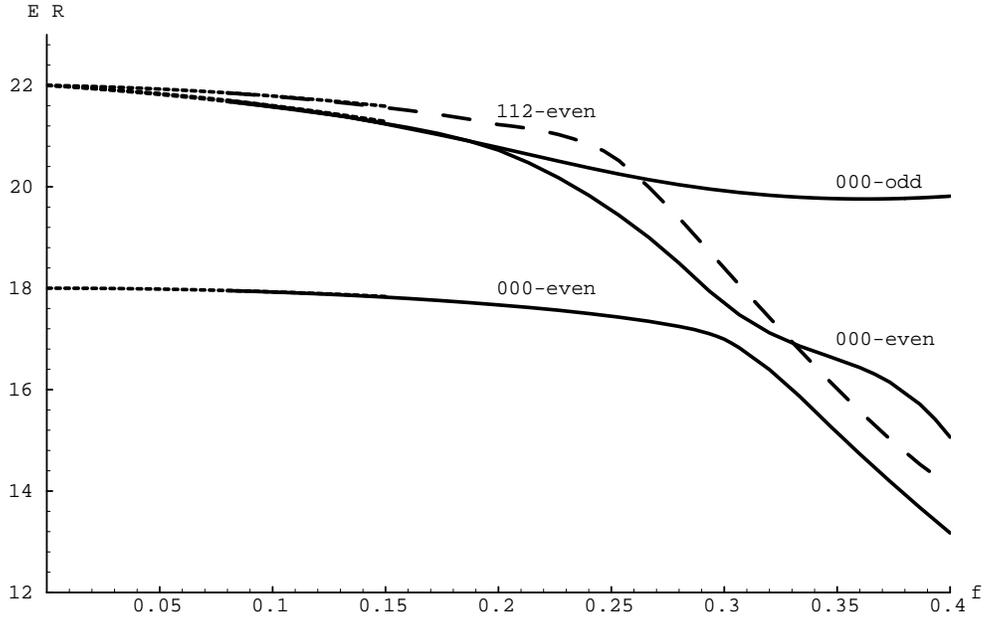} \\
  \epsfxsize=0.8\textwidth
  \leavevmode
  \epsfbox[72 222 540 570]{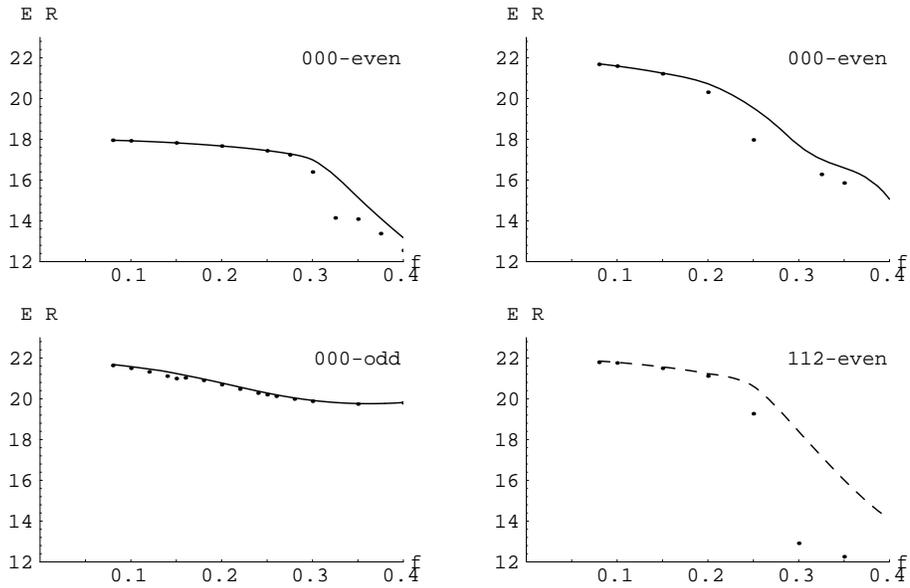}
  \caption{One-loop results. 
  Lowest energy levels for $\theta = 0$. 
  Drawn curves correspond to levels
  in the $(0,0,0)$ sector. The dashed curve denotes the ground level in the
  $(1,1,2)$-even sector, the short-dashed curves are the
  perturbative expansions. The four small plots also show the lower bounds 
  (individual dots) as obtained by Temple's inequality.}
  \label{niveauxone}
\end{figure}

The fact that $m_{0^+} R$ is decreasing, down from $m_{0^+} R = 4 $,
is what should be expected for the following reason.
A rough estimate for where one would expect instantons to
become relevant, based on what one finds on T$^3$, would be
$m_{0^+} R \simeq 1.4$.
Here we equate the largest geodesic distance on a torus
of size $L$, $\sqrt{3} L/2$, to its value on \Sdr, $\pi R$, or
\beq
  L = \frac{2 \pi}{\sqrt{3}} R,
\label{RLdef}
\eeq
and we use that on T$^3$ instantons are relevant for 
$z \equiv m_{0^+} L > 5$~\cite{baa8,hoe}.
Furthermore, we assume that the scalar glueball masses are roughly 
equal in both geometries at these volumes.
These low values of $m_{0^+} R$ are not reached in the tree level
approach, but we do reach them in the one-loop case which shows
that it was necessary to include the one-loop corrections.
This regime of masses occurs for values of the coupling where we expect our
model to still be valid. Specifically, $m_{0^+} R = 1.4$ corresponds to
$f = 0.28$. At larger couplings, we clearly see that
the wave functional feels too much of the boundary of
the fundamental domain. 
For $f= 0.4$ this is dramatically
clear from fig.~\ref{psione4}, where the
wave function is seen to probe unacceptable regions of the
boundary of the fundamental domain
to remain a good approximation to the full wave function.
From fig.~\ref{niveauxone} we see
that the scalar and tensor mass even cross around $f=0.33$,
which is certainly unacceptable for the full theory.
Clearly we have pushed the model passed its
region of validity for $f> 0.3$. 
\par
Of course, at some point $m_{0^+} R$ has to start to rise again,
and when $m_{0^+}$ reaches its asymptotic infinite-volume value,
$m_{0^+} R$ grows linearly with $R$.
Both the truncated and the one-loop results show
that $m_{0^+} R$ exhibits a minimum, after which it starts
to rise again. It rises linearly in $f$ for $f \rightarrow \infty$,
as follows from the strong coupling results in fig.~\ref{niveauxstrong},
which are also valid for the one-loop case.
This however does not mean that we are approaching the infinite-volume 
limit, because our effective model is certainly not fit to describe
this regime. Moreover, it is clear that no statements can be made
on the $R$ dependence of $f$ for these large couplings and volumes.
\par
\begin{figure} \centering
  \epsfxsize=0.8\textwidth
  \leavevmode
  \epsfbox[72 222 540 570]{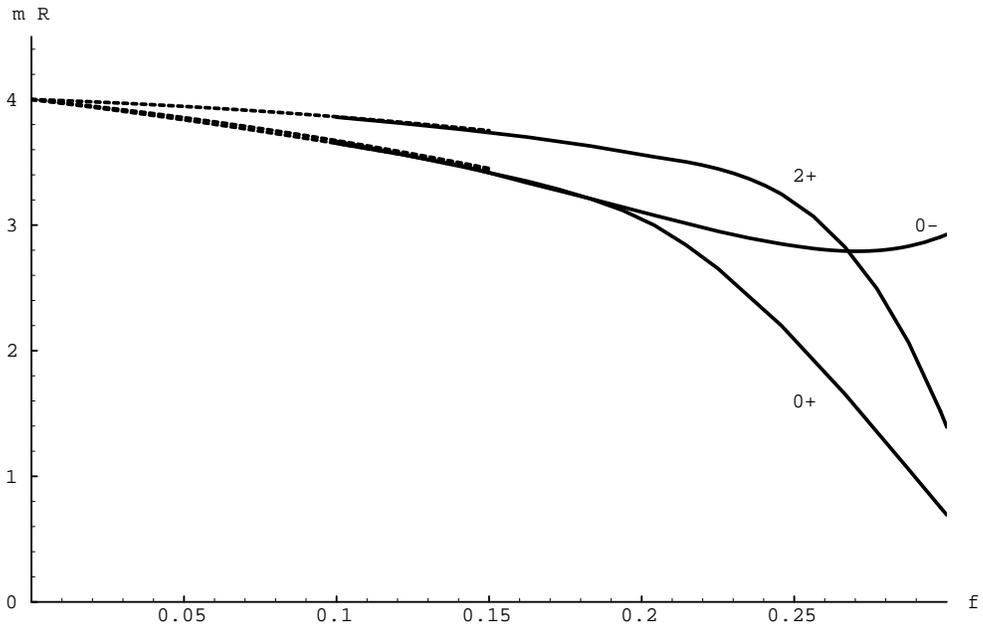}\\ 
  \caption{One-loop results. 
  Glueball masses for $\theta = 0$ as a function of 
  the coupling constant. 
  The drawn curves are the masses of the 
  various scalar and tensor glueballs. The dotted
  lines denote the perturbative results.}
  \label{massasone}
\end{figure}

Other indications that our results are in the domain of expected
validity are that at $f=0.25$ the tensor to scalar mass ratio
is given by $1.5$, rising from $1$ at zero coupling to $1.8$
at $f=0.28$, see fig.~\ref{massasone}.
For a torus geometry one finds a similar range. 
Below $z = m_{0^+} L = 5$ the tensor is split into a doublet at
$\sim 0.9 m_{0^+}$ and a triplet at $\sim 1.7 m_{0^+}$~\cite{baa8,mic}
(due to breaking of rotational to cubic invariance), whereas
these states seem to merge at $z>7$ into a degenerate
quintet with $m_{2^+} = 1.5 m_{0^+}$.
\par
\begin{figure} \centering
  \epsfxsize=0.8\textwidth
  \leavevmode
  \epsfbox[72 222 540 570]{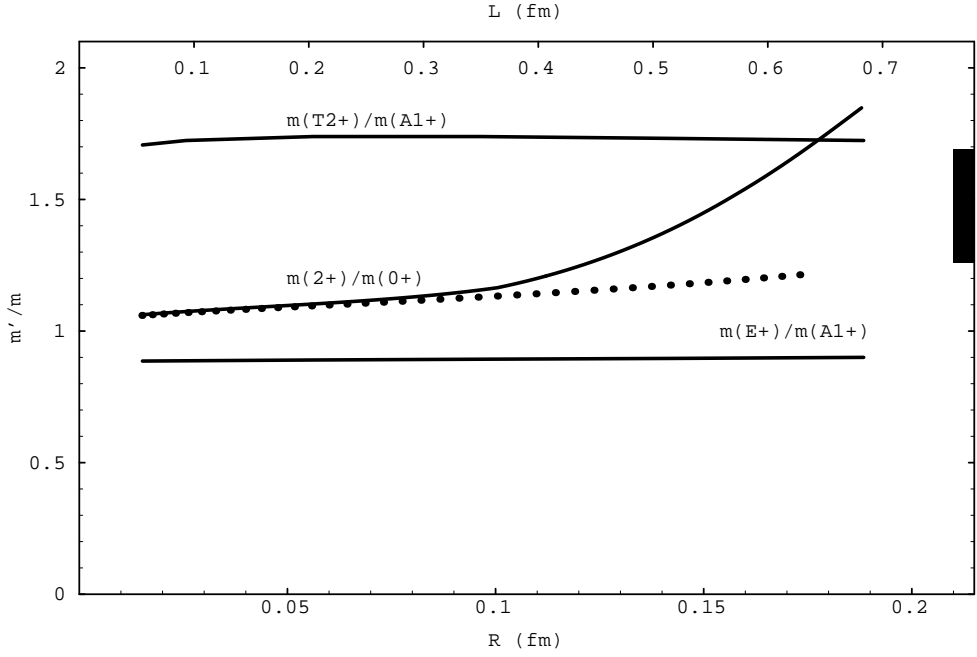}
  \caption{Comparison of the tensor to scalar mass ratio $m_{2^+}/m_{0^+}$
as obtained on the \drs\ to the relevant ratios on the torus. The dotted
line denotes the perturbative expansion, the bar on the right
indicates the range of lattice Monte Carlo values at
$L=1$ to $L = 1.5~\fm$.
}
\label{torus}
\end{figure}

We assume that in intermediate volumes the two-loop $\beta$ function
can be used to convert the dependence on the 
relevant coupling constants to the
dependence on the radius $R$ or the size $L$.
Thus, for $R$ we have ($f \equiv g^2/(2 \pi^2)$)
\beq
 (\Lm_R R)^2 = 
  \left( \beta_1 g^2\right)^{- \frac{\beta_2}{\beta_1^2}}
   e^{- \frac{1}{\beta_1 g^2}} , 
  \quad \beta_1 = \frac{11}{24 \pi^2}, 
  \quad \frac{\beta_2}{\beta_1^2} = \frac{102}{121},
\label{betafun}
\eeq
and the same formula holds for $\Lm_L L$ in terms of the minimal
subtraction coupling $g_{\mbox{\scriptsize MS}}$.
If we, as usual, set the scale by a string tension of 
$\sg = (425 \mbox{MeV})^2$ and use that
$\sqrt{\sg}/m_{0^+} \approx 0.3$ for T$^3$, we have that 
in physical units $z = m_{0^+} L = 5$ corresponds to $L = 0.7~\fm$ and
hence, using eq.~(\ref{RLdef}), to $R = 0.19~\fm$.
As derived earlier, this corresponds 
to $f = 0.28$ on the \drs, and
eq.~(\ref{betafun}) gives $\Lm_R = 1.3~\fm^{-1}$.
On the torus, $z=5$ corresponds to
$g_{\mbox{\scriptsize MS}}= 2.42$~\cite{baa2}. 
Relating this to $L = 0.7~\fm$ gives $\Lm_L = 0.39~\fm^{-1}$.
In this way we can compare our result for $m_{2^+}/m_{0^+}$
as a function of the volume to the doublet $E^+$ and the 
triplet $T_2^+$ on the torus, as is shown in fig.~\ref{torus}.
The vertical
bar on the right indicates the range of lattice Monte Carlo 
values~\cite{mic} for the $E^+$ and $T_2^+$ masses
(equal within errors) at $L=1$ to $1.5~\fm$.
\par
Another way to relate the length scales above to a value of the
coupling constant uses the
definition of the running coupling constant in~\cite{lus3} and
the relation between this coupling constant and the one in the
\MS\ scheme. Proceeding in this way relates $R= 0.19~\fm$
to a value for $f$ of $0.26$, which is yet 
another indication that this regime for the coupling
constant corresponds to volumes where the instanton effects
are important.
To obtain this number we correct for the finite renormalizations
appearing in eq.~(\ref{renorm}) and we assumed that setting
the linear size $L$ of the $\veps$-dimensional torus
to $R$ gives the \MS\ scheme.
This ambiguity could be fixed by computing the effective 
hamiltonian using \eg Pauli-Villars regularization and the known
relation to the minimal subtraction scheme.
In our calculations $f$ is however just a parameter that
allows us to probe different volumes and its precise relation
to the physical scale is not so relevant.
\par
Returning to fig.~\ref{torus}, we can distinguish three regimes in $R$.
For small $R$, we expect the finite-size effects, like the effect
of the curvature of \Sdr, to be large.
The masses in the effective model, although perturbatively
calculable, will therefore not correspond to the masses on the torus.
In this regime the smallest mass in the 000-odd sector $0^-$ 
(see fig.~\ref{massasone})
is actually smaller than the mass of
the scalar glueball $0^+$, an effect that also can be seen from the
perturbative evaluation of the energy levels eq.~(\ref{blocheven}) 
and~(\ref{blochodd}).
This we expect to be a consequence of the strong finite
size effects that are visible at small values of $R$.
\par
Around $R = 0.1~\fm$, corresponding to $L = 0.36~\fm$, 
we see (in the sphere geometry) the masses deviate from the
perturbative expansion, signalling the onset of the instantons.
They drive the tensor to scalar ratio in the right direction, that is,
towards the large-volume value of 1.5.
The mass of the odd glueball is now higher than that of the even
glueball. Apparently the non-perturbative effects
also drove this quantity towards more physical values:
from lattice calculations a value of $m(0^-) / m(0^+) \approx 1.7$
can be extracted~\cite{mic2}, compatible with what is found here.
\par
For $R > 0.2~\fm$, our effective model is no longer valid, 
due to the spreading out of the wave functional over the boundary of the 
fundamental domain, 
as discussed before.
\par
Finally, an important goal of this project was to get results
for glueball masses as a function of $\theta$. The truncated
results showed a pronounced dependence on $\theta$ in the
regime where boundary effects are appreciable (see fig.~\ref{theta}).
Quite remarkably, and unexpectedly, this strong dependence 
disappears when adding the one-loop corrections. In particular,
at $\theta =0$ and $\theta = \pi$, where boundary conditions
can be implemented most accurately, the masses do not differ significantly
(see fig.~\ref{thetaone}). Caution needs to be applied in concluding that the
same will hold at large volumes, but in any case it would be interesting
if glueball masses could be measured at $\theta = \pi$ on the
lattice as comparison.
\par
In conclusion, we should expect our one-loop corrected result
to be a relatively accurate representation of the true masses
on \Sdr\ up to $f=0.28$ corresponding to a circumference of approximately 
$1.3~\fm$, up to where also the variational basis does not
exhibit too much of the problems with the Temple 
bound (cf.~fig.~\ref{niveauxone}).
The approach to infinite-volume values 
of results on the \drs, as compared
to torus results, is slow. We typically have a dependence
on powers of $1/R$ 
as compared to an exponential behaviour in $L$~\cite{lus4}.
As our results should not be expected to be already in the asymptotic
domain, the rough agreement we find with results on the torus is
gratifying. 
\par
When comparing the truncated and one-loop corrected results,
the results for the $\theta$ dependence show that
strong non-linear and non-perturbative effects influence the
spectrum. 
We have shown there is a small, but finite window from $R = 0.1~\fm$
to $R = 0.2~\fm$
(at smaller volumes everything can be described perturbatively)
where these effects can be included reliably, showing 
convincingly how important the global properties of the
field space are for the non-perturbative dynamics of
non-abelian gauge theories.

\section*{Acknowledgments}

The author wishes to thank Pierre van Baal for many
helpful discussions. He also wishes to thank D. Sch\"{u}tte for
useful discussions.
Support by Stichting Nationale Computer Faciliteiten (NCF) 
for use of the Cray Y-MP at SARA was appreciated.

\section*{Appendices}

\appendix
\section{Details on the basis}
\label{appbasis}

In this appendix we will exploit the
relation between \Sdr\ and \SUtw\ 
to rederive the basis~(\ref{scalbasis}).
After this we will make
the link with the definition of functions on S$^n$ in
terms of homogeneous polynomials in the coordinates of the embedding space.

\subsection{The \SUtw\ group structure of \Sdr}

There is a one-to-one correspondence between points of 
\Sdr\ and \SUtw\ given by $g = n \cdot \sg$. 
For instance, the matrix $V^i_j$ of eq.~(\ref{Vdef}) can in this light be seen
as the adjoint representation of \SUtw.
The group of spatial symmetries of \Sdr\ is SO(4). These
symmetries correspond to the left and right symmetries on \SUtw.
Let $\psi$ be
a function on \SUtw\ or equivalently on \Sdr.
\SUtw\ can now act on this function by left or right multiplication:
\beq
  \left(\Ss{D}^L(g_1) \psi \right)(g) = \psi( g_1^{-1} g), \quad
  \left(\Ss{D}^R(g_2) \psi \right)(g) = \psi( g g_2),
   \quad g,g_1,g_2 \in \mbox{SU}(2).
\eeq
These infinite-dimensional representations of the group \SUtw\
induce representations of the Lie algebra \sutw. We will
show that the generators \Ss{L} of these representations correspond
to $\Leen$ and $\Ltw$:
\bea
\left( i \Ss{L}^L_j \psi \right) (g) &=& 
\ddt \psi( \exp( - i t \half \tau_j) g)|_{t=0} \Next
&=& \ddt \psi( n \cdot \sg + \half t \sgbar_j n \cdot \sg )|_{t=0} \Next
&=& \ddt \psi( (n_\al + \half t \etabar^\al_{j \mu}  n_\mu )\sg_\al) |_{t=0}
 \Next
&=& \half \eta^j_{\mu \al} n_\mu \frac{\pr}{\pr x_\al} \psi(n)
   = \left(i L_1^j \psi\right)(n).
\eea
According to the Peter-Weyl theorem~\cite{bar}, a basis of functions on \SUtw\ 
is provided by the collection of matrix elements of all
the finite-dimensional unitary irreducible representations of \SUtw.
The relation of these matrix elements with the basis 
of eq.~(\ref{scalbasis}) is
\beq
  \langle g | l\,m_L\,m_R \rangle = (-1)^{m_L} \sqrt{2 l +1}\,
    D^l_{-m_L m_R}(g). \label{scalbasistwee}
\eeq

\subsection{The eigenfunctions of the laplacian on S$^n$}

In this section we summarize some well-known facts about the spectrum
of the laplacian on an $n$-dimensional sphere. This allows us to 
identify the functions~(\ref{scalbasis}) with explicit functions on
\Sdr. Let $x_0,\ldots,x_{n}$ be coordinates in $\real^{n+1}$.
Using radial coordinates $r,\theta_1,\dots,\theta_n$ one can explicitly solve
for the eigenfunctions of the spherical laplacian $\sphlap_n$,
using separation of variables and recursion in $n$. 
The result of this calculation (cf.\ \eg~\cite{bar}) is that the spectrum
is given by
\beq
 \{-L(L+n-1)\}_{L=0,1,2,\ldots},
\eeq
and the degeneracy of the level $L$ is given by
\beq
  \binom{L+n}{n} - \binom{L-2+n}{n}.
\eeq
We will relate these eigenfunctions to polynomials in $x_\mu$.
Let $V^L$ be the set of polynomials in
$x_\mu$ that are homogeneous of degree $L$. 
To such a polynomial $p$ corresponds a function $Y$ on S$^n$ defined by
\beq
  p(x_0,\ldots,x_n) = r^L\,Y(\hat{x}_0,\ldots,\hat{x}_n).
\eeq
Suppose that $p$ is a harmonic polynomial, \ie $\Dl p = 0$. We then
have
\beq
  0 
=  \left( \frac{1}{r^n} \frac{\pr}{\pr r} r^n \frac{\pr}{\pr r}
   + \frac{1}{r^2} \sphlap_n \right) \left( r^L\,Y \right) 
= L(L+n-1)\,r^{L-2}\,Y + r^{L-2}\,\sphlap_n Y.
\eeq
We have found that every harmonic polynomial of degree $L$ corresponds
to a function on the $n$-sphere that is an eigenfunction of the
spherical laplacian with eigenvalue $-L(L+n-1)$.
The number of monomials of degree $L$ in
the variables $x_0,\dots,x_n$ is given by
\beq
  \mydim(V^L) = \binom{L+n}{n}.
\eeq
Hence we have
\bea
  \mydim(\Ker(\Dl)) &=& \mydim(V^L) - \mydim(\Imag(\Dl)) \Next
  &\geq& \mydim(V^L) - \mydim(V^{L-2}) \Next
  &=&  \binom{L+n}{n} - \binom{L-2+n}{n}.
\eea
Since we know the degeneracy of the eigenvalue $-L(L+n-1)$, we have
shown that all the eigenfunctions of the spherical laplacian 
correspond to harmonic polynomials. Moreover, the inequality in the last
equation is actually an equality. 
\par
For the case of S$^2$ the eigenfunctions are the well-known
spherical harmonics $Y^L_M$.
For the case \Sdr\ the spectrum becomes
$\{-L(L+2)\}_{L=0,1,2,\ldots}$,
with the degeneracy given by
\beq
  \binom{L+3}{3} - \binom{L+1}{3} = (L+1)^2.
\eeq
When comparing this with the results~(\ref{scalbasistwee}), we see
that $L = 2 l$. The four $l=\hf$ modes 
$\langle x|\hf \,m_L\,m_R \rangle$
are linear combinations of the four scalar functions 
$x_\mu,\,(\mu=0,\ldots,3)$, whereas the nine $l=1$ modes 
$\langle x|1\,m_L\,m_R \rangle$ correspond to the nine components of $V^i_j$.
\par
In section~\ref{ch-RR} the results of this appendix are applied to
S$^8$.

\section{The linear term in the path integral}
\label{appJ}

In this appendix we show that the term $J_\nu Q_\nu$ in
eq.~(\ref{Zbos})
does not make a contribution to the effective action up to the
order we are interested in: the lowest order contribution from $J$
to the effective potential will be of the form 
$c^4 d^2$ and $c^2 d^4$. 
\par
The crucial remark is that in 
the path integral the constant modes are excluded from the
integration over $Q_0$, and the $(c,d)$ modes are excluded from the
integration over $Q_i$. This means that one must remember
to put projectors $(1 - P)$ around operators like $W_{\mu \nu}$,
and replace source currents like $J_\mu$ with $((1-P)J)_\mu$.
We will show that after this projection $J$ contains only terms that are
cubic in $B$.
\par
We have $F_{0 i} = - F_{i 0} = \dot{B}_i$ and 
$F_{i j} = - 2 \veps_{i j k} (c_k - \dV_k) + [B_i,B_j]$, 
with $B_i = c_i + \dV_i$.
This implies $J_0 = D_\mu F_{\mu 0} = - [B_i,\dot{B}_i]$ and
$J_j = D_\mu F_{\mu j} = \ddot{B}_j + D_i F_{i j}$.
In our approximation we are only interested in the term
$D_i F_{i j} = J^{(1)}_j + J^{(2)}_j + J^{(3)}_j$, where
$J^{(n)}$ is of order $n$ in the field $B$. From
$D_i F_{i j} = \pr_i F_{i j} - \veps_{i j k} F_{i k} + [A_i,F_{i j}]$,
we obtain 
\bea
  J^{(1)}_j &=& -4 \,c_j, \\
  J^{(2)}_j &=&  -3\, \veps_{i j k} [c_i,c_k] 
    - 3\, \veps_{m p n} [d_m,d_n] V^p_j, \\
  J^{(3)}_j &=& [B_i,[B_i,B_j]] \Next
   &=& [c_i,[c_i,c_j]] + [\dV_i,[\dV_i,c_j]] \Next
   && + [c_i,[c_i,\dV_j]] + [\dV_i,[\dV_i,\dV_j]] \Next
   && + [\dV_i,[c_i,c_j]] + [c_i,[\dV_i,c_j]] \Next
   && + [\dV_i,[c_i,\dV_j]] +[c_i,[\dV_i,\dV_j]]
\eea
This directly implies that $(1-P_V) J^{(1)} = (1-P_V) J^{(2)} = 0$.
The first four terms of $J^{(3)}$ do not contribute since they too
are modes with $(l,k) = (0,1)$ or $(l,k) = (1,0)$.
We write the last four terms of $J^{(3)}$ as follows:
\beq
A^p_{i j} V^p_i + B^{m n}_i V^m_i V^n_j,
\eeq
with the \sutw-valued constants
\beq
  A^p_{i j} = [d_p,[c_i,c_j]] + [c_i,[d_p,c_j]], \quad
  B^{n m}_i = [d_m,[c_i,d_n]] + [c_i,[d_m,d_n]].
\label{ABdef}
\eeq
The matrices $A^p$ and $B_i$ can be decomposed in a
trace part, an antisymmetric part and a symmetric traceless part.
For the $A^p$ terms, this is the decomposition into $(l,k)$
is $(1,0)$, $(1,1)$ and $(1,2)$ respectively.
For the $B_i$ terms, it gives the split in the three possible
$l$ values $(0,1)$, $(1,1)$ and $(2,1)$.
Projecting out the $(c,d)$ modes from $J_i$, we obtain
\beq
((1-P_V) J)_j = \bar{A}^p_{i j} V^p_i + \bar{B}^{m n}_i V^m_i V^n_j,
\eeq
where $\bar{A}$ and $\bar{B}$ are obtained from the $A$ and $B$
of eq.~(\ref{ABdef}) by projecting out the trace part.
\par
Expanding the path integral of eq.~(\ref{Zbos}) gives diagrams with
interactions of the $Q$ field with the source $J(B)$.
If we also take along the $Q^3$, $B Q^3$ and $Q^4$ terms in
the action, 
diagrams with only one occurrence of $J$ can in principle
give a contribution of the order $c^2 d^2$.
However, in order to obtain a $c^2 d^2$ invariant,
we must take the trace part of $\bar{A}$ and $\bar{B}$.
Since these trace parts are zero, we see that
a single occurence of $J$ can not give rise to 
contributions of the order $c^2 d^2$.

\section{Angular matrix elements}
\label{appreduced}

We will perform the reduction of the matrix elements of the
operator \Ss{A} of eq.~(\ref{scrAdef}) to reduced matrix elements.
We use eq.~(\ref{cddef}).
\bea
&& \hspace{- 1cm} 
\langle j',m',i'_1,i'_2 | \Ss{A} | j,m,i_1,i_2 \rangle \Next
&=& \langle j',m',l'_s;L'_1,l'_1,\tau'_1;L'_2,l'_2,\tau'_2 | \Next
&& \quad   \left\{ | 0,0,\ltil_s;\Ltil_1,0,\tautil_1;\Ltil_2,0,\tautil_2 \rangle
    | j,m,l_s;L_1,l_1,\tau_1;L_2,l_2,\tau_2 \rangle \right\} \Next
&=& \sum_{m'_1, m'_2}
(-1)^{l'_1 - l'_2 + m'} \sqrt{2 j' + 1} 
   \left( \begin{array}{ccc}
     l'_1 & l'_2 & j' \\ 
     m'_1 & m'_2 & -m' 
   \end{array} \right) \times \Next
&&  \sum_{m_1, m_2}
   (-1)^{l_1 - l_2 + m} \sqrt{2 j + 1} 
   \left( \begin{array}{ccc}
     l_1 & l_2 & j \\ 
     m_1 & m_2 & -m 
   \end{array} \right) \times \Next
&&    \sum_{m'_s, \tilde{m}_s, m_s}
    (-1)^{l'_s - m'_s} \frac{1}{\sqrt{2 l'_s + 1}} ~
     (-1)^{\ltil_s - \tilde{m}_s} \frac{1}{\sqrt{2 \ltil_s + 1}} ~
     (-1)^{l_s - m_s} \frac{1}{\sqrt{2 l_s + 1}} \times \Next
&&  \langle L'_1;l'_s,l'_1,\tau'_1;m'_s,m'_1 |
    \{ | \Ltil_1;\ltil_s,0,\tautil_1;\tilde{m}_s,0 \rangle  
    | L_1;l_s,l_1,\tau_1;m_s,m_1 \rangle\} \times \Next
&&  \langle L'_2;l'_s,l'_2,\tau'_2;-m'_s,m'_2 |
    \{ | \Ltil_2;\ltil_s,0,\tautil_2;-\tilde{m}_s,0 \rangle 
    | L_2;l_s,l_2,\tau_2;-m_s,m_2 \rangle \}.
\label{reductie2}
\eea
We introduce reduced matrix elements through
\bea
&&  \terug
    \langle L'_1;l'_s,l'_1,\tau'_1;m'_s,m'_1 |
    \{ | \Ltil_1;\ltil_s,0,\tautil_1;\tilde{m}_s,0 \rangle 
    | L_1;l_s,l_1,\tau_1;m_s,m_1 \rangle \} =\Next
&& \quad 
   (-1)^{m'_s + m'_1} \left( \begin{array}{ccc}
     l'_s & \ltil_s      & l_s \\ 
     -m'_s & \tilde{m}_s & m_s 
   \end{array} \right)
   \left( \begin{array}{ccc}
     l'_1  & 0 & l_1 \\ 
    -m'_1 & 0 & m_1 
   \end{array} \right) \tilde{F}(i'_1,\tilde{\imath}_1,i_1).
\label{Ftildef}
\eea
Using
\beq
   \left( \begin{array}{ccc}
     l'_1  & 0 & l_1 \\ 
    -m'_1 & 0 & m_1 
   \end{array} \right) = (-1)^{l_1 - m_1} \frac{1}{\sqrt{2 l_1 + 1}}
    \dl_{l'_1 l_1} \dl_{m'_1 m_1},
\label{somtriv}
\eeq
the summations over $m'_1$ and $m'_2$ become trivial.
The remaining summations in eq.~(\ref{reductie2}) can then be
dealt with using twice the completeness relation
\bea
 \sum_{m_1 m_2} 
   \left( \begin{array}{ccc}
     j_1 & j_2 & j_3 \\ 
     m_1 & m_2 & m_3 
   \end{array} \right) 
   \left( \begin{array}{ccc}
     j_1 & j_2 & j'_3 \\ 
     m_1 & m_2 & m'_3 
   \end{array} \right)
=   \frac{1}{ 2 j_3 + 1} 
   \dl_{j'_3 j_3} \dl_{m'_3 m_3} \dl(j_1, j_2, j_3),
\eea
where $\dl(j_1 ,j_2, j_3) = 1$ if $j_1$, $j_2$ and $j_3$ satisfy
the triangular condition, and is zero otherwise.
We obtain
\bea
&& \terug \langle j',m',i'_1,i'_2 | \Ss{A} | j,m,i_1,i_2 \rangle =  \Next
&& \dl_{l'_1 l_1} \dl_{l'_2 l_2} \dl_{j' j} \dl_{m' m} \dl(l_1, l_2, j)
    \dl(l'_s , \ltil_s, l_s) 
    \frac{F(i'_1,\tilde{\imath}_1,i_1) F(i'_2,\tilde{\imath}_2,i_2)}
    {\left[ (2 l'_s + 1) (2 \ltil_s + 1)  (2 l_s + 1)\right]^\hf}.
\label{reductie}
\eea
We absorbed the factors $(-1)^{l_i} ( 2 l_i + 1)^{-\halfje}$
for $i=1,2$ in the $\tilde{F}$ functions.
We can evaluate the $F$ functions
by setting in eq.~(\ref{Ftildef})
\beq
m'_1 = m_1 = l_1, \quad
m_s  = l_s, \quad
m'_s = l'_s, \quad
\tilde{m}_s = l'_s - l_s.
\eeq
Thus we find
\bea
&& \terug 
   F(i',\tilde{\imath},i) = (-1)^{l'_s}    \left( \begin{array}{ccc}
     l'_s & \ltil_s      & l_s \\ 
     -l'_s & l'_s- l_s & l_s 
   \end{array} \right)^{-1} \times \Next
&& \quad  \langle L';l'_s,l'_r,\tau';l'_s,l'_r | \left\{
    | \Ltil;\ltil_s,0,\tautil;l'_s- l_s ,0 \rangle  
    | L;l_s,l_r,\tau;l_s,l_r \rangle \right\}.
\label{Fdef}
\eea
Note that the only dependence of the matrix elements on $j$ is
through the triangular condition on $l_1$, $l_2$ and $j$.
Also note that the explicit $\dl$ functions
\beq
  \dl_{l'_1 l_1} \, \dl_{l'_2 l_2} \, \dl(l'_s , \ltil_s, l_s) 
\eeq 
are superfluous: they are also contained within the $F$ functions.
The $\dl(l_1, l_2, j)$ function can be deleted too: the
corresponding triangular condition is satisfied 
from the beginning.
\par
Specializing to the case where \Ss{A} only depends on $c$,
we have $\ltil_s =0$ and we can also absorb the factors
 $(-1)^{l_s} ( 2 l_s + 1)^{-\halfje}$ in $F$.
We have for instance
\beq
 \langle j',m',i'_1,i'_2 | \inv_3(\chat) | j,m,i_1,i_2 \rangle 
 = \dl_{j' j} \dl_{m' m} F_3(i'_1,i_1) \dl(i'_2,i_2),
\eeq
with
\beq
F_3(i',i) = \langle L';l'_s,l'_r,\tau';l'_s,l'_r | \inv_3(\chat)
    | L;l_s,l_r,\tau;l_s,l_r \rangle.
\eeq
We similarly treat $\inv_4(\chat)$, $\inv_6(\chat)$, 
$\inv_7(\chat)$ and $\inv_8(\chat)$
which give rise to the reduced matrix elements $F_4$, $F_6$, $F_7$ 
and $F_8$ respectively. Operators that depend only on $d$ and
operators that are products of a function of $c$ and of a
function of $d$ pose no problems either. We have for instance
\bea
 \langle j',m',i'_1,i'_2 | \inv_3(\dhat) | j,m,i_1,i_2 \rangle
 &=& \dl_{j' j} \dl_{m' m} \dl(i'_1,i_1) F_3(i'_2,i_2), \\
 \langle j',m',i'_1,i'_2 | \inv_3(\chat) \inv_3(\dhat) | j,m,i_1,i_2 \rangle
 &=& \dl_{j' j} \dl_{m' m} F_3(i'_1,i_1) F_3(i'_2,i_2).
\eea
The operator $\inv_2(c,d)$ is treated with the general
formulae eq.~(\ref{reductie}) and~(\ref{Fdef}).
One can show
$
\inv_2(\chat,\dhat) = \frac{1}{9} - \frac{1}{99} \sqrt{5} 
 \langle \chat \dhat | 0,0,2;2,0,1;2,0,1 \rangle$.
Using eq.~(\ref{Fdef}) we define
\bea
&& \terug 
  F_{22}(i',i) = (-1)^{l'_s}    \left( \begin{array}{ccc}
     l'_s & 2      & l_s \\ 
     -l'_s & l'_s- l_s & l_s 
   \end{array} \right)^{-1} \times \Next
&& \quad \langle L';l'_s,l'_r,\tau';l'_s,l'_r | \left\{
    | 2;2,0,1;l'_s- l_s ,0 \rangle 
    | L;l_s,l_r,\tau;l_s,l_r \rangle \right\},
\eea
and we obtain
\bea
&& \terug
  \langle j',m',i'_1,i'_2 | \inv_2(\chat,\dhat) | j,m,i_1,i_2 \rangle = \Next
 &&  \quad \dl_{j' j} \dl_{m' m} 
   \left\{ \sfrac{1}{9} \dl(i'_1,i_1) \dl(i'_2,i_2) -\sfrac{1}{99}
    \frac{F_{22}(i'_1,i_1) F_{22}(i'_2,i_2)}
    {\left[ (2 l'_s + 1) (2 l_s + 1)\right]^\hf}\right\}.
\eea
For products like  $\inv_3(c) \inv_2(c,d)$ we obtain
\bea
&& \terug \langle j',m',i'_1,i'_2 | \inv_3(\chat) \inv_2(\chat,\dhat) 
    | j,m,i_1,i_2 \rangle =  \Next
&& \quad  \dl_{j' j} \dl_{m' m} 
   \left\{ \sfrac{1}{9} F_3(i'_1,i_1) \dl(i'_2,i_2) -\sfrac{1}{99}
    \frac{F_{52}(i'_1,i_1) F_{22}(i'_2,i_2)}
    {\left[ (2 l'_s + 1) (2 l_s + 1)\right]^\hf}\right\},
\eea
with 
\bea
&& \terug 
  F_{52}(i',i) = (-1)^{l'_s}    \left( \begin{array}{ccc}
     l'_s & 2      & l_s \\ 
     -l'_s & l'_s- l_s & l_s 
   \end{array} \right)^{-1} \times \Next
&& \quad \langle L';l'_s,l'_r,\tau';l'_s,l'_r | \left\{ \inv_3(\chat) 
    | 2;2,0,1;l'_s- l_s ,0 \rangle 
    | L;l_s,l_r,\tau;l_s,l_r \rangle \right\}.
\eea
Similarly the operator $\inv_4(c) \inv_2(c,d)$ leads to the 
reduced matrix element $F_{62}$.
\par
The only operator in $\Ss{V}^2$ that we have yet to deal with is
$\inv_2^2(c,d)$. For this operator we need to construct the
spectral decomposition. We write
\bea
\inv_2^2 
&=&  \hphantom{+}
     a   \, | 0,0,4;4,0,1;4,0,1 \rangle + 
     b   \, | 0,0,2;4,0,1;4,0,1 \rangle \Next
&& + c   \, | 0,0,0;4,0,1;4,0,1 \rangle
   + d   \, | 0,0,2;2,0,1;2,0,1 \rangle \Next
&& + e   \,(| 0,0,2;4,0,1;2,0,1 \rangle + 
         \, | 0,0,2;2,0,1;4,0,1 \rangle ) \Next
&& + f   \,(| 0,0,0;4,0,1;0,0,1 \rangle + 
         \, | 0,0,0;0,0,1;4,0,1 \rangle )\Next
&& + g   \, | 0,0,0;0,0,1;0,0,1 \rangle.
\eea
Using the explicit formulae for the corresponding polynomials,
we solve for the coefficients $a, \ldots, g$. The new reduced
matrix elements we need are $F_{44}$, $F_{42}$ and $F_{40}$ for
the operators with $(\Ltil,\ltil_s,\ltil_r)$
respectively $(4,4,0)$, $(4,2,0)$ and $(4,0,0)$. 
The function $F_{40}$ can be obtained from $F_4$ using the relation
\beq
\inv_4(\chat) = \sfrac{1}{22} + \sfrac{1}{22} \sqrt{\sfrac{5}{39}} 
  \langle \chat |4;0,0,1;0,0 \rangle.
\eeq
The result is
\bea
&& \terug
\langle j',m',i'_1,i'_2 |  \inv^2_2(\chat,\dhat) | j,m,i_1,i_2 \rangle = 
\dl_{j' j} \dl_{m' m} 
   \left\{ \frac{A}{\left[ (2 l'_s + 1) (2 l_s + 1) \right]^\hf} 
   + B \right\},
\eea
with
\bea
A &=&  \sfrac{2}{11583} F_{44}(i'_1,i_1) F_{44}(i'_2,i_2) 
\Next && 
+ \sfrac{4}{50193} F_{42}(i'_1,i_1) F_{42}(i'_2,i_2) 
-   \sfrac{4}{1859} F_{22}(i'_1,i_1) F_{22}(i'_2,i_2)
\Next &&  
- \sfrac{1}{1859} \sqrt{\sfrac{56}{2187}}
      ( F_{42}(i'_1,i_1) F_{22}(i'_2,i_2) +
        F_{22}(i'_1,i_1) F_{42}(i'_2,i_2) ), \\
B &=& \sfrac{16}{45} F_4(i'_1,i_1) F_4(i'_2,i_2)
  - \sfrac{4}{135} (F_4(i'_1,i_1) \dl(i'_2,i_2)
    + \dl(i'_1,i_1) F_4(i'_2,i_2) )   
\Next && + \sfrac{2}{135} \dl(i'_1,i_1) \dl(i'_2,i_2).
\eea

\newpage

\clearpage \newpage 

\clearpage \newpage 

\clearpage \newpage 

\clearpage \newpage 

\clearpage \newpage 

\clearpage \newpage 

\clearpage \newpage 

\clearpage \newpage 

\clearpage \newpage 

\clearpage \newpage 

\clearpage \newpage 

\end{document}